\documentclass[]{pasj01}

\Received{}
\Accepted{}
 
\usepackage{bm}
\usepackage{natbib}
\usepackage[switch,mathlines]{lineno}

\newcommand{\cs}{c_\mathrm{s}}
\newcommand{\au}{\mathrm{au}}
\newcommand{\Tmid}{T_\mathrm{mid}}
\newcommand{\OmegaK}{\Omega_\mathrm{K}}

\newcommand{\epsilonmid}{\epsilon_\mathrm{mid}}
\newcommand{\etaO}{\eta_{\mathrm{O}}}
\newcommand{\etaA}{\eta_{\mathrm{A}}}
\newcommand{\LambdaO}{\Lambda_{\mathrm{O}}}
\newcommand{\LambdaA}{\Lambda_{\mathrm{A}}}

\newcommand{\tK}{t_{\mathrm{K0}}}
\newcommand{\zb}{z_{\mathrm{b}}}
\newcommand{\zatm}{z_{\mathrm{atm}}}
\newcommand{\Rmri}{R_{\mathrm{MRI}}}
\newcommand{\Rbuf}{R_{\mathrm{buf}}}
\newcommand{\Rch}{R_{\mathrm{ch}}}
\newcommand{\Reta}{R_{\eta}}

\newcommand{\avphi}[1]{\langle #1 \rangle}

\newcommand{\dmachR}{ \langle \delta {\cal M}_R\rangle }
\newcommand{\dmachphi}{ \langle \delta {\cal M}_\phi\rangle }
\newcommand{\dmachz}{ \langle \delta {\cal M}_z\rangle }

\begin{document} 

\title{ 
Dynamics Near the Inner Dead-Zone Edges in a Proprotoplanetary Disk
}

\author{Kazunari \textsc{Iwasaki}\altaffilmark{1}%
}
\altaffiltext{1}{Center for Computational Astrophysics, 
National Astronomical Observatory of Japan, Mitaka, Tokyo 181-8588, Japan}
\email{kazunari.iwasaki@nao.ac.jp}

\author{Kengo \textsc{Tomida}\altaffilmark{2}}
\altaffiltext{2}{
Astronomical Institute, Tohoku University, Sendai, Miyagi, 980-8578, Japan
}

\author{Shinsuke \textsc{Takasao}\altaffilmark{3}}
\altaffiltext{3}{
Department of Earth and Space Science, Osaka University, 
Machikaneyama-cho, Toyonaka, Osaka 560-0043, Japan}

\author{Satoshi \textsc{Okuzumi}\altaffilmark{4}}
\altaffiltext{4}{Department of Earth and Planetary Sciences, Tokyo Institute of Technology, 2-12-1 Ookayama, Meguro,
Tokyo 152-8551, Japan}

\author{Takeru K. \textsc{Suzuki}\altaffilmark{5}}
\altaffiltext{5}{School of Arts \& Sciences, University of Tokyo, 3-8-1, Komaba, Meguro, Tokyo, 153-8902, Japan}


\maketitle

\begin{abstract}
We perform three-dimensional global non-ideal magnetohydrodynamic 
simulations of a protoplanetary disk 
containing the inner dead-zone edge.
We take into account realistic diffusion coefficients of the Ohmic resistivity and 
ambipolar diffusion based on detailed chemical reactions with single-size dust grains.
We found that
the conventional dead zone identified by the Els\"asser numbers of the Ohmic resistivity and ambipolar diffusion 
is divided into two regions: "the transition zone" and "the coherent zone".
The coherent zone has the same properties as the conventional dead zone, 
and extends outside of the transition zone in 
the radial direction. Between the active and coherent zones, 
we discover the transition zone whose inner edge is identical to that of the conventional dead-zone.
The transition zone extends out over the regions where 
thermal ionization determines diffusion coefficients. 
The transition zone has completely different physical properties 
than the conventional dead zone, the so-called undead zone, and zombie zone.
The combination of amplification of the radial magnetic field
owing to the ambipolar diffusion 
and a steep radial gradient of the Ohmic diffusivity 
causes the efficient evacuation of the net vertical magnetic flux from the transition zone
within several rotations.
Surface gas accretion occurs in the coherent zone but not in the transition zone. 
The presence of the transition zone prohibits mass and magnetic 
flux transport from the coherent zone to the active zone.
Mass accumulation occurs at both edges of the transition zone as a result of mass supply
from the active and coherent zones.

\end{abstract}


\section{Introduction}

The evolution of protoplanetary disks (PPDs) is controlled by 
angular momentum transfer.
One of the important angular momentum transfer mechanisms 
is the magnetorotational instability \citep[MRI,][]{Velikhov1959,Cha1961,BalbusHawley1991}.
The turbulence driven by MRI transfers angular momentum radially.
MRI has been extensively investigated in local shearing-box 
simulations \citep[e.g.,][]{Hawley1995} and in global 
simulations \citep{Armitage1998,Hawley2011,Sorathia2012,Suzuki2014,Takasao2018,ZhuStone2018,Jacquemin-Ide2021}.

Low ionization degrees in PPDs induce the non-ideal MHD effects 
and affect the growth of the MRI.
There are three non-ideal MHD effects, Ohmic resistivity (OR), 
Hall effect (HE), and ambipolar diffusion (AD).
The regions where non-ideal MHD effects suppress the MRI are called 
dead zones \citep{Gammie1996}.
Ionization fractions sufficient to drive the MRI are provided 
only in innermost and outermost radii and upper atmosphere of PPDs owing 
to thermal ionization, far-ultraviolet and X-ray 
radiation from the central star, and cosmic rays.

The angular momentum transfer in the inner regions ($R\lesssim 10~$au) of PPDs where 
OR plays an important role was 
investigated in \citet{Gammie1996}. He showed that 
so-called layered accretion is driven around the disk surfaces where MRI is active
while the MRI is completely suppressed in the disk interior.
Detailed linear analyses of the MRI modified by OR were done by 
\citet{Jin1996} and \citet{Sano1999}.
MRI-driven turbulence above the dead zone was also studied in local simulations 
\citep{FlemingStone2003,TurnerSano2008,HiroseTurner2011}

When AD operates in addition to OR, 
the MRI is suppressed even in the disk surface and 
the layered accretion disappears 
because AD works in higher latitudes than OR \citep{BaiStone2013}.
Instead of the layered accretion, 
the $\phi z$ component of the magnetic stress owing to global coherent magnetic fields drives 
laminar surface gas accretion just above the dead zone in 
the inner region of PPDs where both OR and AD are important \citep{BaiStone2013,Bai2013,Gressel2015}.

The inner dead-zone edge, inside which the thermal ionization activates the MRI, 
has attracted attention as a dust accumulation site.
In such a region, a local pressure bump is created
because the outer part of the active zone moves outward owing 
to a negative turbulent viscosity gradient.
A pressure bump at the inner dead-zone edge traps dust particles 
in the dead zone that are moving to the pressure bump \citep{Kretke2009,Ueda2019}.

The inner dead-zone edge has been investigated with global simulations 
taking OR into account.
\citet{Dzyurkevich2010} found that a pressure bump develops at the inner dead-zone edge.
\citet{Lyra2012} pointed out that the Rossby-wave instability \citep{Lovelace1999} induces 
the formation of vortices at the pressure bump. 
\citet{Flock2017} performed realistic non-ideal MHD simulations with radiation transfer, and 
found that a vortex forms at the inner dead-zone edge that casts a nonaxisymmetric shadow on the outer disk.

In the global simulations containing the inner dead-zone edge performed by 
\citet{Dzyurkevich2010} and \citet{Lyra2012}, 
the angular momentum of the disk is transferred by 
the MRI turbulence above the dead zone \citep{Gammie1996}.
They observed that the height-averaged mass accretion stress does not change 
significantly across the dead-zone edge.
Although sudden decreases in the accretion stresses are found across the inner 
dead-zone edge in \citet{Flock2017}, 
their range of the polar angle may not be wide enough for the MRI to occur above the dead zone.
In addition, in their simulations, the spatial distributions of the OR coefficient are 
given by simplified analytic formulae with a sharp transition across the inner dead-zone edge.

Inclusion of AD changes the angular momentum transfer mechanism in the dead zone from the MRI turbulence.
The MRI above the dead zone 
will be suppressed once AD is considered \citep{BaiStone2013,Gressel2015,Bai2017}.
In a realistic situation where both OR and AD work, 
the angular momentum transfer mechanism changes from the MRI turbulence to 
the magnetic stress due to global coherent magnetic field across the inner dead-zone edge.
When both OR and AD are considered, the spatial distribution of the accretion stress across the inner dead-zone edge is still unclear.

In this paper, we perform global three-dimensional non-ideal MHD simulations 
of a protoplanetary disk around an intermediate mass star
to reveal the structure around the inner dead-zone edge by 
taking into account both OR and AD whose coefficients are given based on detailed chemistry.
In the previous global simulations, 
a limited range of $\theta$ in the spherical polar coordinates of $(r,\theta,\phi)$ 
is considered, which 
may affect surface gas accretion expected to be driven above the dead zone \citep{BaiStone2013} and 
launching of disk winds both from the active zone \citep{Suzuki2009,Suzuki2014} 
and the dead zone \citep{BaiStone2013}.
Our simulation box covers a full solid angle ($0\le \theta \le \pi$ and $0\le \phi <2\pi$) to 
capture the angular momentum transfer in the upper atmosphere and vortex formation.

This paper is organized as follows.
We will describe the numerical setup and method in Section \ref{sec:method}. 
The main results are presented in  Section \ref{sec:results}.
In Section \ref{sec:mass_angmom_flux},
we will present 
detailed analyses on transfer of the mass, angular momentum, and 
magnetic flux in the disk.
Astrophysical implications are discussed in Section \ref{sec:discuss}.
Our findings are summarized in Section \ref{sec:summary}.

\section{Numerical Setup and Methods} \label{sec:method}

\subsection{Basic Equations} \label{sec:basiceq}

The resistive MHD equations with OR and AD are given by 
\begin{equation}
        \frac{\partial \rho}{\partial t} + \bm{\nabla} \cdot \left( \rho\bm{v} \right)=0
\label{eoc}
\end{equation}
\begin{equation}
\frac{\partial \rho \bm{v}}{\partial t} 
+ \bm{\nabla} \cdot \left( \rho \bm{v}\otimes \bm{v} + {\bf T} \right) = 
- \rho \bm{\nabla} \psi,
\end{equation}
\begin{equation}
\frac{\partial \bm{B}}{\partial t} = 
\mathbf{\nabla}\times \left( \bm{v}\times \bm{B} \right) 
 - \mathbf{\nabla}\times \left[ \etaO \bm{J}
+ \etaA \bm{J}_\perp\right]
\label{induc}
\end{equation}
and
\begin{equation}
        {\bf T} = 
        \left( P + \frac{B^2}{8\pi} \right){\bf I} - \frac{\bm{B}\otimes \bm{B}}{4\pi},
     \label{Tij}
\end{equation}
where $\rho$ is the gas density, $\bm{v}$ is the gas velocity, 
$\bm{B}$ is the magnetic field, 
$\psi$ is the gravitational potential of the central star,
$\bm{J}$ is the current density, 
$\bm{J}_\perp \equiv  \bm{J} - \bm{J}\cdot\bm{B}/|\bm{B}|$ 
is the electric current density perpendicular to the local magnetic field,
$P$ is the gas pressure, ${\bf T}$ is the stress tensor, 
${\bf I}$ is the identity matrix,
$\etaO$ and $\etaA$ are the diffusion coefficients 
for OR and AD, respectively.  
They are defined in Section \ref{sec:diffcoeff}.
For simplicity, the Hall effect is neglected in this work.

In this paper, for simplicity we use the locally isothermal equation of state where the gas 
temperature remains the same as the initial one \citep{Suzuki2014,ZhuStone2018}.
To achieve this, we solve the MHD equations including the energy equation with $\gamma=5/3$, and 
the gas pressure is derived from the locally isothermal condition where the temperature distribution is shown in Equation (\ref{tem}).
We note that a locally isothermal equation of state potentially drives the vertical shear
instability (VSI) \citep{Nelson2013}. 
In our simulations, the VSI does not grow because of insufficient 
simulation time as discussed in Section \ref{sec:VSI}.

\subsection{Models and Methods}
\subsubsection{Initial Condition}
The initial surface density profile of the disk is given by 
\begin{equation}
    \Sigma_\mathrm{ini}(R) = \Sigma_0 \left( \frac{R}{1~\mathrm{au}} \right)^{-1}
        \label{sigma}
\end{equation}
where $\Sigma_0 = 500~\mathrm{g~cm^{-2}}$ and $R$ is the cylindrical radius.
The disk mass integrated from $R=0$ to $R$ 
is given by 
\begin{equation}
M_\mathrm{d}(R)=2.35\times 10^{-4}~M_\odot~\left(\frac{\Sigma_0}{500~\mathrm{g~cm^{-2}}}\right)
\left(\frac{R}{1~\mathrm{au}}\right)^{1.5}.
\end{equation}

The mid-plane gas temperature $\Tmid$ is determined by assuming radiative equilibrium between 
the gas and radiation from the central star.
We consider a central star with $M_*=2.5~M_\odot$ of mass, $R_*=2.5~R_\odot$ of radius, 
and $T_*=10^4~\mathrm{K}$ of effective temperature, which corresponds to a Herbig Ae star.
The mid-plane gas temperature is given by 
\begin{equation}
        \Tmid(R) = 760~\mathrm{K}~f_\mathrm{boost}\left( \frac{L_*}{56L_\odot}\right)^{1/4}\left(\frac{R}{1~\mathrm{au}}\right)^{-1/2},
        \label{temmid}
\end{equation}
where $f_\mathrm{boost}$ is a coefficient which is set to 2 
to increase the aspect ratio of the disk, 
$L_*=4\pi R_*^2 \sigma_\mathrm{SB} T_*^4$ is the luminosity of the central star, and $\sigma_\mathrm{SB}$
is the Stefan-Boltzmann constant.
The adopted stellar luminosity is typical in Herbig Ae stars with ages larger than 1~Myr although 
the luminosities of Herbig Ae stars with ages less than 1~Myr are highly uncertain \citep[e.g.,][]{Montesinos2009}.
The scale height is given by $H(R) = \sqrt{2}c_\mathrm{s,mid}/\Omega_\mathrm{K}(R)$,
where $c_\mathrm{s,mid}=\sqrt{k_\mathrm{B} T_\mathrm{mid}/\mu m_\mathrm{H}}$ is the sound speed at the mid-plane,
$\mu=2.33$ is the mean molecular weight, $m_\mathrm{H}$ is the hydrogen mass, and 
$\OmegaK = \sqrt{GM_*/R^3}$ is the Kepler angular velocity of the disk.
The aspect ratio of the disk is given by 
\begin{equation}
    \epsilonmid(R) = \frac{H(R)}{R} = 0.049 f_\mathrm{boost}^{1/2}\left(\frac{R}{1~\mathrm{au}}\right)^{1/4},
        \label{emid}
\end{equation}
The initial mid-plane density is given by 
\begin{eqnarray}
    \rho_\mathrm{mid,ini}(R) &=& 5.4\times 10^{-10}~\mathrm{g~cm^{-3}} \nonumber \\
   && \times f_\mathrm{boost}^{-1/2} \left(\frac{\Sigma_0}{500~\mathrm{g~cm^{-2}}}\right) \left(\frac{r}{1~\au}\right)^{-9/4}.
\end{eqnarray}

We consider a similar temperature distribution as \citet{Bai2017}.
The whole region is divided into a cold gas disk and warm atmospheres 
sandwiching the cold disk.
Inside the disk, the temperature is given by 
$T_\mathrm{d}(r) = (\mu m_\mathrm{H}/k_\mathrm{B})(\epsilonmid(r)^2/2) r^2\OmegaK(r)^2$
which is identical to $T_\mathrm{mid}$ (Equation (\ref{temmid})) at the mid-plane.
In the warm atmosphere well above the disk, the gas is heated up mainly by 
the radiation from the central star.
We define $\Sigma_\mathrm{FUV}$ as the column density below which 
the FUV photons from the central star penetrate.
If the radial column density $\Sigma_r(r,\theta) = \int_{r_\mathrm{min}}^r \rho(r',\theta)dr'$ is smaller than 
$\Sigma_\mathrm{FUV}$, the gas temperature increases to 
$T_\mathrm{atm}(r) = (\mu m_\mathrm{H}/k_\mathrm{B})(\epsilon_\mathrm{atm}(r)^2/2) r^2\OmegaK(r)^2$,
where $\epsilon_\mathrm{atm}(r)=0.2 (r/1~\au)^{1/4}$.
The initial temperature profile smoothly connects between $T_\mathrm{d}$ and $T_\mathrm{atm}$ 
around $\Sigma_r \sim  \Sigma_\mathrm{FUV}$
as follows:
\begin{equation}
T(r,\theta) = (\mu m_\mathrm{H}/k_\mathrm{B})(\epsilon(r)^2/2) r^2\OmegaK(r)^2
        \label{tem}
\end{equation}
and
\begin{eqnarray}
    \epsilon(r,\theta) &=& \frac{\epsilonmid(r)+\epsilon_\mathrm{atm}(r)}{2}\nonumber \\
    &+& \frac{\epsilonmid(r) - \epsilon_\mathrm{atm}(r)}{2} \tanh\left[ 0.3\log \left( \frac{\Sigma_\mathrm{FUV}}{\Sigma_r(r,\theta)} \right) \right].
\end{eqnarray}
The values of $\Sigma_\mathrm{FUV}$ are highly uncertain, and 
\citet{Perez-BeckerChian2011} found that 
FUV layers have a thickness of $\sim 0.01-0.1$~g~cm$^{-2}$
in plane-parallel models. 
We adopt $\Sigma_\mathrm{FUV}=0.3~\mathrm{g~cm^{-2}}$. 
It is three times larger than the upper limit of that of \citet{Perez-BeckerChian2011} 
because the radial column density is considered instead of the vertical column density \citep{Bai2017}.

The aspect ratio for $R<1~$au is too small to resolve the MRI structure.
As explained below Equation (\ref{temmid}), we artificially 
increase the aspect ratio by increasing the gas temperature by a factor of two, or $f_\mathrm{boost}=2$.
As mentioned later, for the calculation of $\etaO$ and $\etaA$, 
$T f_\mathrm{boost}^{-1}$ is used as the gas temperature.

\begin{figure}
    \begin{center}
    \includegraphics[width=8.5cm]{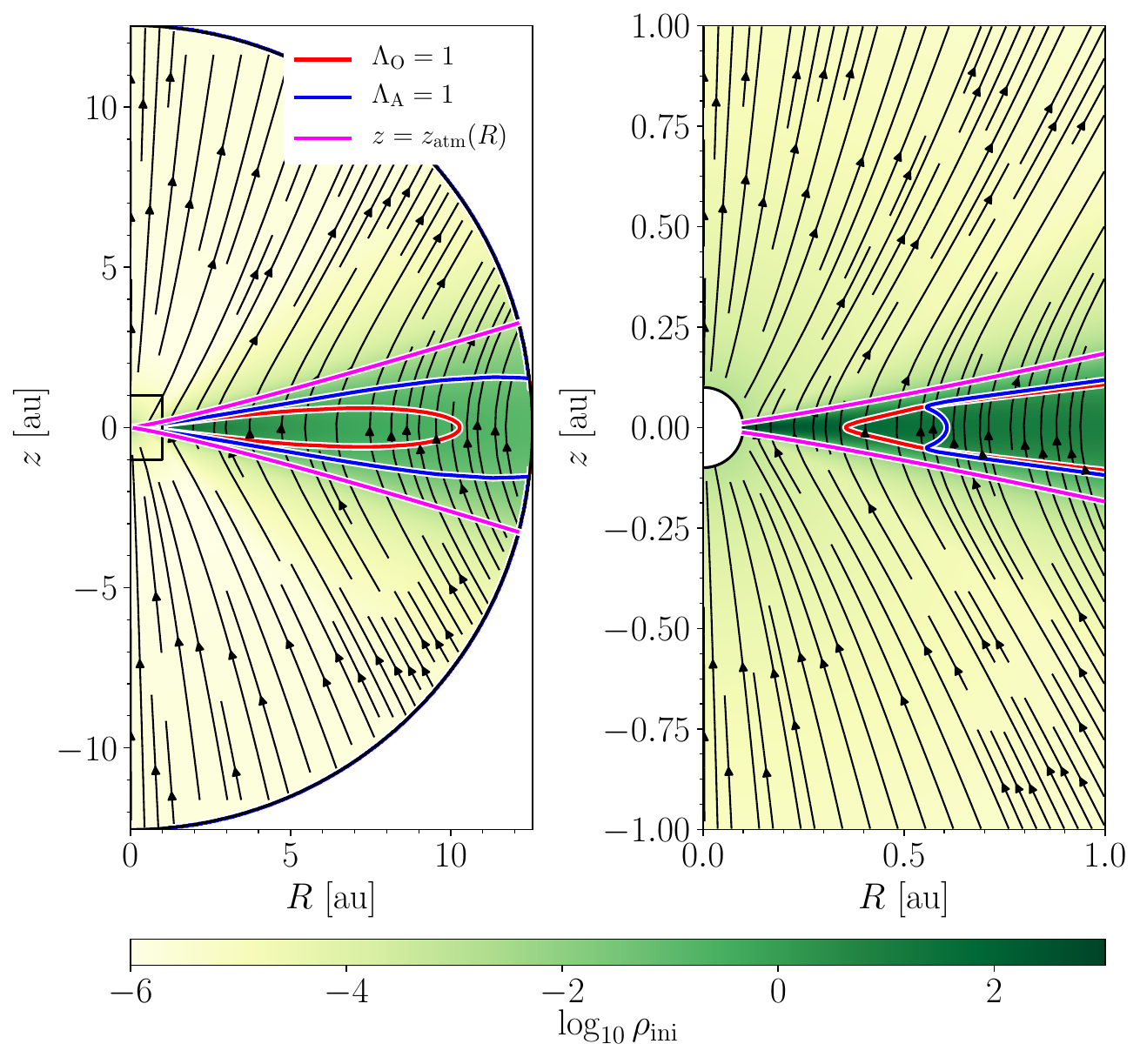}
    \end{center}
    \caption{
        (Left) Initial setup of the simulation.
        The color map indicates the initial density profile in the code units, and the black lines indicate 
        the streamlines of the magnetic fields.
        The red (blue) line corresponds to the condition $\Lambda_\mathrm{O}=1$ ($\Lambda_\mathrm{A}=1$), which 
        show the OR (AD) dead zone boundary.
        The magenta lines show $z=\pm\zatm(R)$ above which the warm atmospheres exist.
        The right panel is the same as the left panel but it shows the zoom-in of the region enclosed by the 
        black rectangle shown in the 
        left panel.
    }
    \label{fig:init}
\end{figure}

The initial profiles of the density and $\phi$ component of the velocity 
are constructed using an iterative manner to satisfy equations (\ref{sigma}) and 
(\ref{tem}) simultaneously in the hydrostatic equilibrium between 
the pressure-gradient force, gravitational force, and centrifugal force without the Lorentz force.
The initial poloidal velocities are set to be zero.
The initial magnetic field has only the poloidal components which have 
an hourglass-shape field given by \citet{Zanni2007}.
It is described by the $\phi$ component of the vector potential,
\begin{equation}
    A_\phi(r,\theta) = 
    Cr^{-a}\left[ m^2 + \left( \frac{z}{r} \right)^{2} \right]^{-5/8},
\end{equation}
where the coefficient $C$ is determined so that the mid-plane plasma beta 
takes a constant value of $\beta_0=10^{4}$.
The initial condition of our simulation is displayed in Figure \ref{fig:init}.

A height of the base of the atmospheres at a given $R$ is denoted by $z=\zatm(R)$ above which 
the gas temperature is more than twice the mid-plane temperature (Figure \ref{fig:init}).

\subsubsection{Diffusion Coefficients}\label{sec:diffcoeff}

The diffusion coefficients $\etaO$ and $\etaA$ are given by 
\begin{equation}
        \etaO = \frac{c^2}{4\pi} \frac{1}{\sigma_\mathrm{O}}\;\;\mathrm{and}\;\;
        \etaA = \frac{c^2}{4\pi} \left(\frac{\sigma_\mathrm{P}}{\sigma_\mathrm{P}^2 + \sigma_\mathrm{H}^2} - \frac{1}{\sigma_\mathrm{O}}\right),
\end{equation}
where $\sigma_\mathrm{O}$, $\sigma_\mathrm{H}$, and $\sigma_\mathrm{P}$ are the Ohmic, Hall, and Pedersen conductivities, respectively, and $c$ is the speed of light.
The three conductivities are given by 
\begin{equation}
        \sigma_\mathrm{O} = \frac{ec}{B} \sum_j n_j |Z_j| \beta_j,
        \label{sigmaO}
\end{equation}
\begin{equation}
        \sigma_\mathrm{H} = -\frac{ec}{B} \sum_j \frac{n_j Z_j \beta_j^2}{1+\beta_j^2},
        \label{sigmaH}
\end{equation}
and 
\begin{equation}
        \sigma_\mathrm{P} = \frac{ec}{B} \sum_j \frac{n_j |Z_j|\beta_j}{1+\beta_j^2},
        \label{sigmaP}
\end{equation}
where the subscript ``$j$'' denotes the species of charged particles, $Z_j e$ is the charge and $n_j$ is the 
density of the charged particle $j$.
The Hall parameter $\beta_j$ is the ratio between the gyrofrequency of the charged particle $j$ and 
its collision frequency with the neutral particles, and is defined as
\begin{equation}
    \beta_j = \frac{|Z_j| eB}{m_j c}\frac{1}{\gamma_j \rho},
\end{equation}
where $m_j$ is the mass of the charged particle $j$ and 
$\gamma_j$ is the rate coefficient for momentum transfer with the neutral particles through collisions.

The three conductivities are computed by the summation between the conductivities calculated with thermal and non-thermal ionization 
instead of calculating a chemical network including both thermal and non-thermal ionization,
as follows:
\begin{equation}
        \sigma_i = \sigma_{i,\mathrm{T}} + \sigma_{i,\mathrm{NT}},
\end{equation}
where $i=(\mathrm{O,H,P})$ and the subscripts ``T'' and ``NT'' indicate 
the thermal and non-thermal ionization limits of the conductivities, respectively.

Since the thermal ionization is important in the regions of sufficiently high temperatures, 
dust grains are assumed to be sublimated in the calculation of chemical reactions.
We consider the thermal ionization of potassium, which has an ionization energy of 3.43~eV. 
The potassium abundance relative to the H$_2$ number density is $3\times 10^{-7}$.
The thermal ionization degree $(x_e)_\mathrm{T}$ is derived by the Saha equation where 
the electron densities are assumed to be equal to the potassium ion densities. 
As mentioned before, $T$ is replaced with $Tf_\mathrm{boost}^{-1}$ in the Saha equation.
When the gas temperature exceeds $\sim 1000~$K, the thermal ionization 
of potassium provides a sufficient high ionization and the region becomes MRI active.

In order to take into account the photo-ionization of carbon and sulfur due to FUV photons in the 
disk atmosphere, we replaced the ionization degree calculated in the thermal ionization limit by 
\begin{equation}
        \max\left[ \left( x_e \right)_\mathrm{T}, 
        10^{-4} \exp\left\{ -\left( \frac{\Sigma_r(r,\theta)}{\Sigma_\mathrm{FUV}} \right)^4 \right\} \right],
\end{equation}
where $\left( x_e \right)_\mathrm{T}$ is the ionization degree obtained from the Saha equation for potassium ionization and 
the second term indicates that the electrons are provided by
photo-ionization of carbon and sulfur when $\Sigma_r < \Sigma_\mathrm{FUV}$ \citep{Bai2017}.

For the non-thermal ionization, we use the data table of the three conductivities based on 
a chemical reaction network of e$^-$, H$^+$, He$^+$, C$^+$, H$_3^+$, HCO$^+$, Mg$^+$ in the 
gas phase and the charged dust grains using the methods described by  
\citet{Nakano2002} and \citet{Okuzumi2009}. The table is a function of 
$\rho/B$, $T$, and $\rho/\zeta$, where $\zeta$ is the ionization rate that will be defined in 
Equation (\ref{zeta}) and $T$ is replaced with $Tf_\mathrm{boost}^{-1}$.
For simplicity, the single size dust grain with a radius of 
$a=0.1~\mu$m 
and an internal density of $2~\mathrm{g~cm^{-3}}$ is considered. 
In this paper, the dust-to-gas mass ratio is set to $10^{-4}$, assuming that dust growth 
reduces the amount of dust grains smaller than the interstellar value $\sim 0.01$.
The following three kinds of non-thermal ionization sources are taken into account,
\begin{equation}
    \zeta(r,\theta) = \zeta_\mathrm{CR}(r,\theta)
    + \zeta_\mathrm{X}(r,\theta)
    + \zeta_\mathrm{radio}.
    \label{zeta}
\end{equation}
The first source is the cosmic rays whose ionization rate is given by 
\begin{equation}
        \zeta_\mathrm{CR} = 10^{-17}~\mathrm{s}^{-1} \left[ \exp\left( -
        \frac{\Sigma_\theta^+}{\Sigma_\mathrm{CR}}\right)
+ \exp\left( - \frac{\Sigma_\theta^-}{\Sigma_\mathrm{CR}} \right)\right],
\end{equation}
where $\Sigma_\mathrm{CR}=96~$g~cm$^{-2}$ \citep{Umebayashi1981},
and $\Sigma_\theta^+(r,\theta)$ ($\Sigma_\theta^-(r,\theta)$) is the column density
integrated along the $\theta$-direction from $\theta=0$ ($\theta=\pi$) to $\theta$ at a given $r$.
Secondly, 
we consider ionization due to the X-ray radiation from the central star 
using the fitting formula \citep[][based on the calculation done by \citet{IgeaGlassgold1999}]{BaiGoodman2009} 
as follows:
\begin{eqnarray}
  \zeta_\mathrm{X} &=& \left( \frac{r}{2.2~\au} \right)^{-2.2}
  \left[ \zeta_1 e^{-(\Sigma_r/5\Sigma_\mathrm{X,a})^\alpha} + \right. \nonumber \\
          && \hspace{10mm} \left. \zeta_2 \left\{ e^{-(\Sigma_\theta^+/\Sigma_\mathrm{X,s})^\beta} +
                          e^{-(\Sigma_\theta^-/\Sigma_\mathrm{X,s})^\beta} 
  \right\}
  \right],
\end{eqnarray}
where $\zeta_1 = 6.0\times 10^{-11}$~s$^{-1}$, $\Sigma_\mathrm{X,a}=3.6\times10^{-3}$~g~cm$^{-2}$, 
$\alpha=0.4$, 
and $\Sigma_r(\theta,r)$ is the column density
integrated along the $r$-direction from 
the innermost radius of the simulation box to $r$ at a given $\theta$.
The third ionization source is provided by radioactive nuclei,
\begin{equation}
    \zeta_\mathrm{radio} = 6.0\times 10^{-19}~\mathrm{s}^{-1}
\end{equation}
\citep{FinocchiGail1997}.
To reduce the computational cost, $\zeta(r,\theta)$ is calculated for the initial condition and fixed during the simulations.

\subsection{Basic Properties of the Diffusion Coefficients}
\subsubsection{Definition of the Dead Zone Boundaries}

The non-ideal MHD effects weaken the growth rate of the MRI when
the dimensionless Els\"asser numbers 
\begin{equation}
    \Lambda_\mathrm{O} \equiv \frac{v_\mathrm{A}^2}{\eta_\mathrm{O}\Omega_\mathrm{K}},\;\;\;
    \Lambda_\mathrm{A} \equiv \frac{v_\mathrm{A}^2}{\eta_\mathrm{A}\Omega_\mathrm{K}},\;\;\;
\end{equation}
are smaller than unity, where $v_\mathrm{A}=B/\sqrt{4\pi \rho}$ is the Alfv\'en speed
\citep{Sano1999,BlaesBalbus1994,Wardle1999,KunzBalbus2004,BaiStone2011}.

In order for the MRI to occur at a given height $z$, the wavelength of the maximum growing mode 
measured locally should be smaller than the scale height \citep{Sano1999}.
Using the expressions of the most unstable wavelengths $\lambda_\mathrm{max,O}$ 
and $\lambda_\mathrm{max,A}$ given by \citet{Sano1999} and \citet{BaiStone2011}, respectively,
we confirmed that $\lambda_\mathrm{max,O}>H$ and 
$\lambda_\mathrm{max,A}>H$ are satisfied in most regions where 
$\LambdaO<1$ and $\LambdaA<1$, respectively, 
because of steep spatial gradients of $\etaO$ and $\etaA$.
This indicates that in the regions just outside the dead zone, 
the MRI turbulence is partly suppressed as seen in 
Section \ref{sec:structureform_transition}.

Hereafter,
the dead-zone boundaries for OR and AD are defined as $\LambdaO=1$ and $\LambdaA=1$, respectively.

\subsubsection{Spatial Distributions of the Diffusion Coefficients at the Mid-plane}

\begin{figure}
    \begin{center}
    \includegraphics[width=8cm]{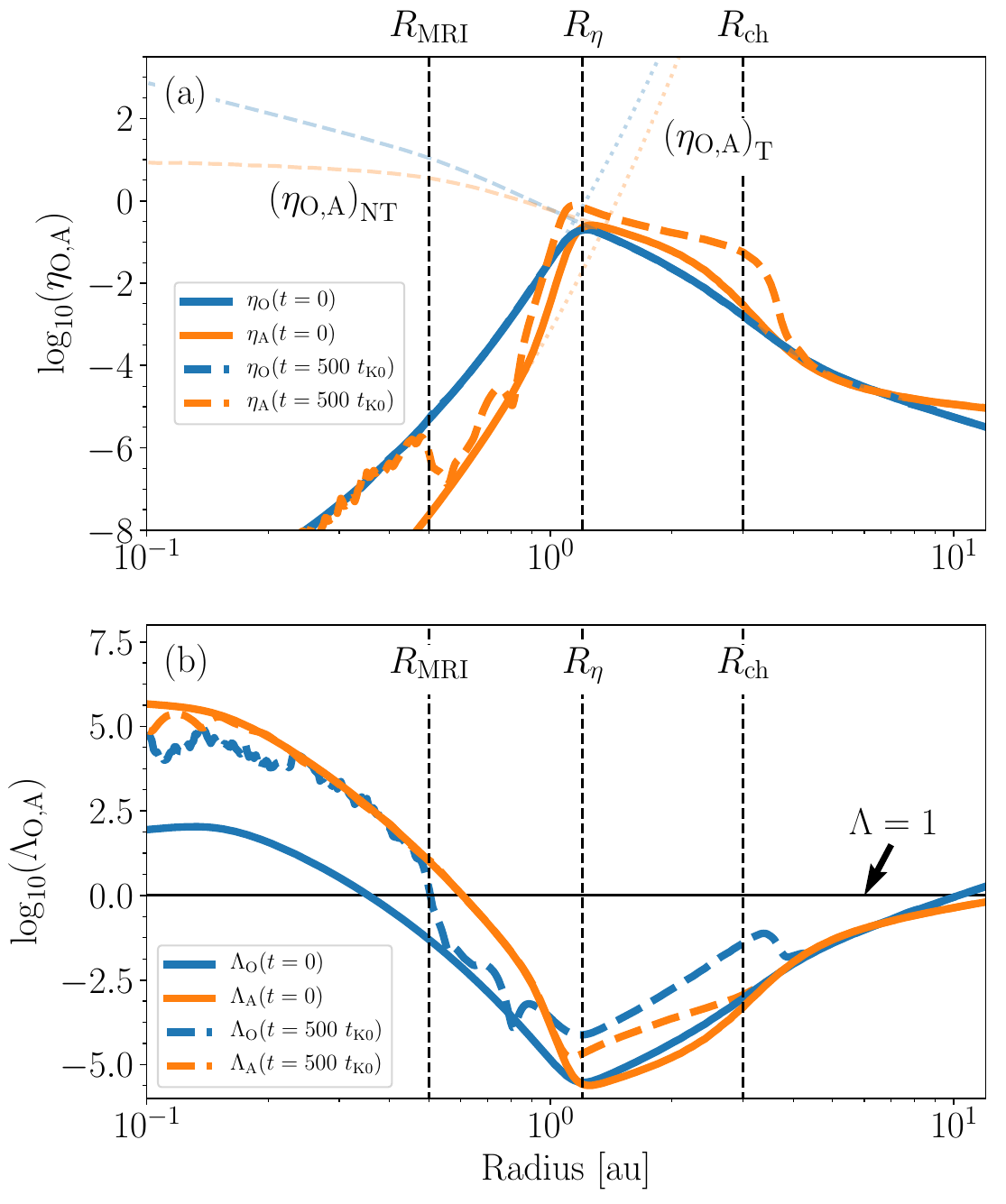}
    \end{center}
    \caption{
            (a) Radial profiles of the diffusion coefficients  
            $\etaO$  and $\etaA$ at 
            the mid-plane at  $t=0$ (solid)
            and  $t=500~\tK$ (dashed) when the MRI turbulence has been saturated.
            The diffusion coefficients are shown in the code units.
            The thin dashed and dotted lines show the diffusion coefficients at $t=0$ derived by 
            chemical reactions with thermal and non-thermal ionization, respectively.
            Panel (b) is the same as Panel (a) but for the Els\"asser numbers $\LambdaO$ and $\LambdaA$ 
            at the mid-plane at $t=0$ and $t=500~\tK$.
            For both panels, 
            the three vertical dashed lines correspond to $\Rmri$, $R_\eta$, and $R_\mathrm{ch}$ from 
            left to right.
    }
    \label{fig:profinimid}
\end{figure}

We present the spatial distributions of $\etaO$ and $\etaA$ in the initial condition.
Figure \ref{fig:profinimid}a shows the radial profiles of $\etaO$ and $\etaA$ at the mid-plane.
Both resistivities show a similar behavior; 
they increase steeply as a function of the radius, 
reach maxima around $R\sim 1~$au, and then decrease steeply.

Here we define three characteristic radii from the radial profiles of $\etaO$, $\etaA$, $\LambdaO$, and 
$\LambdaA$.
The first characteristic radius is $\Rmri$ which is the inner edge of the 
OR dead zone at the mid-plane.
As shown in Figure \ref{fig:profinimid}b, in the initial condition, 
the inner dead-zone edges for OR and AD are 
located in the region where the thermal ionization dominates over the non-thermal ionization.
The inner dead-zone edge for OR 
moves outward until the MRI turbulence is saturated while that for AD
do not since $\LambdaO\propto B^2$ and $\LambdaA\propto B^0$, where 
we use the fact that $\etaO\propto B^0$ and $\etaA\propto B^2$ (Section \ref{sec:etaA_Bdep}).
In Figure \ref{fig:profinimid}b, 
we plot the radial profiles of $\LambdaO$ and $\LambdaA$ at $t=500~\tK$ when the MRI turbulence has been saturated. 
At $R=\Rmri=0.5~\au$, $\LambdaO$ becomes unity, and $\Rmri$ does not change after the MRI turbulence is saturated.
Hereafter, the inner dead-zone edge for OR is simply called the inner dead-zone edge.

The second characteristic radius is $R=\Reta=1.2~\au$ around which 
$\etaO$ and $\etaA$ take the largest values 
and $\LambdaO$ and $\LambdaA$ take the lowest values (Figures \ref{fig:profinimid}a and \ref{fig:profinimid}b).
This corresponds to the radius around which the main ionization process switches from 
the thermal to non-thermal ionization.

The third characteristic radius is $R_\mathrm{ch}=3~$au beyond which 
the main negative charge carrier changes from dust grains to electrons (Section \ref{sec:etaA_Bdep}).

\begin{figure}
    \begin{center}
    \includegraphics[width=9cm]{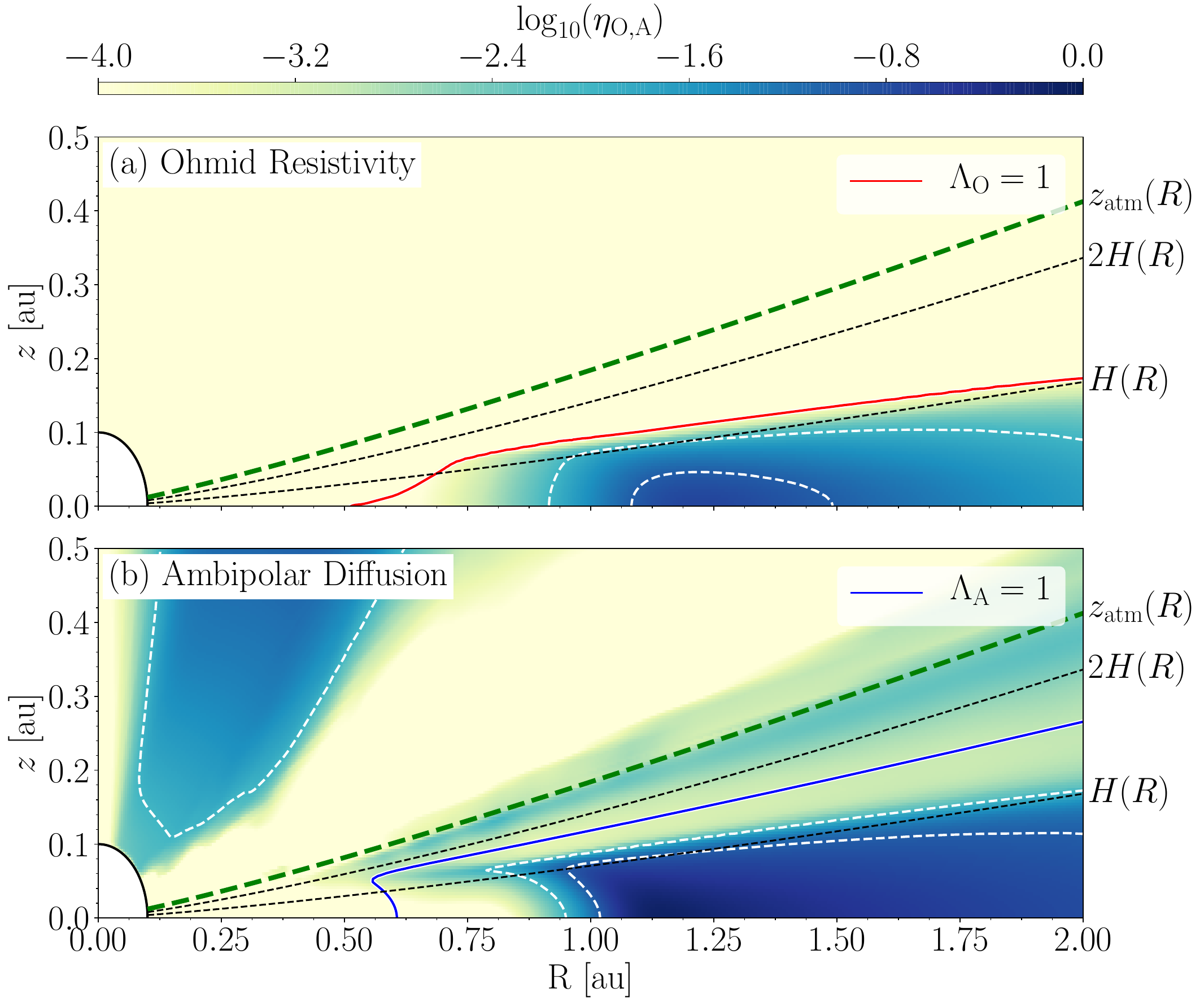}
    \end{center}
    \caption{
        Two-dimensional distributions of (a) $\etaO$ and (b) $\etaA$ in the code units
        at $t=500~\tK$ when the MRI turbulence is saturated.
        The white dashed lines show the contours of $\etaO$ and $\etaA$ at $10^{-4}$ and $10^{-2}$.
        The conditions of $\LambdaO=1$ and $\LambdaA=1$ are displayed by the red and blue lines in Panel (a) and (b), respectively.
        In each panel, the two black dashed lines correspond to $z=H(R)$ and $z=2H(R)$, and 
        the green dashed line show $z=\zatm(R)$ above which the warm atmospheres exist.
    }
    \label{fig:eta2d}
\end{figure}

\subsubsection{Two Dimensional Distributions of the Diffusion Coefficients}

The two-dimensional distributions of $\etaO$ and $\etaA$ in the ($R$,$z$) plane at $t=500~\tK$ 
are displayed in Figures \ref{fig:eta2d}a and \ref{fig:eta2d}b, respectively.
$\etaO$ decreases from the mid-plane toward upper latitudes
because the density decreases.
The spatial distribution of $\etaA$ is different from that of $\etaO$ especially for $R<1~$au, 
where $\etaA$ increases from the mid-plane toward upper latitude in the AD dead zone.
This comes from the fact that AD is important for strong $B$ and/or low 
$\rho$ when dust grains do not play an important role in the resistivities.

The shapes of the dead-zone boundaries around the inner edges reflect 
the physical properties of OR and AD.
As the radius increases from the inner boundary of the simulation box, OR makes the disk dead at the mid-plane first around 
$R\sim \Rmri$ while high latitude regions are the first place which become dead for AD.

The vertical dead-zone boundaries for OR and AD lie between $z=H(R)$ and $z=2H(R)$, 
and the AD dead-zone is more extended vertically than the OR dead-zone.
The base of the warm atmospheres $z=\zatm(R)$ is well above the dead-zone boundaries.

\subsubsection{  The Dependence of the Diffusion Coefficients on the Field Strength}\label{sec:etaA_Bdep}

For OR,  $\etaO$ is independent of $B$ because $B$ in the factor $(ec/B)$ is cancelled 
with $\beta_j \propto B$ in Equation (\ref{sigmaO}).
By contrast, $\etaA$ depends on $B$ in a more complex manner 
than $\etaO$ \citep{XuBai2016}.
If there are no dust grains,
using the fact that $\beta_i \ll \beta_e$, 
one finds that $\etaA
= \beta_i \times c B/(4\pi e n_e)\propto B^2$
\citep{SalmeronWardle2003}, where $\beta_i$ and $\beta_e$ are the Hall parameters of ions and electrons, respectively, and $n_e$ is the electron number density.

\begin{figure}
    \begin{center}
    \includegraphics[width=8cm]{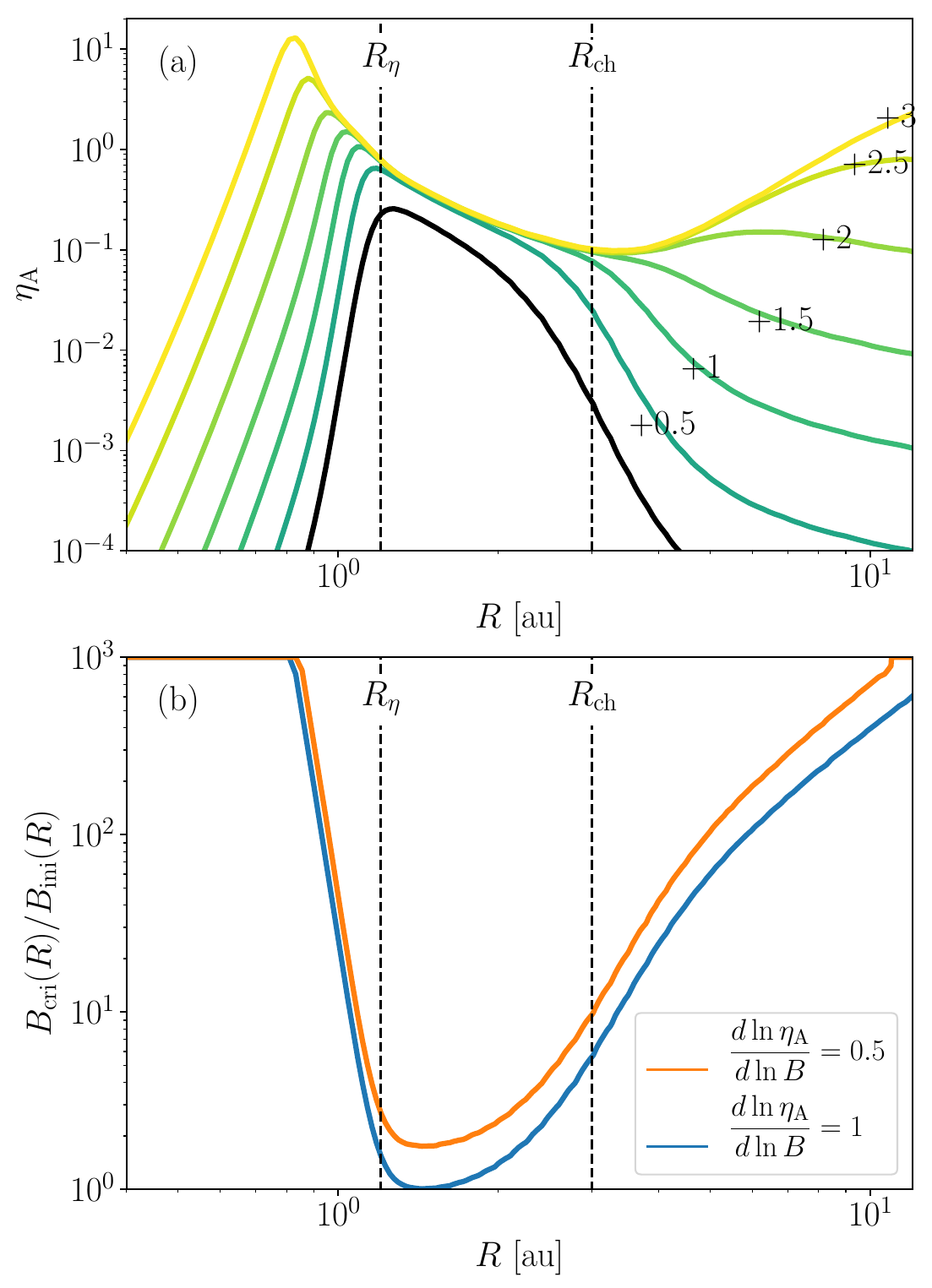}
    \end{center}
    \caption{
        (a) Dependence of $\etaA$ on the field strength.
        The radial profiles of $\etaA$ at the mid-plane are displayed by changing
        the field strengths.
        The values of $\log_{10}(B/B_\mathrm{ini}(R))$ are labelled on the lines.
        The vertical axis is shown in the code units.
        The black line shows the radial profile of $\etaA$ in the initial condition.
  (b) The field strengths $B_\mathrm{cri}$ where $d\ln \etaA/d \ln B = 1.0$ and $0.5$.
  For field strengths larger than $B_\mathrm{cri}$, the $B$-dependence of $\etaA$ becomes weak.
  }
    \label{fig:etadep}
\end{figure}

The dependence of $\etaA$ on the field strength 
is critical in the structure formation of magnetic fields as will be discussed in 
Section \ref{sec:tr_co_discuss}.
Development of sharp structures in the magnetic null due to AD
has reported in \citet{BrandenburgZweibel1994}.
For such a structure to develop, $\etaA$ should increase with $B$ so that 
diffusion near the magnetic null is much more inefficient than in strongly magnetized regions.

Existence of dust grains changes $\etaA$ significantly.
The dependence $\etaA\propto B^2$
is realized at the weak field limit $\beta_e\ll 1$ \citep{XuBai2016}. 
For $\beta_e\gtrsim 1$, $\etaA$ is not necessarily proportional to $B^2$ if the contribution from 
dust grains to the number density of charged particles is non-negligible.
The dependence of $\LambdaA$ on the field strength for various grain sizes and 
dust-to-gas mass ratios are shown in Figure 4 of \citet{XuBai2016}.

Figure \ref{fig:etadep}a shows how the 
radial profile of $\etaA$ at the mid-plane 
changes as the field strength is increased from the initial value $B_\mathrm{ini}(R)$.
As $B/B_\mathrm{ini}$ increases, $\eta_\mathrm{A}$ does not increase in proportional to $B^2$, but 
the increase in $\eta_\mathrm{A}$ is saturated especially in the inner region of the 
dead zone $\Reta \lesssim R \lesssim \Rch$, where 
the dust grains are the main negative charge carrier.
Figure \ref{fig:etadep}b shows the 
field strengths where $d\ln \eta_\mathrm{A}/d\ln B=1$ and $0.5$.
For field strengths larger than these critical field strength,  
the $B$ dependence of $\etaA$ become weak, leading to inefficient formation of substructures 
due to AD.

\subsection{Methods}

To solve the basic equations (\ref{eoc})-(\ref{induc}), 
we use {Athena++} \citep{Stone2020} 
which is a complete
rewrite of the Athena MHD code \citep{Stone2008}. 
The HLLD (Harten-Lax-van Leer-Discontinuities) method is used as the MHD Riemann solver \citep{MK2005}.
Magnetic fields are integrated by the constrained transport method 
\citep{EH1988,Gardiner2008}.
The second-order Runge-Kutta-Legendre super-time stepping 
technique is used to calculate the magnetic diffusion processes \citep{Meyers2014},
where 
the time step is limited so that it is not greater than 30 times the time step
determined by OR and AD.

\begin{table}
    \tbl{List of the models.}{
    \begin{tabular}{|c|c|c|}
    \hline
    Model Name  & Effective Resolution\footnotemark[$*$]  &  $t_\mathrm{end}$\footnotemark[$\dag$] \\
    \hline
       LowRes & $1532\times 1024\times 1024$ & $2000t_\mathrm{K0}$ \\
    \hline
       HighRes & $3064\times 2048\times 2048$ & $500t_\mathrm{K0}$ \\
    \hline
    \end{tabular}
}
    \begin{tabnote}
        \footnotemark[$*$] The total cell number in each direction if the whole computational domain is resolved with the finest level. 
        \footnotemark[$\dag$] The simulations are conducted up to $t_\mathrm{end}$.
    \end{tabnote}
    \label{tab:model}
\end{table}

A spherical-polar coordinate system  $(r,\theta,\phi)$
is adopted in the simulation box 
$r_\mathrm{in}\le r \le r_\mathrm{out}$, 
$0\le \theta \le \pi$, and $0\le \phi < 2\pi$,
where $r_\mathrm{in}=0.1~\mathrm{au}$ and $r_\mathrm{out} = 12.5~\mathrm{au}$. 
The radial cell width increases with radius by a constant factor so that 
the radial cell width is almost identical to the meridional one. 

\begin{figure}
    \begin{center}
    \includegraphics[width=0.45\textwidth]{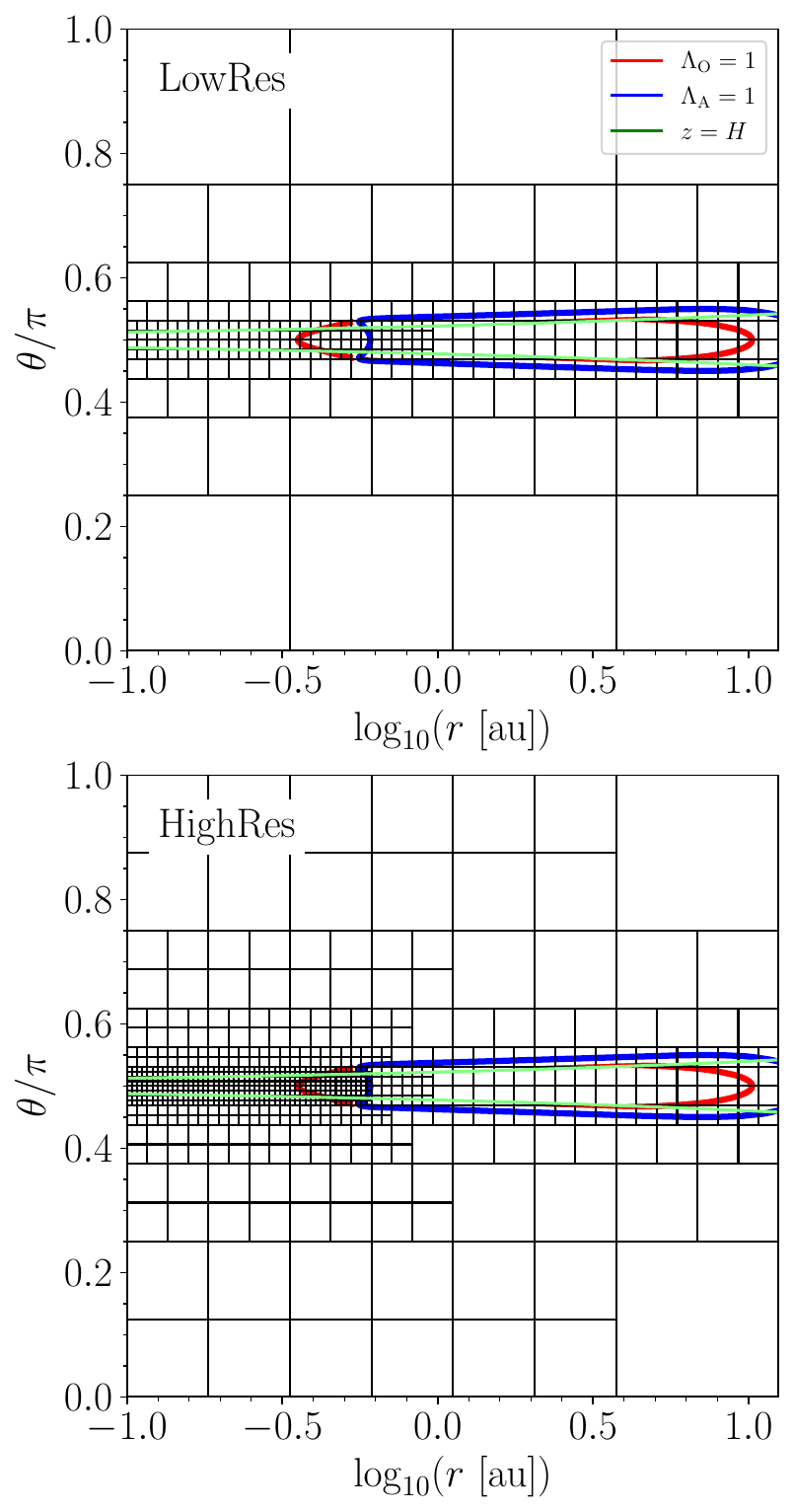}
    \end{center}
    \caption{
    The mesh structures of LowRes and HighRes runs in the $(r,\theta)$ plane.   
    Each cell corresponds to a mesh block with $(N_r,N_\theta)=(24,16)$.
    The red and blue lines represent the boundaries of the dead zones 
    for OR and AD ($\LambdaO=1$ and $\LambdaA=1)$, respectively. 
    The green line indicates the scale height $z=\pm H$.
    }
    \label{fig:init_mesh}
\end{figure}

The static mesh refinement technique is used to resolve 
the disk region which requires high resolution. 
We adopt two mesh configurations as shown in Figure \ref{fig:init_mesh}.
Each rectangle enclosed by the black lines in Figure \ref{fig:init_mesh}
consists of $(N_r,N_\theta)=(24,16)$.
Table \ref{tab:model} lists the models considered in this paper, where 
$\tK=2\pi \sqrt{r_\mathrm{in}^3/GM_*}$
is the Kepler rotation period at the inner boundary of the computation box.
LowRes run corresponds to the lower resolution simulation where 
the meshes are refined by 5 levels. 
If the whole computation domain was divided by the finest cell, 
the resolution would be 
$(N_r,N_\theta,N_z)=(1536,1024,1024)$.
The disk scale height $H(R)$ is resolved by $19~(R/\Rmri)^{1/4}~\mathrm{cells}$.

LowRes run was conducted until $2000~\tK$ (Table \ref{tab:model}).
As will be shown in Section \ref{sec:QF}, 
the resolution of LowRes run is not high enough to obtain the converged 
MRI turbulence.
For comparison, we conduct HighRes run, where 
the cells in the active zone are further refined in one more higher level 
(the lower panel of Figure \ref{fig:init_mesh}).
HighRes run could only be calculated to a 
short time, 500 rotations at the inner boundary, 
which is long enough for the MRI turbulence to be a saturated state in 
the active zone.

We note that the resolution of LowRes run
is high enough to drive turbulence in the active zone although 
the saturated $\avphi{\alpha_{R\phi}}_H$ is 
underestimated by a factor of 2 
compared 
with HighRes run that is expected to give the converged 
$\avphi{\alpha_{R\phi}}_H$ (Appendix \ref{sec:QF}).
In this paper, the long-term evolution is shown on the basis of LowRes run, 
referring the results of HighRes run to keep the resolution issue in mind.

To prevent the time step $\Delta t$ from being extremely small due to the Alfv\'en speed,
we set the following spatially dependent density floor,
\begin{equation}
    \rho(r,\theta,\phi) = \max\left( \rho(r,\theta,\phi), 10^{-6}\rho(R,z=0) \right).
\end{equation}
This density floor works only in the polar regions.
In the other regions, the disk winds keep the densities greater than the floor value.

\subsection{Boundary Conditions and Buffer Layer}\label{sec:buffer}

In Athena++, the boundary conditions are applied by setting 
the primitive variables in the so-called ghost zones located outside of the computation 
domain.
We impose the following inner boundary conditions.
The density and pressure in the ghost cells are fixed to be the initial values, 
the toroidal velocity follows the rigid rotation $v_\phi=\min(R\Omega_\mathrm{K}(r_\mathrm{in}),v_{\phi,\mathrm{ini}})$, and 
$v_r$ and $v_\theta$ are set to zero, where $v_{\phi,\mathrm{ini}}$ is the initial toroidal velocity.
We set the boundary conditions for $B_r$ and $B_\phi$ 
so that $B_r\propto r^{-2}$ and $B_\phi\propto r^{-1}$.

The outer boundary conditions are as follows.
For the density, pressure, $v_\theta$ and $B_\theta$, zero-gradient boundary conditions are imposed.
We set the boundary conditions for $v_\phi$, $B_r$, and $B_\phi$ 
so that $v_\phi\propto r^{-1/2}$, $B_r\propto r^{-2}$, and $B_\phi\propto r^{-1}$.
The radial velocity is set by using the zero-gradient boundary condition only for $v_r>0$ otherwise set to zero.

In order to mitigate the artificial effect of the inner boundary conditions, 
we introduce a buffer region in $r_\mathrm{in}\le r\le r_\mathrm{buf}$, where 
$r_\mathrm{buf}-r_\mathrm{in}=0.025$~au is the width of the buffer region.

The treatment in the buffer region is similar to that of \citet{Takasao2018}.
The poloidal velocities are damped according to the following equation:
\begin{equation}
        \left( \frac{\partial v_{r,\theta}}{\partial t} \right)_\mathrm{damp} = - \frac{v_{r,\theta}}{\tau_\mathrm{d}(r,\rho)},
\end{equation}
where $\tau_\mathrm{d}$ is the damping timescale, $\tau_\mathrm{d}^{-1} = f_\mathrm{rad}(r) f_v(v_r) \tau_\mathrm{d0}^{-1}$,
\begin{equation}
        f_\mathrm{rad}(r) = \frac{1}{2}\left[ 1-\tanh\left( \frac{r-r_\mathrm{in}-d/2}{0.13d} \right) \right],
\end{equation}
\begin{equation}
        f(v_r) = \frac{1}{2}\left[ \tanh\left( \frac{v_r - v_{r,c}}{0.1|v_{r,c}|} \right) +1\right],
\end{equation}
and 
\begin{equation}
        \tau_\mathrm{d0} = \max\left[ \min\left( f_\mathrm{d,min}\frac{\rho}{\rho_\mathrm{in}},f_{\mathrm{max}} \right),f_\mathrm{d,min} \right] \Omega_\mathrm{in}^{-1}.
\end{equation}
$f_\mathrm{d,min}=1$, $f_\mathrm{d,max}=100$, and $v_{r,c} = - 0.05 v_\mathrm{K,in}$.
The other physical quantities are unchanged in the damping layer.

\subsection{Normalization Units}
In this paper, all the physical quantities are normalized by the following three quantities,
the length scale $R_0=1~\au$, the time scale $t_0 = \sqrt{R_0^3/GM_*}=0.1~$yr, and 
the surface density $\Sigma_0=500$~g~cm$^{-2}$.
The gas temperature is normalized by $T_0 = (\mu m_\mathrm{H}/k_\mathrm{B})(R_0/t_0)^2=6.3\times 10^5~\mathrm{K}$.
Hereafter, we use normalized physical quantities unless otherwise noted
except for $R$ and $z$ that are shown in unit of the astronomical unit.

\begin{figure*}
    \begin{center}
    \includegraphics[width=17.0cm]{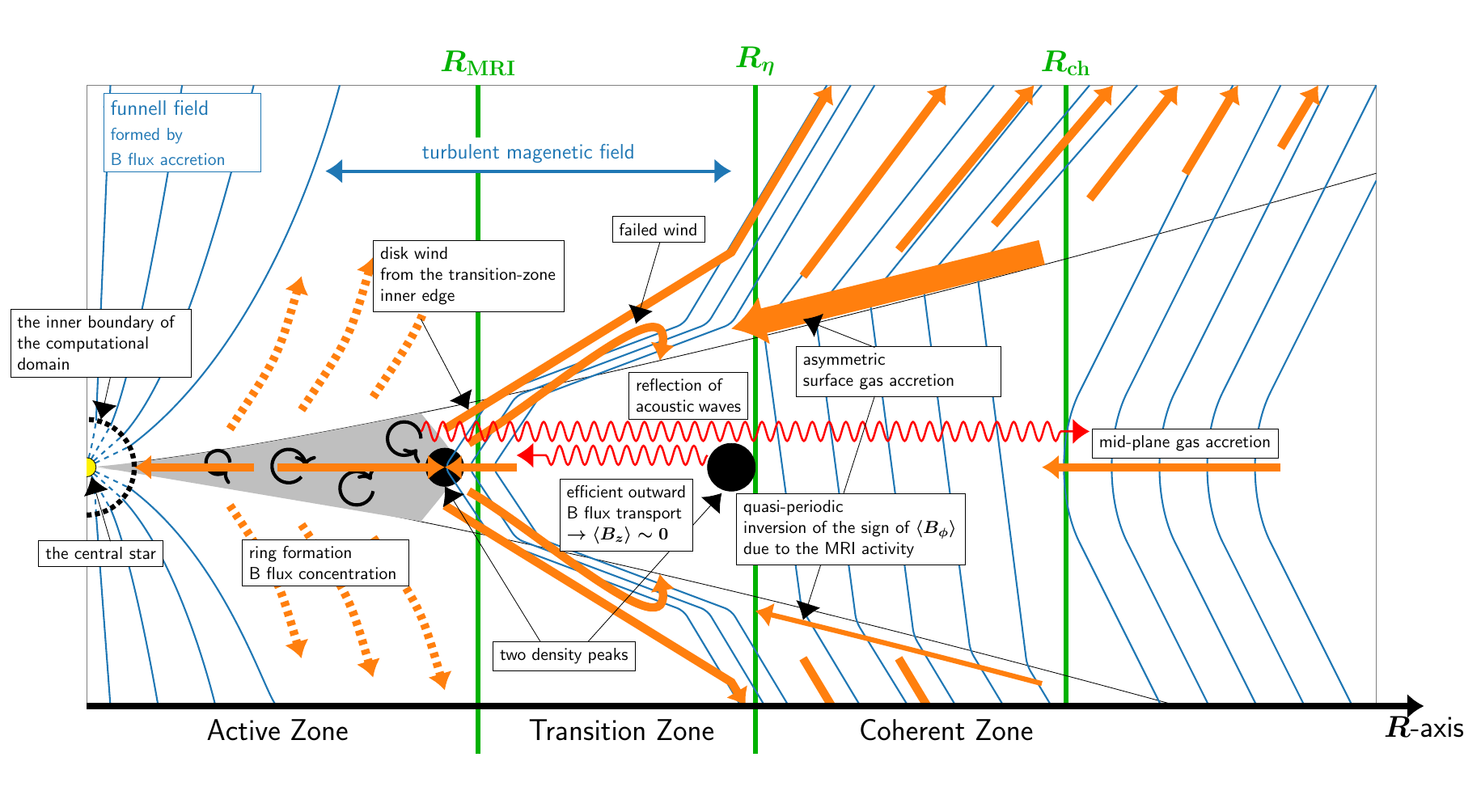}
    \end{center}
    \caption{
    A schematic picture showing our main findings with respect to the time-averaged gas dynamics of the disk.
    The orange arrows represent the gas motion, and the blue lines show the poloidal 
    magnetic field lines.
    }
    \label{fig:ppdisc}
\end{figure*}

\begin{figure*}
    \begin{center}
    \includegraphics[width=16cm]{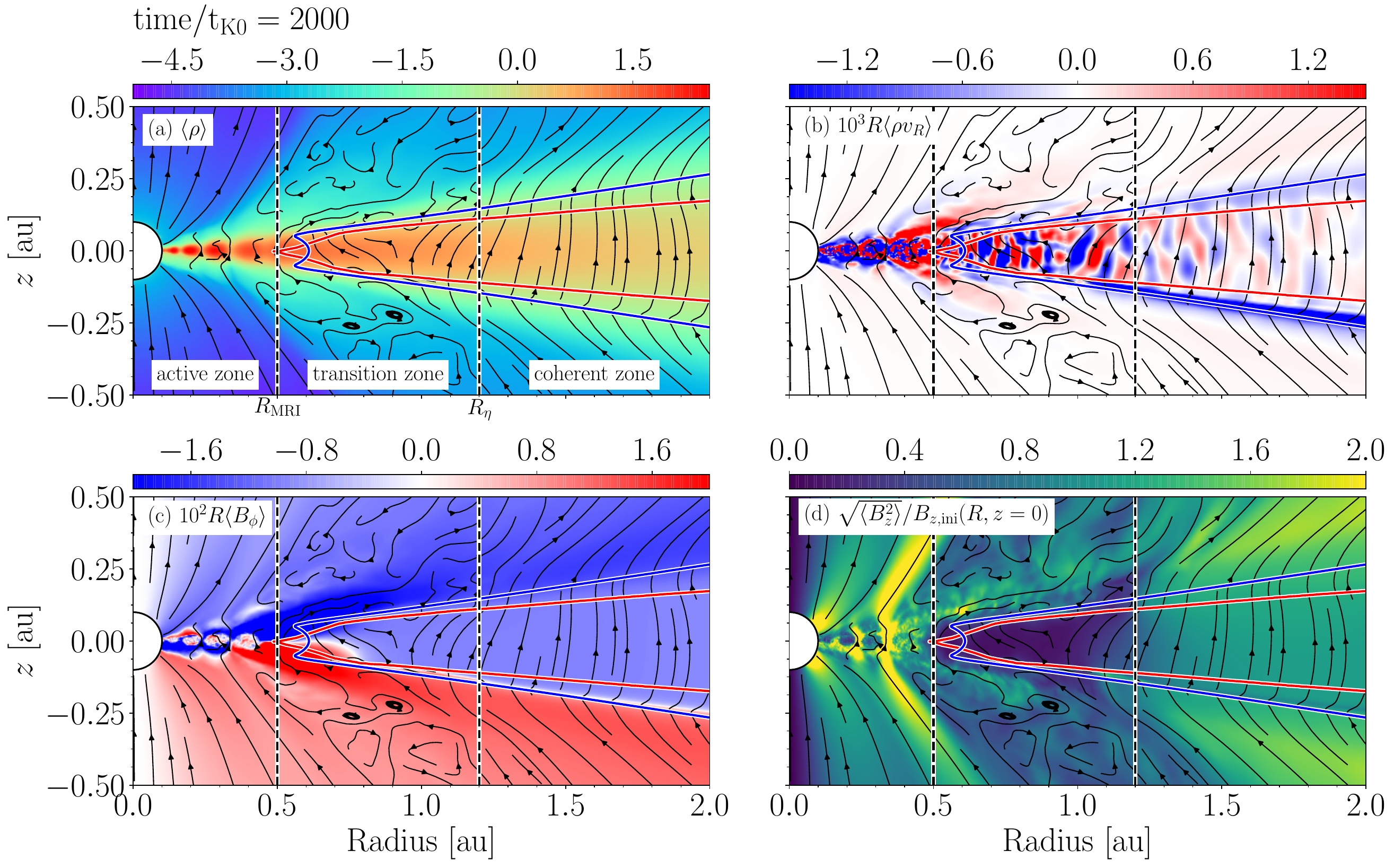}
    \end{center}
    \caption{
        Color maps of (a) $\langle \rho \rangle$, (b) $\avphi{\rho v_R}$, 
        (c) $R\avphi{B_R}$, and (d) $\sqrt{\avphi{B_z}^2}/B_{z,\mathrm{ini}}(R,z=0)$
        in the inner region 
        in the code units at $t=2000t_\mathrm{K0}$, where $B_{z,\mathrm{ini}}(R,z=0)$ is the initial 
        vertical field at the mid-plane.
        The red (blue) lines correspond to 
        the dead-zone boundaries for OR (AD).
        The streamlines of the poloidal magnetic fields averaged over $\phi$ 
        are plotted by the black lines.
        The two vertical dashed lines show $R=\Rmri$ and $R_\eta$ from left to right.
        An animation of this figure is available.
    }
    \label{fig:overall_slice}
\end{figure*}

\begin{figure*}
    \begin{center}
    \includegraphics[width=16cm]{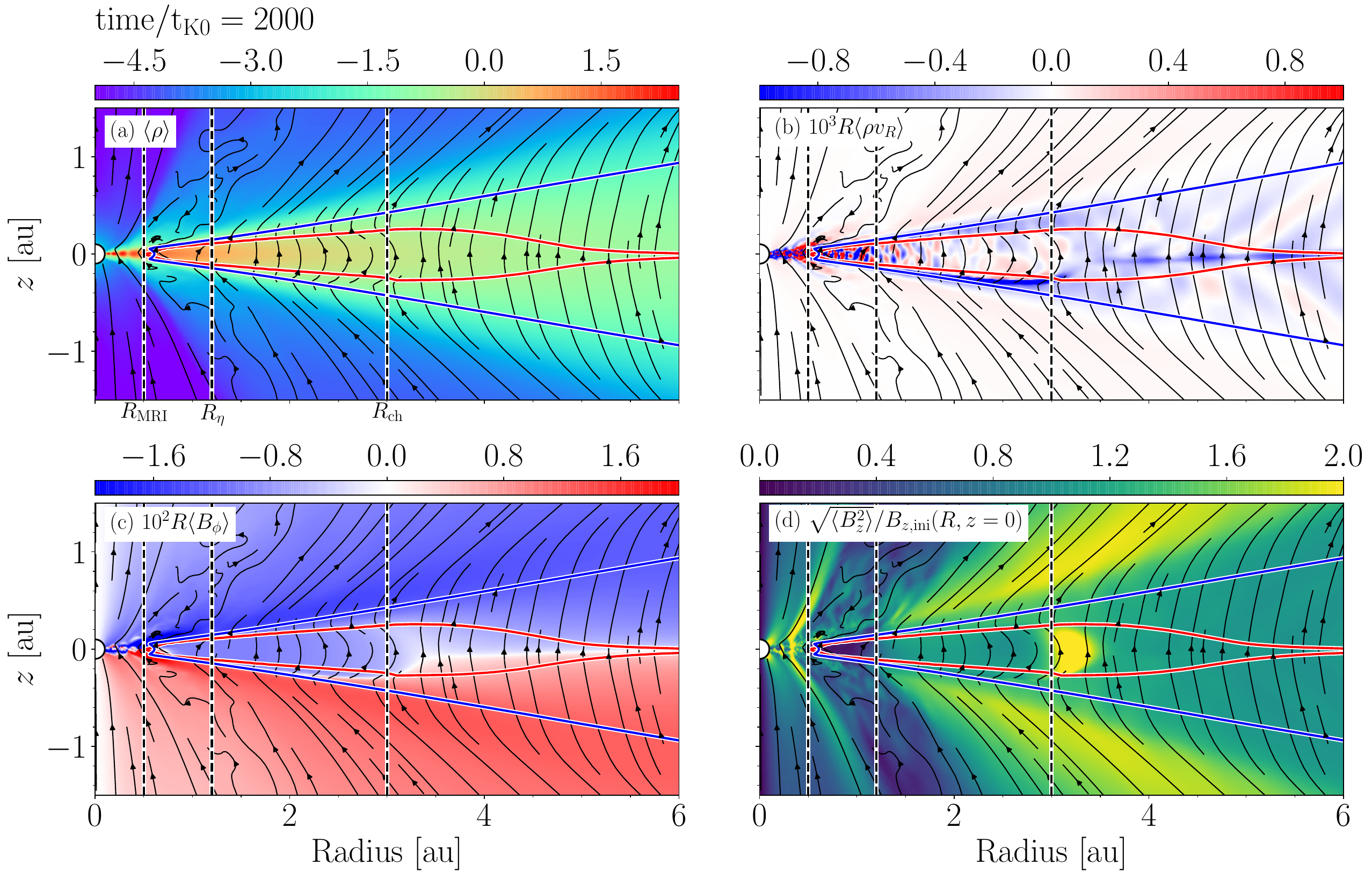}
    \end{center}
    \caption{
        The same as Figure \ref{fig:overall_slice} but the radial range $0\le R\le 6~\mathrm{au}$.
        The three vertical dashed lines show $R=\Rmri$, $R_\eta$, and 
        $R_\mathrm{ch}$ from left to right.
        An animation of this figure is available.
    }
\label{fig:overall_slice_outer}
\end{figure*}

\begin{figure*}
    \begin{center}
   \includegraphics[width=17cm]{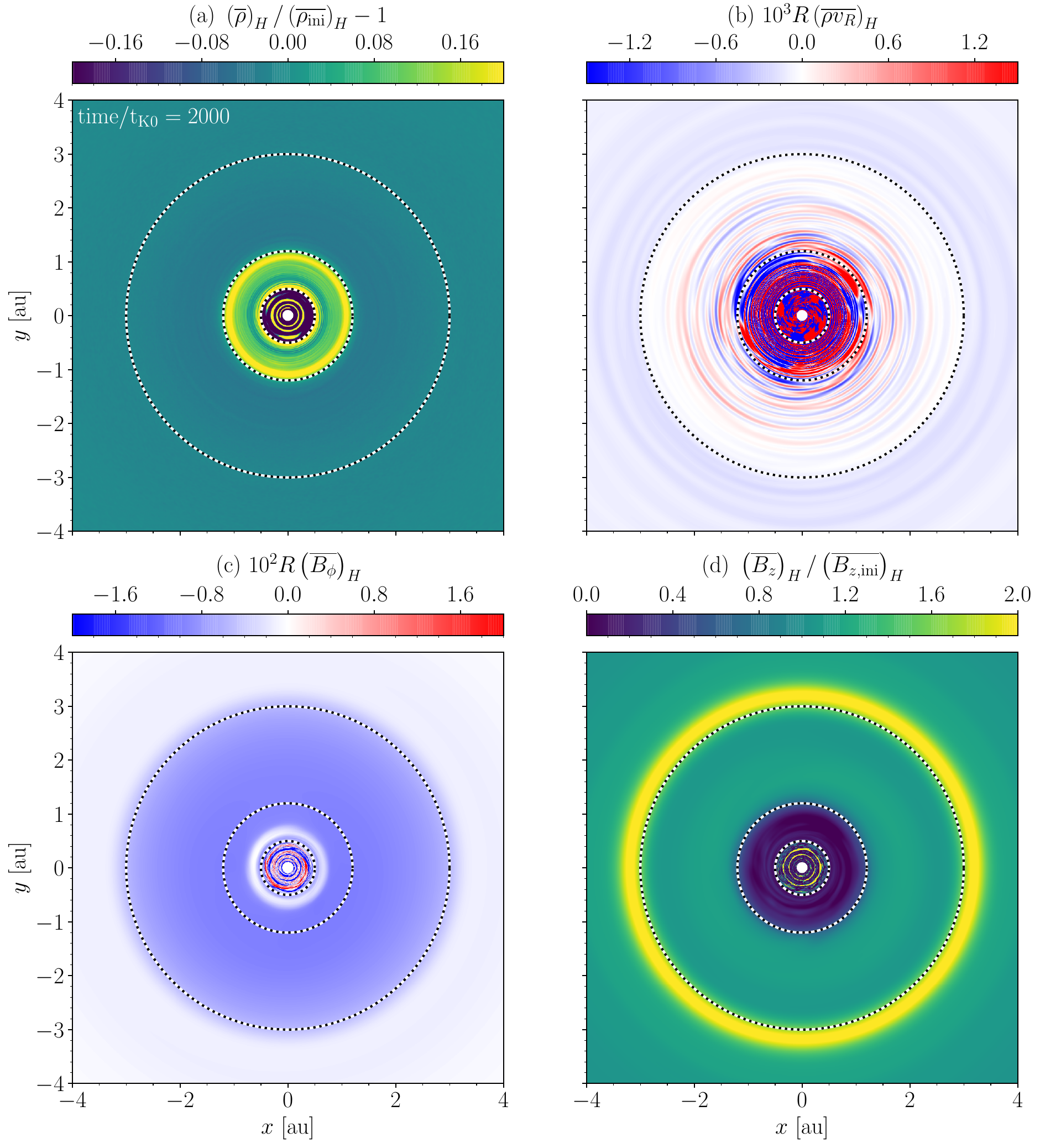}
    \end{center}
    \caption{
        Color maps of (a) the density, (b) the radial mass flux, (c) the toroidal magnetic field, (d) the 
        vertical magnetic field averaged over $|z|\le H$ at $t=2000~\tK$.
        All the quantities are shown in the code units.
        In each panel, 
        the three dashed circles correspond to $\Rmri$, $\Reta$, and $\Rch$ from the center outward.
        An animation of this figure is available.
    }
\label{fig:faceon}
\end{figure*}

\subsection{Averaged Quantities}\label{sec:average}

We define
averaged quantities used to analyse the simulation results
in this section.
In the subsequent sections, we present the data converted from 
the spherical polar coordinates $(r,\theta,\phi)$ to the 
cylindrical coordinates $(R,\phi,z)$.

We define the data averaged over $\phi$, 
which are denoted by using angle brackets, or $\avphi{A}_\phi$.
Different weight functions are used for different physical quantities in 
taking averages.
For quantities with the dimensions of mass density, momentum density, 
energy density, and field strength, we take 
the simple volume-weighted average 
\begin{equation}
    \avphi{A}(t,R,z) \equiv \frac{1}{2\pi} \int_0^{2\pi} Ad\phi,
\end{equation}
where a quantity $A$ can be $\rho$, $\rho v_z$, $B_\phi^2/8\pi$, or $B_R$.
For velocities and kinetic energies per unit mass, we take 
the density-weighted average
\begin{equation}
    \avphi{A}(t,R,z) \equiv \left( \int_0^{2\pi}\rho d\phi \right)^{-1}
    \int_{0}^{2\pi} \rho A d\phi,
\end{equation}
where a quantity $A$ can be $v_z$ or $v_R^2/2$.

In order to eliminate stochastic features originating from the MRI turbulence, 
$\avphi{A}_\phi$ is averaged over $t_1\le t\le t_2$ as follows:
\begin{equation}
    \avphi{A}_{t}(R,z) \equiv \frac{1}{t_2 -t_1}\int_{t_1}^{t_2} \avphi{A}dt.
\end{equation}
When the radial dependence of the disk structure is investigated,
we take the following vertical average of the disk,
\begin{equation}
    \avphi{A}_{z_\mathrm{b}}(t,R) \equiv \frac{1}{2z_\mathrm{b}}
    \int_{-z_\mathrm{b}}^{z_\mathrm{b}} \avphi{A} dz ,
\end{equation}
where $z_\mathrm{b}$ is a thickness over which $\avphi{A}$ is averaged.

When the $\phi$ dependence of the physical quantities is investigated, 
we take the following vertical average over $|z|\le z_\mathrm{b}$, 
\begin{equation}
    \left(\overline{A}\right)_{z_\mathrm{b}}(t,R,\phi)
    \equiv \frac{1}{2z_\mathrm{b}} \int_{-z_\mathrm{b}}^{z_\mathrm{b}} A dz.
\end{equation}

Using $\avphi{\bm{v}}$ and $\avphi{\bm{B}}$,
we define turbulent components of velocities and magnetic fields 
as variations from the $\phi$-averaged values as follows;
\begin{equation}
    \delta \bm{v} = \bm{v}- \avphi{\bm{v}}\;\;\mathrm{and}\;\;
    \delta \bm{B} = \bm{B} - \avphi{\bm{B}},
    \label{vBaveturb}
\end{equation}
where the turbulent components ($\delta \bm{v}$ and $\delta \bm{B}$)
satisfy $\avphi{\rho \delta \bm{v}}=0$ and $\avphi{\delta \bm{B}}=0$.

\section{ Main Results}\label{sec:results}

Figure \ref{fig:ppdisc} summarises our findings 
about the gas dynamics and magnetic field properties
in our disks.
Details are provided in the subsequent sections.

\subsection{Overall Features}\label{sec:overall}

In this section, we briefly describe overall features of our results.
Figures \ref{fig:overall_slice} and \ref{fig:overall_slice_outer} show 
the spatial distributions of the four $\phi$-averaged variables
in the inner region of $R\le 2$~au and outer region of $R\le 6$~au at $t=2000t_\mathrm{K0}$ for LowRes run,
respectively.
We note that the outer region has not reached a quasi-steady state 
since 2000 rotations at the inner boundary corresponds to only $12$ rotations at $R=3~\au$.
Figure \ref{fig:faceon} shows the face-on color maps of the four vertically-averaged variables.

From Figure \ref{fig:overall_slice}, 
in the polar regions of the northern and southern hemispheres, 
we identify the so-called funnel magnetic fields that 
are connected with the inner boundary and 
the magnetic energy dominates over the thermal energy. 
The radial size of this region increases with time 
owing to magnetic flux accretion \citep{Beckwith2009,Takasao2019}.
The size of the funnel regions may be overestimated because in reality the funnel magnetic fields 
come from open fields around the central star, 
whose size is much smaller than the inner boundary in our simulations.

We found that the conventional dead zone identified by $\LambdaO\le 1$ and $\LambdaA\le 1$ 
can be divided into two regions separated by $R=\Reta$.
We call the inner region of $\Rmri \le R \le \Reta$ "the transition zone" which has different 
properties from the conventional dead zone.
We call
the outer region of $R > \Reta$ "the coherent zone" which has the same 
properties as the conventional dead zone.

The overall properties of the active, transition, and coherent zones are briefly 
summarized in Sections \ref{sec:overall_active}, \ref{sec:overall_trans}, and 
\ref{sec:overall_dead}, respectively.
In Section \ref{sec:overall_influence_of_active}, 
we discuss
the influence of the active zone on the outer regions.

\subsubsection{Active Zone ($R\le \Rmri$)}\label{sec:overall_active}

The physical properties of the MRI turbulence in the active zone 
are consistent with those found in the previous studies. 
We briefly summarise the physical properties of the MRI turbulence, and 
detailed descriptions are found in Appendix \ref{app:MRIturbulence}.

The vertical structure of the magnetic fields changes around $|z|\sim 2H$.
The MRI turbulence generates turbulent magnetic fields for $|z|<2H$ while 
the magnetic fields become coherent in the upper atmosphere \citep{Suzuki2009}.
We observe a so-called butterfly structure in the $t$-$z$ diagram of $B_\phi$ at 
a fixed radius (Figure \ref{fig:active_ver}). 
The signs of the toroidal field change quasi-periodically and 
$B_\phi$ drifts toward upper atmospheres.

The MRI turbulence drives gas accretion in 
the inner region of the active zone.
The outer region expands outward by receiving 
the angular momentum from the inner region \citep{Lynden-BellPringle1974}.
In the upper atmospheres ($|z|>2H$), 
the magnetic torque of the coherent magnetic field drives coherent gas motion near the 
surface layers (Figure \ref{fig:active_ver}d).

One of the striking features of our simulations is the 
occurrence of multiple ring structures in the density field in the long-term evolution, 
as shown in Figures \ref{fig:overall_slice}a and \ref{fig:faceon}a 
(also see Figure \ref{fig:Ring}).
The spatial distributions of the density and $B_z$ are anti-correlated;
the magnetic fluxes are concentrated in the density gaps.
Possible formation mechanisms of these ring structures are investigated in 
Section \ref{sec:structureform_active} and are discussed in Section \ref{sec:ring_discuss}.

\subsubsection{Transition Zone ($\Rmri < R < R_\eta$)}\label{sec:overall_trans}

We discover a distinct region, the transition zone, in this simulation.
Although it was traditionally classified as a dead zone since 
$\LambdaO<1$ and $\LambdaA<1$ are satisfied in most regions, 
it has many characteristics not found in conventional dead zones.
An interesting feature is found in the magnetic field structures 
in Figure \ref{fig:overall_slice}d.
The vertical magnetic fields almost completely disappear in $\Rmri \lesssim R\lesssim \Reta$
(Section \ref{sec:loopform}). 
Such a region has not been found in the previous studies \citep{Lyra2012,Dzyurkevich2010,Flock2017}
because AD, which was neglected in their work, plays an essential role (Section \ref{sec:loopform}).
The gap in the vertical magnetic fields is almost concentric as shown in Figure \ref{fig:faceon}d.

The disappearance of the vertical magnetic field 
suppresses surface gas accretion expected in the conventional dead zone 
(Section \ref{sec:structureform_transition}).
Figure \ref{fig:overall_slice}b shows disturbances in the radial mass flux 
although there are no turbulence driving mechanisms.
These disturbances originate from the active zone \citep{Pucci2021}, and the net radial mass flux is 
extremely low when $\avphi{\rho v_R}$ is averaged over time (Section \ref{sec:structureform_transition}).
Thus, in the transition zone, gas accretion does not occur either around the mid-plane or
on the surface of the disk.

Figure \ref{fig:faceon}a clearly shows that 
a density peak appears at each edge of the transition zone, or $R\sim \Rmri$ and $R\sim \Reta$.
This structure is formed by a combination of no net radial gas motion in the transition zone and 
the gas supply from the active zone ($R\le \Rmri$) and 
the coherent zone ($R\ge \Reta$).
Details will be investigated in Sections 
\ref{sec:angmomtrans} and \ref{sec:masstransition}.
No $B_z$ concentration occurs while 
the density peak forms around the outer-edge of the 
transition zone (Figure \ref{fig:faceon}d).
This is because the magnetic flux drifts outward from the transition zone 
to the coherent zone as shown in Sections \ref{sec:fluxtransport_transition} and 
\ref{sec:mag_dead}.

\subsubsection{Coherent Zone ($R\ge R_\eta$)}\label{sec:overall_dead}
The coherent zone with $R\ge \Reta$ has properties consistent with 
those of the conventional dead zone found in the literature.
The magnetic field lines are smooth and coherent both inside and outside the disk.

Just above the coherent zone, 
the toroidal magnetic field is amplified by the differential rotation 
because the magnetic field is relatively coupled with the gas.
The magnetic tension force $-B_z \partial B_\phi/\partial z$ extracts the angular 
momentum from the gas efficiently,
driving surface gas accretion between 
the OR and AD dead-zone boundaries as shown in 
Figure \ref{fig:overall_slice}b \citep{BaiStone2013,Gressel2015}.
The electric current sheet where the sign of $\langle B_\phi\rangle$ is reversed 
is located not at the mid-plane but around the lower AD dead-zone boundary, 
indicating that the $z$-symmetry adopted in the initial condition is broken
(Figure \ref{fig:overall_slice}c).
This is because inside the disk, 
$\eta_\mathrm{O}$ and $\eta_\mathrm{A}$ are so large that a current sheet cannot exist 
inside the coherent zone
(Figure \ref{fig:profinimid}a).
As a result, a current sheet is lifted either upward and downward to the height where 
$\LambdaO$ is larger than unity \citep{BaiStone2013,Bai2017}.
The off-mid-plane current sheet causes the surface gas 
accretion to be asymmetric with respect to the mid-plane 
because the magnetic torque exerted in the lower side is stronger than  that in the upper side.

The behaviors of the magnetic field change around $R\sim \Rch$.
For $R\lesssim \Rch$, the current sheet is located at the lower disk surface as explained 
before. Beyond $R\sim \Rch$, $\etaO$ and $\etaA$ are small enough for the current sheet 
to remain at the mid-plane.
A similar feature was reported in \citet{Lesur2021} who demonstrated that 
the surface gas accretion (mid-plane gas accretion) occurs for lower (higher) $\LambdaO$.
Figure \ref{fig:overall_slice_outer}c shows that the toroidal field around the mid-plane at 
$R>\Rch$ is amplified, 
resulting in the OR dead-zone shrinking vertically toward the mid-plane. 
Around the mid-plane, the toroidal magnetic fields are amplified 
by AD. A similar amplification has been found in \citet{Suriano2018,Suriano2019} 
and also around the inner edge of the transition zone 
as discussed in Section \ref{sec:loopform}.
The magnetic tension force $-B_z \partial B_\phi/\partial z$ drives gas accretion at the mid-plane
in $R>\Rch$ (Figure \ref{fig:overall_slice_outer}b).

\subsubsection{Influence of the Active Zone on the Transition 
and Coherent Zones}\label{sec:overall_influence_of_active}

The MRI activity in the active zone affects both the velocity and magnetic fields 
in the outer regions.
As mentioned in Section \ref{sec:overall_trans}, 
spatial variations in the radial mass flux inside the disk seen in Figures \ref{fig:overall_slice}b
and \ref{fig:overall_slice_outer}b 
are caused by outward propagation of disturbances generated by MRI turbulence
(Section \ref{sec:sound}).

The MRI activity induces quasi-periodic inversion of the sign of the toroidal magnetic field 
in the transition zone and the inner part of the coherent zone (Section \ref{sec:magevo}).
This leads to a quasi-periodic switching of the current sheet position between the top and bottom dead-zone 
boundaries.
At $t=2000\tK$, the toroidal field in the disk is negative
(Figures \ref{fig:overall_slice}c and \ref{fig:overall_slice_outer}c), but at another time it can be positive.
The quasi-periodic disturbance of the field structures does not penetrate beyond $R\sim \Rch$.
Comparison between 
Figures \ref{fig:overall_slice_outer}c and 
\ref{fig:overall_slice_outer}d shows that 
$\avphi{B_z}$ has a concentration around $R\sim \Rch$.
Concentrations in $\avphi{B_z}$ propagate outward following 
quasi-periodic inversion of the sign of the toroidal magnetic field.
Details will be investigated in Section \ref{sec:magevo}.

\subsection{Main Findings}\label{sec:structureform}

In this section, we present detailed analyses about main findings in our simulations.

\begin{figure}
    \begin{center}
   \includegraphics[width=8cm]{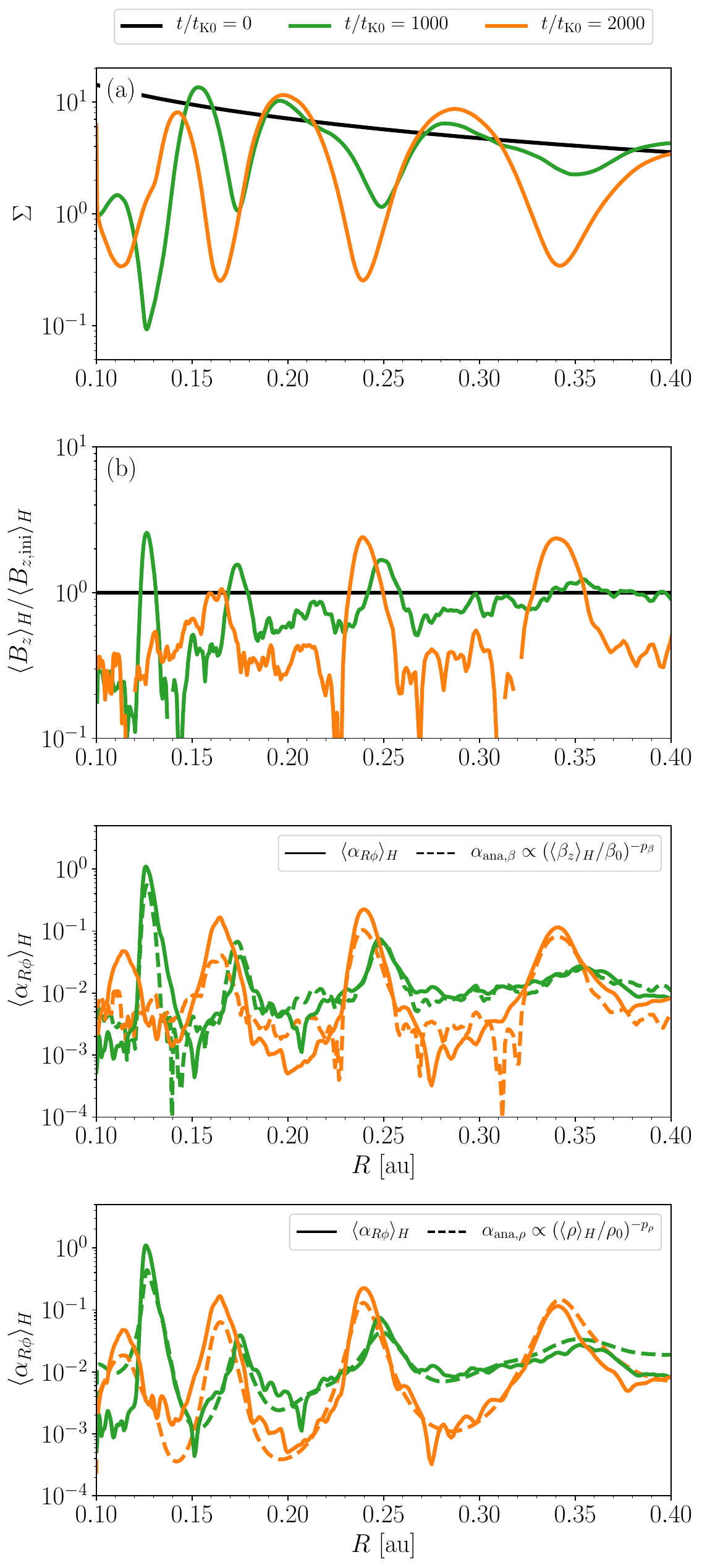}
    \end{center}
    \caption{
        Time evolution of the radial profiles of (a) $\Sigma$, (b) $\langle B_z\rangle_H/\langle B_{z,\mathrm{ini}}\rangle_H$, and (c,d) $\langle \alpha_{R\phi}\rangle$ 
        at $t=1000\tK$ and $t=2000\tK$.
        In Panels (a) and (b), the initial profiles are shown by the black lines.
      In Panels (c) and (d), the dashed lines show the best-fit functions 
      $\alpha_\mathrm{ana,\beta} \propto \avphi{\beta_{z}}_H/\beta_0)^{-p_\beta}$ 
      and 
      $\alpha_\mathrm{ana,\rho} = \alpha_{0,\rho} (\avphi{\rho}_H/\rho_0)^{-p_\rho}$, 
      respectively,
      where $\beta_0=10^4$ and $\rho_0=\rho_\mathrm{mid,ini}(R=0.3~\au)$.
      The best-fit functions are shown in Table \ref{tab:bestfitalpha}.
      All the quantities are shown in the code units.
    }
    \label{fig:sigevo}
\end{figure}

\subsubsection{Ring Formation in the Active Zone}\label{sec:structureform_active}

We observe the formation of multiple rings and gaps in the active zone as shown 
in the radial profiles of the column densities $\Sigma$ (Figure \ref{fig:sigevo}a).
Comparison between Figures \ref{fig:sigevo}a and \ref{fig:sigevo}b shows that 
the ring structures in $\Sigma$ are anti-correlated with the radial profile of 
the vertical magnetic flux averaged over the scale height
$\langle B_z\rangle_{H}$;
the rings (gaps) of $\Sigma$ correspond to the gaps (rings) of 
$\langle B_z \rangle_{H}$.

\begin{figure}
    \begin{center}
    \includegraphics[width=8cm]{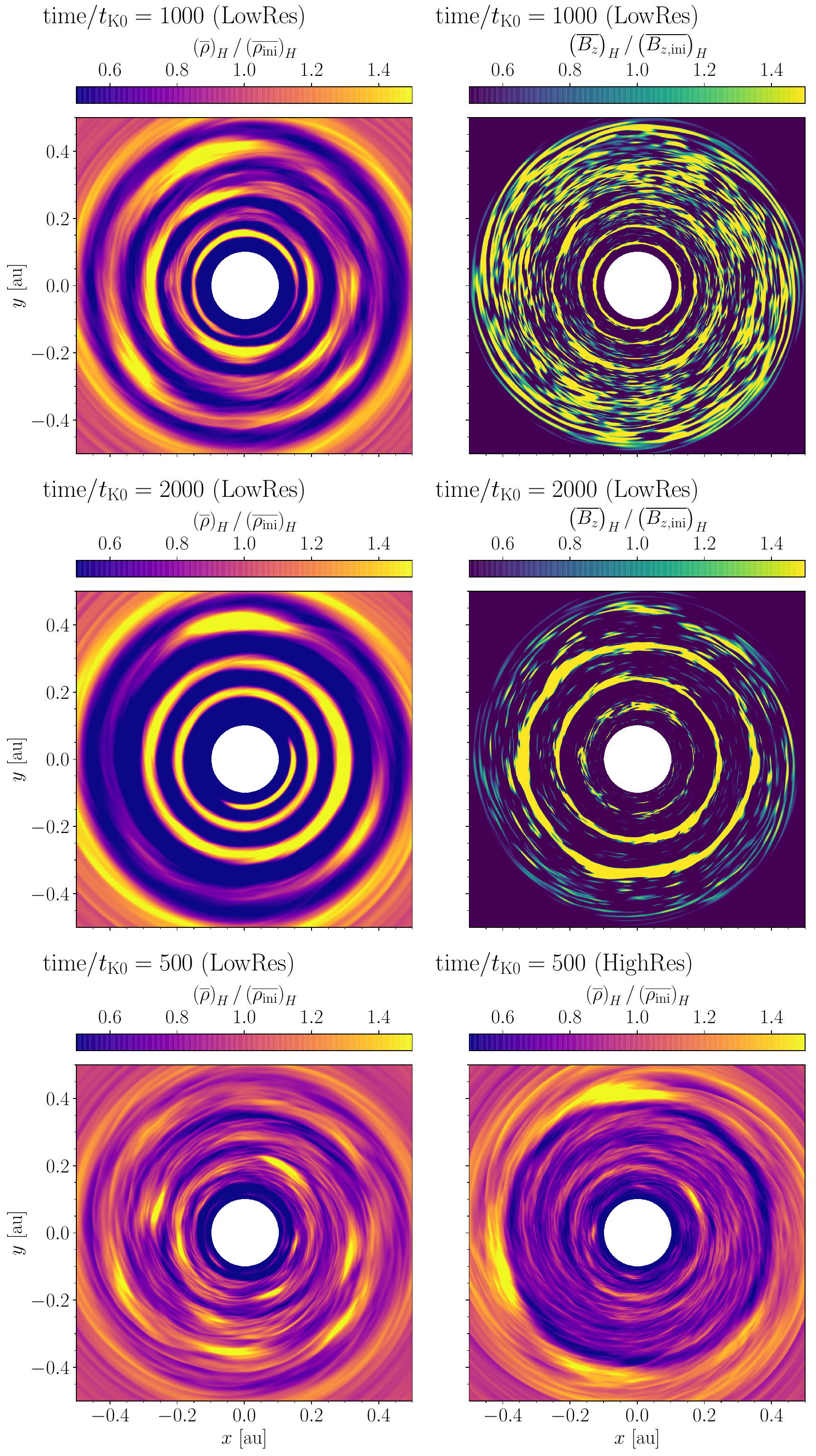}
    \end{center}
    \caption{
    Face-on views of  the density (left) and  $B_z$ (right) 
    averaged over $|z|<H(R)$ at
    $t/t_\mathrm{K0}=1000$ (top) and 2000 (middle).
    (Bottom) Face-on views of the density averaged over $|z|<H(R)$ at $t=500\tK$ for 
    LowRes (left) and  HighRes (right) runs.
    All the quantities are shown in the code units.
    }
    \label{fig:Ring}
\end{figure}

The top and middle panels of Figure \ref{fig:Ring} show 
the face-on views of the density and 
$B_z$ averaged over $|z|<H(R)$.
At $t=1000\tK$, multiple-rings are found in the density map.
Thin magnetic flux concentrations are found around $R\sim 0.2~$au
although the magnetic flux concentrations are not as significant as 
the variations of the density map.
At $t=2000\tK$, the density contrasts between the rings and 
gaps become significant, and 
the magnetic flux concentrations at the density gaps are 
more prominent than those at $t=1000\tK$.

The resolution dependence on the density distributions are shown 
in the bottom panels of Figure \ref{fig:Ring}.
The density is lower for HighRes run than for LowRes run in the active zone
because HighRes run exhibits faster viscous evolution (Figure \ref{fig:resolution}).
Despite the difference in the density structures, ring structures are also found in HighRes run.
This suggests that if the long term simulation performed for HighRes run, 
the ring structures would develop.

Several formation mechanisms of ring structures in the ideal MHD has been proposed.
In this section, we examine two mechanisms:
the viscous instability \citep{LightmanEardley1974,Suzuki2023} and
the wind-driven ring formation \citep{RoilsLesur2019}. 
As shown below, the viscous instability is a possible mechanism of the ring formation 
found in our simulations while the wind-driven instability may not occur.
In Section \ref{sec:ring_discuss}, we further discuss the origin of the ring structures in 
our simulations.

\noindent
\\
{\bf Viscous Instability}

Viscous instability in MRI-active disks has been investigated in \citet{RoilsLesur2019} and 
\citet{Suzuki2023}.
\citet{RoilsLesur2019} assumed that the $\alpha$ parameter \citep{ShakuraSunyaev1973} 
is proportional to 
$\beta_z^{-p_\beta}$, where $\beta_z = 8\pi\avphi{\avphi{P}}/\avphi{\avphi{B_z}}^2$ is the 
plasma beta measured by the net vertical field $\avphi{\avphi{B_z}}$ 
since $\alpha$ is anti-correlated with $\beta_z$
in local shearing-box simulations \citep{Hawley1995,Suzuki2010,Salvesen2016,Scepi2018A&A...620A..49S}, 
where $\avphi{\avphi{A}}$ is a volume-average of $A$.
The instability criterion derived by \citet{RoilsLesur2019} is $p_\beta>1$.
\citet{Suzuki2023} extended the linear analysis in \citet{RoilsLesur2019} by considering 
the dependence on $\avphi{\avphi{\rho}}$ and $\avphi{\avphi{B_z}}$ 
separately, or $\alpha \propto \avphi{\avphi{\rho}}^{-p_\rho} \avphi{\avphi{B_z}}^{-p_{B_z}}$.
The instability criterion derived by \citet{Suzuki2023} is $p_\rho >1$. $p_{B_z}$ does not contributes 
to the instability criterion, but it changes the growth rate.
We note that the instability criteria derived in \citet{RoilsLesur2019} and \citet{Suzuki2023}
are consistent since $\beta_z^{-p_\beta} \propto \avphi{\avphi{\rho}}^{-p_\beta} \avphi{\avphi{B_z}}^{2p_\beta}$.

We investigate whether our results satisfy the instability criteria 
given by \citet{RoilsLesur2019} and \citet{Suzuki2023}.
The $\alpha$ parameter is defined by using the thermal pressure, 
Reynolds stress, and Maxwell stress averaged over
$|z|\le H$ as follows:
\begin{equation}
\avphi{\alpha_{R\phi}}_H = \frac{1}{\avphi{P}_H} 
\left\{
\avphi{\rho \delta v_R \delta v_\phi}_H
+
\left\langle-\frac{B_RB_\phi}{4\pi}\right\rangle_H
\right\}.
\label{alpha}
\end{equation}
The plasma beta averaged over $z\le H$ is defined as 
$\avphi{\beta_{z}}_H = 8\pi \avphi{P}_H/\avphi{B_z}_H^2$.
The radial profiles of $\avphi{\alpha_{R\phi}}_H$ are fitted by 
the two fitting functions 
$\alpha_\mathrm{ana,\beta}\propto (\avphi{\beta_z}_H/\beta_0)^{-p_\beta}$ and
$\alpha_\mathrm{ana,\rho}\propto (\avphi{\rho}_H/\rho_0)^{-p_\rho}$
\footnote{
The fitting is performed in the range $0.2~\au\le R\le 0.35~\au$ 
to avoid the influence of the buffer layer (Section \ref{sec:buffer}) and the transition zone. 
The least square methods are applied to the scattered data in the 
($\ln \avphi{\beta_z}_H$, $\ln \avphi{\alpha_{R\phi}}_H$, )  and 
($\ln \avphi{\rho}_H$, $\ln \avphi{\alpha_{R\phi}}_H$) planes. 
}, 
where $\beta_0=10^4$ and $\rho_0 = \rho_\mathrm{mid,ini}(R=0.3~\au)$.
The best-fit functions are shown in Table \ref{tab:bestfitalpha}.
Figure \ref{fig:sigevo}c (Figure \ref{fig:sigevo}d) compares 
$\avphi{\alpha_{R\phi}}_H$ and
the best-fit functions $\alpha_\mathrm{ana,\beta}$ ($\alpha_\mathrm{ana,\rho}$) 
at both $t=1000~\tK$ and $2000~\tK$.
Figures \ref{fig:sigevo}c and \ref{fig:sigevo}d show 
that both fitting functions $\alpha_\mathrm{ana,\beta}$ and $\alpha_\mathrm{ana,\rho}$
reproduce $\avphi{\alpha_{R\phi}}_H$ reasonably well.

\begin{table}
    \tbl{The best-fit functions.}{
\begin{tabular}{|c|c|c|}
\hline
    &  $\alpha_\mathrm{ana,\beta}$ & $\alpha_\mathrm{ana,\rho}$\\
\hline
\hline
 $t=1000t_\mathrm{K0}$ &  $0.013 (\avphi{\beta_z}_H/\beta_0)^{-0.41}$ & 
                          $0.0091 (\avphi{\rho}_H/\rho_0)^{-1.16}$ \\
 $t=2000t_\mathrm{K0}$ &  $0.011 (\avphi{\beta_z}_H/\beta_0)^{-0.32}$ &  
                          $0.0022 (\avphi{\rho}_H/\rho_0)^{-1.38}$ \\
\hline
\end{tabular}} \label{tab:bestfitalpha}
\end{table}

Table \ref{tab:bestfitalpha} shows that 
the results of LowRes run do not satisfy 
the \citet{RoilsLesur2019} instability criterion while they satisfy the \citet{Suzuki2023} 
instability criterion since 
$p_\beta \sim 0.3-0.4$\footnote{
The power-law index $p_\beta$ may be lower than 
those obtained in local shearing-box simulations that span from $p_\beta \sim 0.5$ to $p_\beta\sim 1$.
\citet{Suzuki2010} and \citet{OkuzumiHirose2011} found that $p_\beta\sim 1$, and 
the results of \citet{Sano2004} show $p_\beta \sim 0.7$.
Recently, \citet{Salvesen2016} found that $p_\beta\sim 0.5$.
It is unclear what causes 
$p_\beta$ in our simulation to be smaller than those in local shearing-box simulations, 
but we note that $p_\beta$ obtained from the spatial variations of $\avphi{\alpha_{R\phi}}_H$ 
need not necessarily consistent with $p_\beta$ obtained from a volume average of 
the $\alpha$ parameter in local shearing-box simulations.
} 
and $p_\rho > 1$.
This suggests that the viscous instability may contribute to the ring formations in our simulations
if the $\alpha$ parameter is determined by the gas density.

Although a quantitative argument why $p_\rho$ becomes greater than unity is still missing, 
an anti-correlation between $\avphi{\rho}_H$ and $\avphi{B_z}_H$ seen in Figures \ref{fig:Ring}a and 
\ref{fig:Ring}b may be the key to understand the strong density dependence of the $\alpha$ parameter.
Similar anti-correlations between the density and the vertical field were reported, for instance, in 
\citet{BaiStone2014}, \citet{Suriano2019}, \citet{Jacquemin-Ide2021}, and \citet{Suzuki2023}.
Using $\alpha_\mathrm{ana,\beta} \propto \avphi{\beta_z}_H^{-p_\beta} \propto 
\avphi{\rho}_H^{-p_\beta} \avphi{B_z}^{2p_\beta}_H$ and assuming $\avphi{B_z}_H\propto \avphi{\rho}_H^{-q_{B_z}}$,
we obtain $\alpha_\mathrm{ana,\beta}\propto \avphi{\rho}_H^{-p_\beta (2q_{B_z}+1)}$.
When $\avphi{B_z}_H$ and $\avphi{\rho}_H$ are anti-correlated ($q_{B_z}>0$), 
the density dependence of $\alpha_\mathrm{ana,\beta}$ is apparently stronger than when 
$\alpha_\mathrm{ana,\beta}$ is assumed to be a function of $\avphi{\beta_z}_H$.
However this simple argument does not quantitatively explain our results.
LowRes run shows that $q_{B_z} \sim 0.43$ for  $t=1000~\tK$ and $\sim 0.95$ for $t=2000~\tK$. 
One obtains that $-\partial \ln \alpha_\mathrm{ana,\beta}/\partial \ln \avphi{\rho}_H = p_\beta (2q_{B_z}+1) \sim 0.76$ 
for $t=1000~\tK$ and $\sim 0.93$ for $t=2000~\tK$, and both values are not consistent with 
$p_\rho$ shown in Table \ref{tab:bestfitalpha}.

\noindent
\\
{\bf Wind-driven Instability}

\citet{RoilsLesur2019} proposed a wind-driven instability 
where disk winds destabilize disks when 
the amount of gas removed from gap regions due to winds is greater than that 
supplied by viscous diffusion from ring to gap regions.
Local simulations found that both the $\alpha$ parameter and 
mass loss rate depend negatively on the plasma beta; 
more strongly magnetized disks yield faster viscous diffusion and 
more efficient mass loss.
Thus, in order for the wind-driven instability to occur, 
the mass loss rate should has a sensitive dependence on the plasma beta 
than the viscous diffusion rate. 

We here define the normalized mass loss flux due to the disk wind averaged over time as 
\begin{equation}
    C_\mathrm{w} = 
    \frac{ \avphi{\rho v_n}_t(z=\zatm) + (-\avphi{\rho v_n}_t(z=-\zatm))}{2\avphi{\rho_\mathrm{mid}c_\mathrm{s,mid}}_t},
    \label{Cw}
\end{equation}
where $v_n$ stands for the velocity component perpendicular to the $\pm \zatm$ surfaces (Figures \ref{fig:init} and 
\ref{fig:eta2d}),
$\rho_\mathrm{mid}$ is the mid-plane density, and $c_\mathrm{s,mid}$ is the mid-plane sound speed.

Assuming that both $C_\mathrm{w}$ and $\alpha$ are anti-correlated with $\beta_z$ and they 
follow the relations $C_\mathrm{w}\propto \beta_z^{-p_\mathrm{w}}$ and $\alpha\propto \beta_z^{-p_\beta}$,
\citet{RoilsLesur2019} found that the wind-driven instability occurs when $p_\mathrm{w}>p_\beta$.
Since $p_\beta$ is roughly equal to $0.3-0.4$ in our simulations 
(Figure \ref{fig:sigevo}c and Table \ref{tab:bestfitalpha}), 
$p_\mathrm{w}$ needs to be larger than $0.3-0.4$ in order for the wind-driven instability to occur.

\begin{figure}
    \begin{center}
    \includegraphics[width=8.5cm]{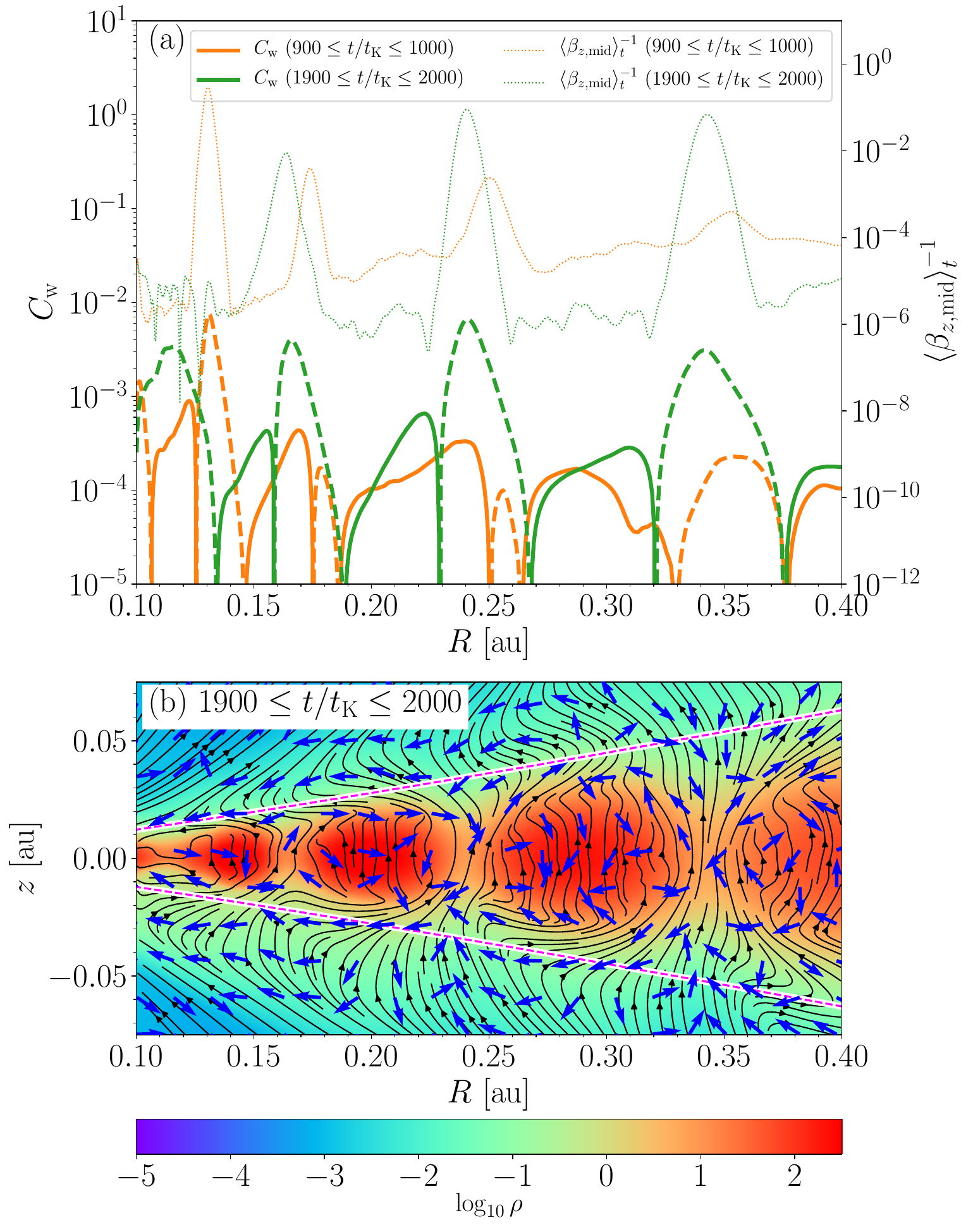}
    \end{center}
    \caption{
    (a) Normalized vertical mass loss rate in the active zone $C_\mathrm{w}$ defined in Equation (\ref{Cw})
    as a function of radius.
    The solid and dashed lines correspond to the radial profiles of $C_\mathrm{w}$ where 
    $C_\mathrm{w}>0$ and $C_\mathrm{w}<0$, respectively.
    The thin dashed lines show the radial profiles of $\avphi{\beta_{z,\mathrm{mid}}}_t^{-1}$ whose
    value is shown on the right vertical axis.
    The data are averaged over 
    $900\le t/\tK \le 1000$
    and
    $1900\le t/\tK \le 2000$.
    (b) Color maps of $\avphi{\rho}$ averaged over $1900~\tK\le t\le 2000~\tK$
    The black lines show the 
    streamlines of the poloidal magnetic field averaged over $\phi$.
    The blue arrows represent the directions of the poloidal velocities averaged over $\phi$.
    The two dotted magenta lines correspond to $z=\pm\zatm$.}
    \label{fig:wind_ac}
\end{figure}

Figure \ref{fig:wind_ac}a compares
the radial profiles of $C_\mathrm{w}$ with
the inverse of the time-averaged mid-plane plasma beta 
$\avphi{\beta_{z,\mathrm{mid}}}_t \equiv 8\pi\avphi{P}_t(z=0)/\avphi{B_z}_t^2(z=0)$.
$C_\mathrm{w}$ is poorly correlated with $\avphi{\beta_{z,\mathrm{mid}}}^{-1}$, 
indicating that the wind-driven instability
is not caused by the disk wind at least in our simulations.

Figure \ref{fig:wind_ac}a shows that in the gap regions where the radial profiles of 
$\avphi{\beta_{z,\mathrm{mid}}}_t^{-1}$ have local maxima, 
$C_\mathrm{w}$ is negative, indicating that the gas flows into the density gap regions
rather than being ejected from the disk in the vertical direction.
The vertical inflow into the density gap regions is evident from the blue arrows in Figure \ref{fig:wind_ac}b.
Why are there no outflows from the gap regions where the magnetic field is strong?
\citet{RoilsLesur2019} pointed out the gas in the gap regions is ejected by 
"wind plumes" where the magnetic fields are strong enough to be coherent and 
are tilted with respect to the $z$-axis
\citep[see also][]{RiolsLesur2018}.
Figure \ref{fig:wind_ac}b shows that the poloidal magnetic field structures originating from 
the gap regions do not maintain a large tilt with respect to the $z$-axis 
as $|z|$ increase, suggesting that the gas is not continuously accelerated 
along the field lines.

\begin{figure*}
    \begin{center}
   \includegraphics[width=17cm]{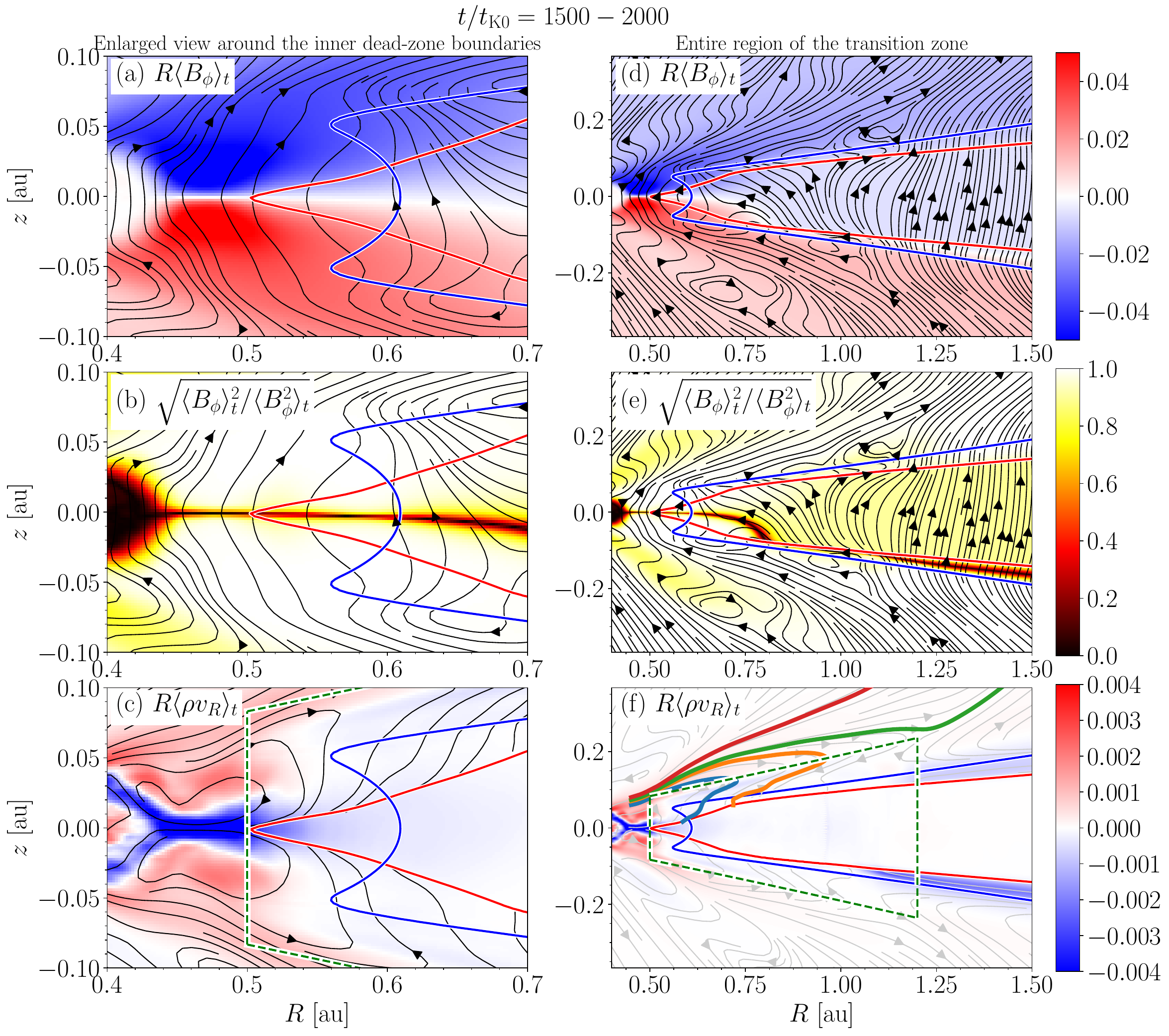}
    \end{center}
    \caption{
    Spatial Structures of $R\langle B_\phi\rangle_{t}$ (top),
     $\sqrt{\langle B_\phi\rangle_{t}^2/\langle B_\phi^2\rangle_{t}}$ (middle),
    and $R\langle \rho v_R\rangle_{t}$ (bottom) in the transition zone.
    The left panels show the enlarged view around the inner dead-zone boundaries, and 
    the right panels display the entire region of the transition zone.
    The results are averaged over $0\le \phi < 2\pi$ and $1500\le t/t_\mathrm{K0} \le 2000$, and 
    all the quantities are shown in the code units.
    In each panel, the red and blue lines represent the OR and AD dead-zone boundaries, respectively.
    The black lines show the streamlines of the poloidal magnetic fields 
    (the top and middle panels) 
    and poloidal velocity fields (the bottom panels).
    In the bottom panels, the streamlines of the poloidal velocity fields are not shown 
    inside the OR dead-zone, and 
    the region enclosed by the green dashed lines is 
    used to evaluate the mass transfer in the transition zone 
    (Section \ref{sec:masstrans}).
    In Panel (f), the four streamlines originating 
    from $(R,z)=(0.45~\au,0.06~\au)$, $(0.45~\au,0.07~\au)$, $(0.45~\au,0.073~\au)$, 
    and $(0.45~\au,0.08~\au)$ are shown by the thick lines.
    }
    \label{fig:BvR_tr}
\end{figure*}

\subsubsection{Gas Dynamics Around the Transition-zone 
Inner Edge and Non-existence of Gas Accretion in the Transition Zone}\label{sec:structureform_transition}

The transition zone is disturbed 
by the influence from the turbulence in the active zone, and 
exhibits time variations
as will be discussed in Sections \ref{sec:magevo} and \ref{sec:sound}. 
In this section, we investigate the quasi-steady structure of 
the transition zone by taking time average.
We will show that AD plays an important role in forming 
the hourglass poloidal magnetic fields and driving the mid-plane gas accretion 
around the inner edge of the transition zone (Figure \ref{fig:ppdisc}).
Disk winds are launched from higher 
latitudes around the inner edge of the transition zone.
In addition, we will also see that almost no gas accretion is driven in the transition zone.
We will quantitatively discuss the gas flows driven by the torques 
in Section \ref{sec:angmomtrans} and \ref{sec:masstrans}.

The left panels of Figure \ref{fig:BvR_tr} 
show the close-up views of the magnetic fields and velocities 
around the dead-zone inner edge.
The MRI turbulence appears to be suppressed around $R\sim 0.45~$au,
which is slightly different from $R=\Rmri=0.5~\au$ defined by $\LambdaO=1$.
This is because MRI turbulence is partially suppressed 
even where $\LambdaO$ is slightly greater than unity.

\begin{figure}
    \begin{center}
   \includegraphics[width=7cm]{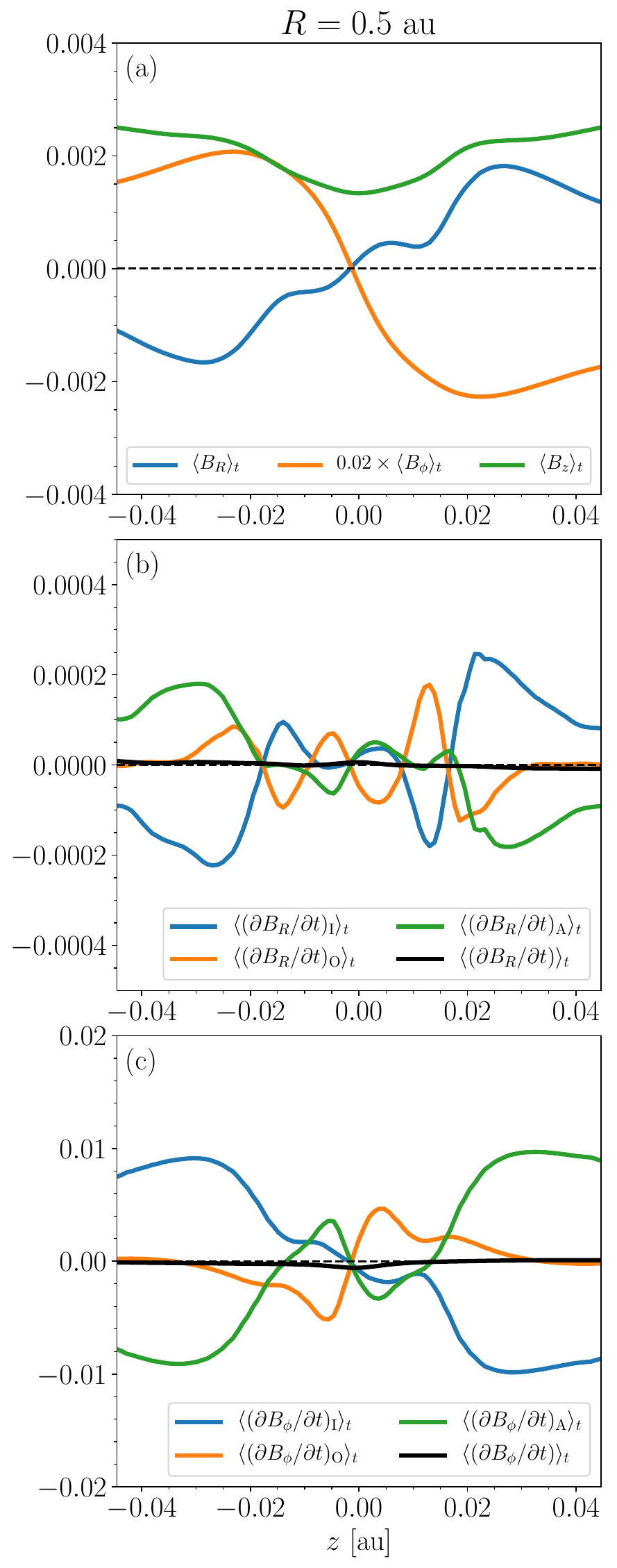}
    \end{center}
\caption{
Vertical profiles of (a) $\avphi{B_R}_t$,
$0.02\avphi{B_\phi}_t$, and 
$\avphi{B_z}_t$,
(b) 
$\avphi{(\partial B_R/\partial t)_\mathrm{I}}_t$,
$\avphi{(\partial B_R/\partial t)_\mathrm{O}}_t$,
$\avphi{(\partial B_R/\partial t)_\mathrm{A}}_t$,
$\avphi{(\partial B_R/\partial t)}_t$, and
(c) 
$\avphi{(\partial B_\phi/\partial t)_\mathrm{I}}_t$,
$\avphi{(\partial B_\phi/\partial t)_\mathrm{O}}_t$,
$\avphi{(\partial B_\phi/\partial t)_\mathrm{A}}_t$,
$\avphi{(\partial B_\phi/\partial t)}_t$ at $R=\Rmri=0.5~$au.
They are averaged over $1500\tK\le t \le 2000 \tK$.
All the quantities are shown in the code units.
}
    \label{fig:dBphidt_0.5}
\end{figure}

Around the mid-plane in $0.45~\au\lesssim R \lesssim 0.55~\au$, 
hourglass-shaped poloidal magnetic fields develop.
This is also clearly seen as an increasing trend of $B_R$ 
with $z$ for a constant $B_z$ across the mid-plane 
in Figure \ref{fig:dBphidt_0.5}a, 
which presents the vertical 
distribution of each component of the magnetic field 
at $R=\Rmri$. 
The toroidal magnetic field dominates over the other components,
and has a sharp gradient at the mid-plane.

\noindent
\\
{\bf Hourglass-shaped Poloidal Magnetic Field}

AD plays a critical role in the formation of 
the hourglass-shaped poloidal magnetic fields in $0.45~\au\lesssim R \lesssim 0.55~\au$.
To investigate which of the ideal, OR, and AD terms is the 
most effective to produce 
the hourglass-shaped magnetic fields, 
we measure their contributions to 
$-\avphi{\left(\nabla \times \bm{E}\right)_R}$, which 
is equal to $\partial \avphi{B_R}/\partial t$, at a given $z$.
First, we focus on the region of $|z|\lesssim 0.01~\au$ where the mid-plane gas accretion is seen in Figure \ref{fig:BvR_tr}c.
Since $B_z$ is almost constant in $|z|\lesssim 0.01~\au$,
the mechanism that steepens the $B_R$ gradient (Figure \ref{fig:dBphidt_0.5}a)
creates an hourglass-shaped magnetic field.
Figure \ref{fig:dBphidt_0.5}b shows that 
the AD term tilts the poloidal magnetic field toward the radial direction of the disk 
($\avphi{(\partial B_R/\partial t)_\mathrm{A}}>0$ for $z>0$, 
and $\avphi{(\partial B_R/\partial t)_\mathrm{A}}<0$ for $z<0$),
and the ideal MHD term does the same.
The vertical profile of $\avphi{B_R}$ is almost stationary 
by the balance between diffusion due to the OR term and 
amplification due to the AD and ideal MHD terms.
Far from the mid-plane ($|z|\gtrsim 0.02~\au)$, the AD term behaves quite the opposite.
The AD term works as diffusion of $\avphi{B_R}$ amplified by the ideal MHD term.

\noindent
\\
{\bf Mid-plane Gas Accretion}

The mid-plane accretion flow seen in 
$0.45~\au\lesssim R \lesssim 0.55~\au$
(Figure \ref{fig:BvR_tr}c)
is driven by the magnetic torque, 
$-B_z\partial B_\phi/\partial z$, due to
the pinched magnetic field by AD near the mid-plane.
Figure \ref{fig:dBphidt_0.5}c shows that 
the signs of $\avphi{B_\phi}$ and $\avphi{(\partial B_\phi/\partial t)_\mathrm{A}}$
are the same near the mid-plane ($|z|\lesssim 0.01~\au$), indicating that 
the AD term amplifies $\avphi{B_\phi}$ and steepens its gradient. 
The vertical profile of $\avphi{B_\phi}$ is kept
almost stationary by the OR term smoothing $\avphi{B_\phi}$. 
The ideal MHD term partially contributes to the steepening 
of $\avphi{B_\phi}$.
In a similar way as $\avphi{(\partial B_R/\partial t)_\mathrm{A}}$,
Far from the mid-plane, the AD term 
works as diffusion of $\avphi{B_\phi}$ amplified by the ideal MHD term.

\noindent
\\
{\bf Disk Wind Launching}

Just above the thin mid-plane gas accretion layer, the wind-like gas flows directing outward are driven. 
This is also caused by the magnetic torque of 
the coherent magnetic fields \citep[][see Section \ref{sec:angmom_later}]{BlandfordPayne1982,Baietal2016}.
The gas streaming lines at lower latitudes return to the mid-plane.
As a result, meridional flows are formed in $0.45~\au\lesssim R\lesssim 
0.5~\au$ and $|z|\lesssim 0.05~\au$; the streamlines of the gas flows are circulated 
each in the north and south sides of the disk 
as shown in Figure \ref{fig:BvR_tr}c.

The right panels of Figure \ref{fig:BvR_tr} zooms out the 
region shown in the left panels to cover the entire transition zone.
Four streamlines are highlighted by colors with thick lines in Figure \ref{fig:BvR_tr}f.
One can find that the lower two streamlines correspond to failed disk winds; 
the material does not reach the outer boundary of the simulation domain 
but falls back on to the disk surface in $R\lesssim \Reta$ by the central star gravity \citep{Takasao2018}.
By contrast, the disk wind flowing from the higher latitudes ($z=0.073~\au$ and $0.08~\au$, green and red)
reaches the outer boundary of the simulations box.
The directions of the winds are not parallel to the poloidal magnetic fields 
owing to AD (Figure \ref{fig:dBphidt_0.5}b).

\noindent
\\
{\bf Absence of Gas Accretion in the Transition Zone}

Comparison between Figures \ref{fig:BvR_tr}f and \ref{fig:overall_slice}b shows that 
the radial mass flux inside the OR dead-zone disappears
when the time average is taken.
The radial mass flux existing inside the OR dead-zone in Figure \ref{fig:overall_slice}b originates from 
the sound waves generated from the MRI turbulence in the active zone \citep[Section \ref{sec:sound} and][]{Pucci2021}.

\subsubsection{Disappearance of the Vertical Magnetic Flux in the Transition Zone}\label{sec:loopform}

\begin{figure}
    \begin{center}
   \includegraphics[width=8cm]{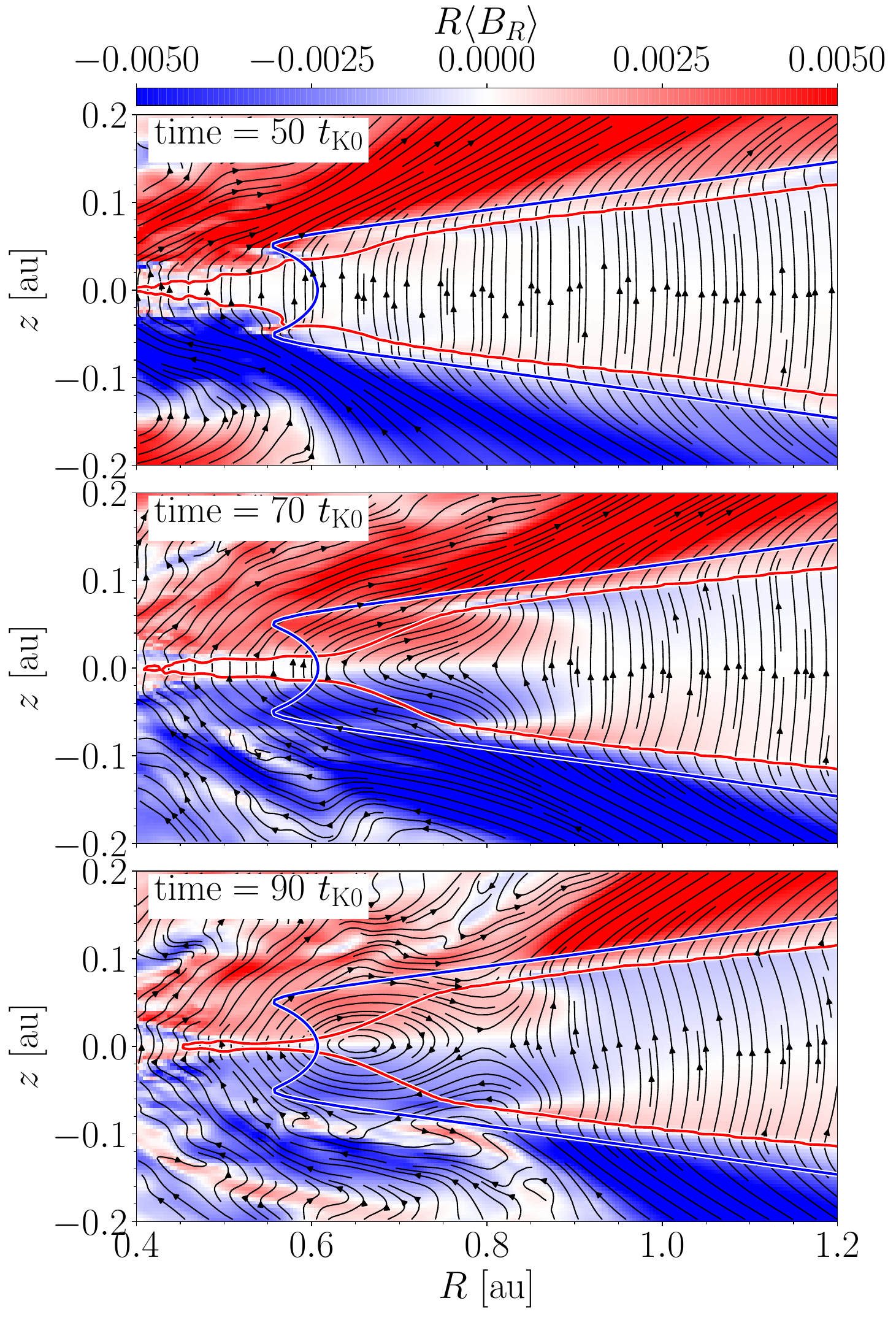}
    \end{center}
    \caption{Time sequence of the $\phi$-averaged field structure near the dead zone boundary 
        from $t/t_\mathrm{K0}=50$ to $90$.
        In each panel, 
        the colors show the $R\avphi{B_R}$ map, and 
        the black lines correspond to the streamline of the poloidal magnetic fields.
        The values of $R\avphi{B_R}$ are shown in the code units.
    }
    \label{fig:B_Rz_loop}
\end{figure}

In this section, we investigate how the vertical flux disappears in the 
transition zone (Figures \ref{fig:overall_slice}d and \ref{fig:faceon}d).
We will present that both a steep spatial gradient of $\etaO$ and 
amplification of the toroidal electric field by AD are necessary for efficient outward 
transfer of the vertical flux.
This flux transport is very efficient, and occurs in less time than 7 rotations at $R=\Rmri$.

Figure \ref{fig:B_Rz_loop} shows the time sequence of the magnetic field structure.
At $t=50t_\mathrm{K0}$, the poloidal magnetic field lines are almost vertical inside the 
AD dead zone at $R>0.6~\mathrm{au}$.
After 1.8 rotations at $R=\Rmri$ ($t=70t_\mathrm{K0}$), the poloidal magnetic fields 
are highly inclined and their directions are almost horizontal in 
$0.6~\mathrm{au}\lesssim R\lesssim 0.8~\mathrm{au}$.
Inside the transition zone, the radial magnetic fields are amplified.
At $t=90~t_\mathrm{K0}$, the magnetic loop structure whose center is at $R=0.62~\mathrm{au}$ and $z=0$ is formed.
The time evolution shown in Figure \ref{fig:B_Rz_loop} occurs on a dynamical timescale at $R=0.6~$au.

\begin{figure}
    \begin{center}
    \includegraphics[width=9cm]{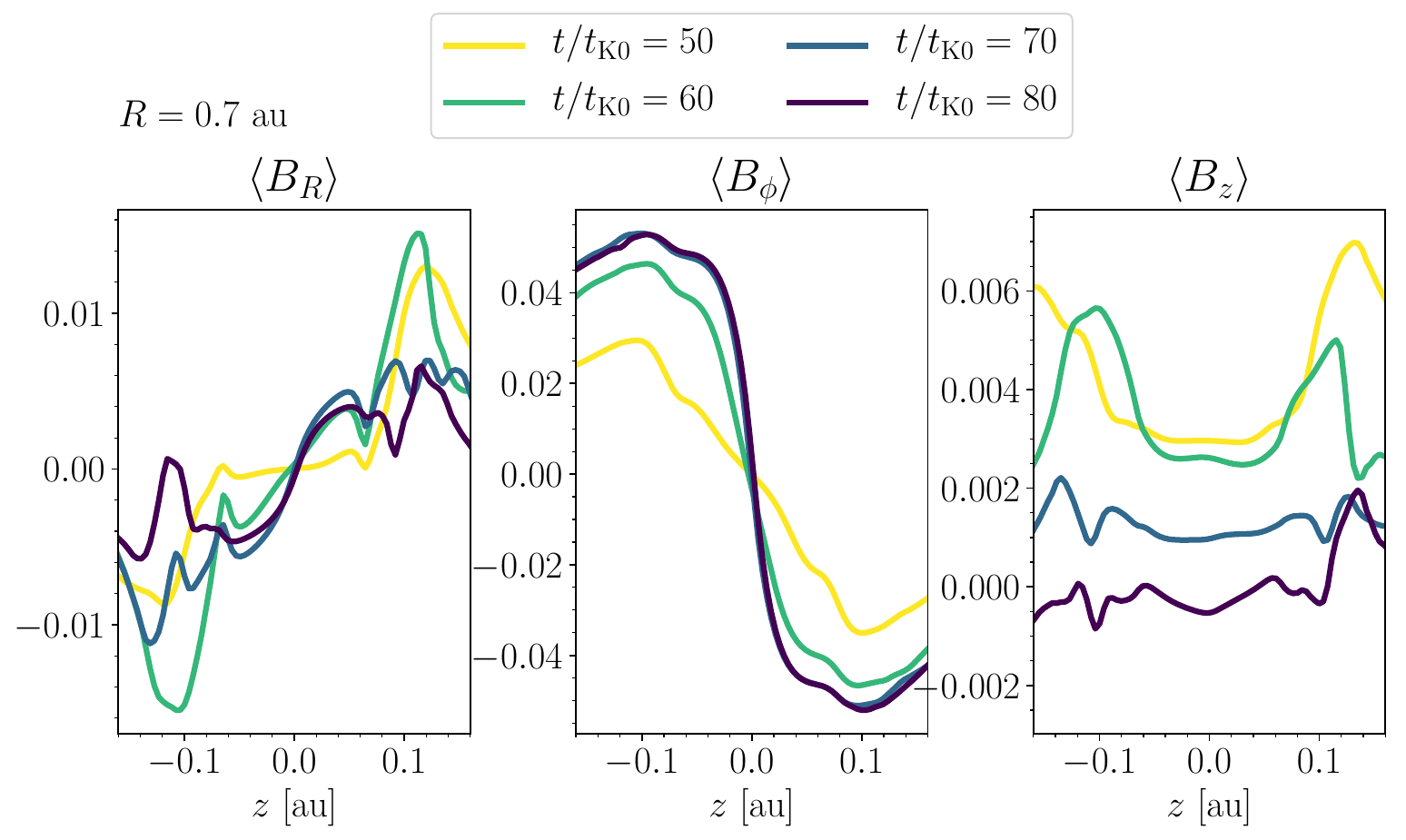}
    \end{center}
    \caption{
        Time evolution of the vertical profiles of $\avphi{B_R}$, $\avphi{B_\phi}$, and $\avphi{B_z}$
        at $R=0.7$~au from $t/t_\mathrm{K0}=50$ to $80$ in an interval of 10.
        All the quantities are shown in the code units.
    }
    \label{fig:B_Rslice0.7}
\end{figure}

In order to investigate the evolution of the magnetic fields, 
the vertical profiles of the $\phi$-averaged magnetic fields at $R=0.7$~au are shown in Figure \ref{fig:B_Rslice0.7}.
As shown in Figure \ref{fig:B_Rz_loop}, $\avphi{B_R}$ is amplified around the mid-plane during $50\le t/t_\mathrm{K0}\le 70$.
At the same time, $\avphi{B_\phi}$ is also amplified. 
Interestingly, $\avphi{B_z}$ decreases while $\avphi{B_R}$ and $\avphi{B_\phi}$ are amplified, 
indicating a rapid transport of the vertical field.

What causes this magnetic field evolution?
The answer can be obtained from the $\phi$-averaged induction equations, which are given by 
\begin{equation}
    \frac{\partial \avphi{B_R}}{\partial t} = \frac{\partial \avphi{E_\phi}}{\partial z},
    \label{inducR}
\end{equation}
\begin{equation}
    \frac{\partial \avphi{B_\phi}}{\partial t} = -\frac{\partial \avphi{E_R}}{\partial z} + \frac{\partial \avphi{E_z}}{\partial R},
    \label{inducphi}
\end{equation}
and
\begin{equation}
    \frac{\partial \avphi{B_z}}{\partial t}
    = -\frac{1}{R} \frac{\partial (R\avphi{E_\phi})}{\partial R}.
    \label{inducz}
\end{equation}
The electric field can be divided into three components,
\begin{equation}
    \bm{ E} = \bm{ E}_\mathrm{I} + \bm{ E}_\mathrm{O} + \bm{ E}_\mathrm{A},
    \label{E}
\end{equation}
where $\bm{E}_\mathrm{I}=-\bm{v}\times \bm{B}$ is the electric field in the ideal MHD, 
$\bm{ E}_\mathrm{O}=\etaO\bm{ J}$ and $\bm{ E}_\mathrm{A}=\etaO\bm{ J}_\perp$
are the electric fields caused by OR and AD, respectively.

\begin{figure}
\vspace{5mm}
    \begin{center}
   \includegraphics[width=8cm]{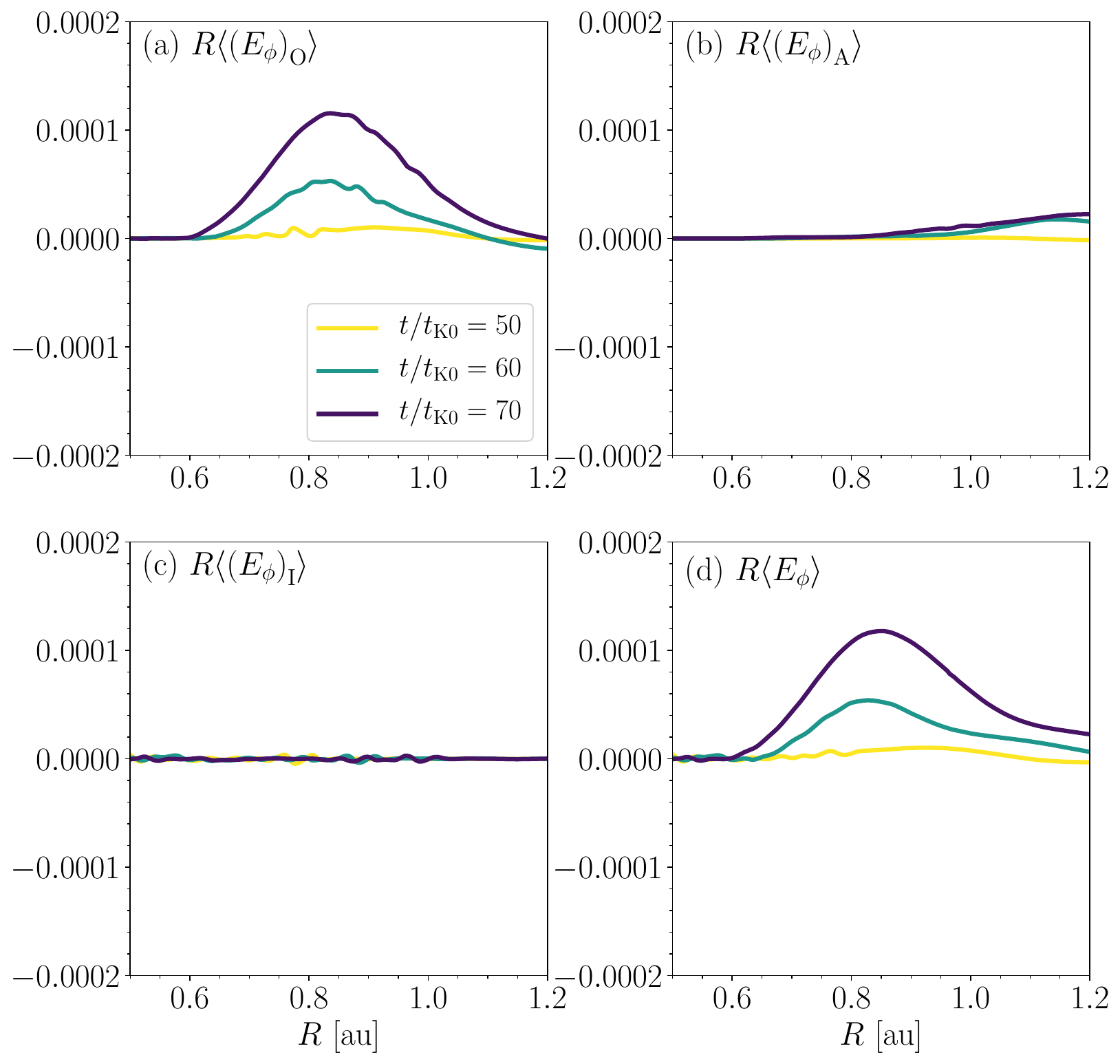}
    \end{center}
    \caption{Time Evolution of the radial profiles of the toroidal electric fields, (a) $R\langle \left( E_\phi \right)_\mathrm{O}\rangle$,
        (b) $R\langle \left( E_\phi \right)_\mathrm{A}\rangle$,
        (c) $R\langle \left( E_\phi \right)_\mathrm{I}\rangle$,
        and (d) $R\langle \left( E_\phi \right)\rangle$ at the mid-plane at $t/\tK=50$, 60, and 70.
        All the quantities are shown in the code units.
    }
    \label{fig:Eph_mid}
\end{figure}

We investigate the radial transport of the vertical field.
Equation (\ref{inducz}) shows that the radial transport velocity of $\langle B_z\rangle $ is estimated by $\langle E_\phi \rangle/\langle B_z\rangle$.
Since $\avphi{B_z}$ is positive in most regions at the mid-plane, the sign of
$\langle E_\phi \rangle$ represents the direction of the vertical field transport.
Figure \ref{fig:Eph_mid} shows the time evolution of the radial profiles of the 
toroidal electric fields at the mid-plane.
OR mainly contributes to $\avphi{E_\phi}$ while
$\avphi{E_\phi}_\mathrm{I}$ is negligible at the mid-plane where OR suppresses the MRI.
AD plays a minor role in $\avphi{E_\phi}$ except for $R\gtrsim 1~\au$.
A sharp increase in $\avphi{E_\phi}$ in $R\lesssim 0.9~\au$ is attributed 
to the strong radial dependence of $\langle \eta_\mathrm{O}\rangle$ in $R<R_\eta$ (Figure \ref{fig:profinimid}).
The radial gradient of $R\avphi{E_\phi}$ around $R\sim 0.8~\au$ 
increases with time from $t\sim 50~\tK$ to $70~\tK$, 
representing the rapid outward transport of $\avphi{B_z}$ (Figure \ref{fig:B_Rslice0.7}c).

\begin{figure}
    \begin{center}
    \includegraphics[width=8cm]{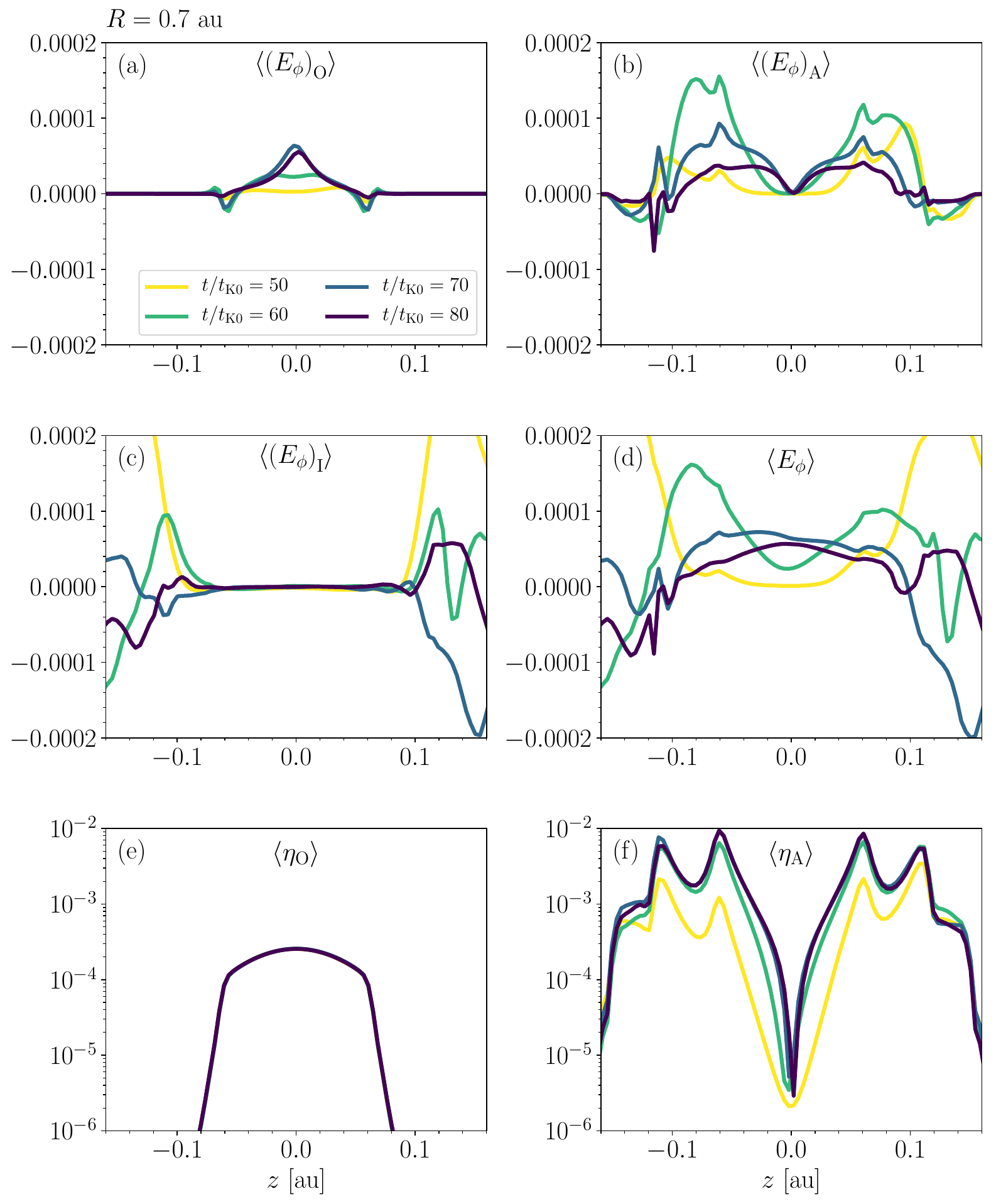}
    \end{center}
    \caption{
        Time evolution of the vertical profiles of (a) $\avphi{\left( E_\phi \right)_\mathrm{O}}$, (b) $\avphi{\left( E_\phi \right)_\mathrm{A}}$, 
        (c) $\avphi{\left( E_\phi \right)_\mathrm{I}}$, (d) $\avphi{E_\phi}$, 
        (e) $\avphi{\eta_\mathrm{O}}$, and (f) $\avphi{\eta_\mathrm{A}}$ 
        at $R=0.7$~au from $t/t_\mathrm{K0}=50$ to $80$ in an interval of 10.
        All the quantities are shown in the code units.
    }
    \label{fig:Eph_Rslice0.7}
\end{figure}

Figure \ref{fig:Eph_Rslice0.7} shows that 
an increase in $E_\phi$ 
results in outward magnetic flux transfer.
We investigate how $\avphi{E_\phi}_\mathrm{O}$ is generated.
Because $|\partial B_R/\partial z|$ dominates over $|\partial B_z/\partial R|$ 
\begin{equation}
\avphi{(E_\phi)_\mathrm{O}} \sim \bigl \langle{\eta_\mathrm{O}\frac{\partial B_R}{\partial z}}\bigr \rangle,
\label{EphO}
\end{equation}
the vertical distribution of $B_R$ is critical for the vertical field transport.

The evolution of $\avphi{B_R}$ is determined by the vertical structure of $\avphi{E_\phi}$ (Equation (\ref{inducR})).
Around the mid-plane, $\avphi{\left( E_\phi \right)_\mathrm{I}}$ is negligible since this region is inside the dead zones.
For $t\le 60~t_\mathrm{K0}$, $\avphi{\left(E_\phi\right)_\mathrm{A}}$ mainly contributes to $\avphi{E_\phi}$.
Around the mid-plane, 
$\partial \avphi{\left(E_\phi\right)_\mathrm{A}}/\partial z$ is positive (negative) for $z>0$ ($<0$), 
leading to amplification of $\avphi{B_R}$ since 
the signs of $\partial \avphi{\left(E_\phi\right)_\mathrm{A}}/\partial z$ and $\avphi{B_R}$ are the same
(the left panel of Figure \ref{fig:B_Rslice0.7}).
This downward-facing convex profile of $\avphi{\left(E_\phi\right)_\mathrm{A}}$ 
is attributed to the profile of $\avphi{\eta_\mathrm{A}}$, which 
increases toward upper low density regions (Figure \ref{fig:Eph_Rslice0.7}f).
Since $\avphi{\etaA}$ is proportional to $B^2$ in this region, the gradient of $\avphi{\etaA}$ becomes steeper and steeper 
owing to the amplification of the magnetic field.
As both $\avphi{B_R}$ and $\avphi{B_\phi}$ increase (Figure \ref{fig:B_Rslice0.7}), 
the current density is enhanced around the mid-plane, and the electric field owing to OR becomes important.
For $t\ge 70t_\mathrm{K0}$, the downward-facing convex  
profiles disappear in the $\avphi{E_\phi}$ profile
since the downward-facing convex part in $\avphi{\left(E_\phi\right)_\mathrm{A}}$ 
is almost compensated by the upward-facing convex part in $\avphi{\left(E_\phi\right)_\mathrm{O}}$.
As a result, OR suppresses the amplification of $\avphi{B_R}$.

The gradient of $\avphi{B_R}$  with respect to $z$ become steeper as the time passes, leading that 
the current sheet around the mid-plane becomes thinner around the central region where the magnetic fields are weak.
Development of such sharp structures in the magnetic null by AD was previously 
reported in \citet{BrandenburgZweibel1994}.

\subsubsection{Quasi-periodic Disturbances of Magnetic Fields 
Beyond the Inner Dead-zone Edge due to the MRI Activity}\label{sec:magevo}

In this section, we will show that 
the MRI activity drives quasi-periodic variations of the magnetic fields 
in the transition and coherent zones.
The sign of $B_\phi$ reverses quasi-periodically. 
Anti-diffusion caused by AD generates a $B_z$ condensation where the sign of $B_\phi$ is flipped. 
The magnetic field disturbances do not propagate beyond $R=\Rch$.

\begin{figure}
    \begin{center}
   \includegraphics[width=8cm]{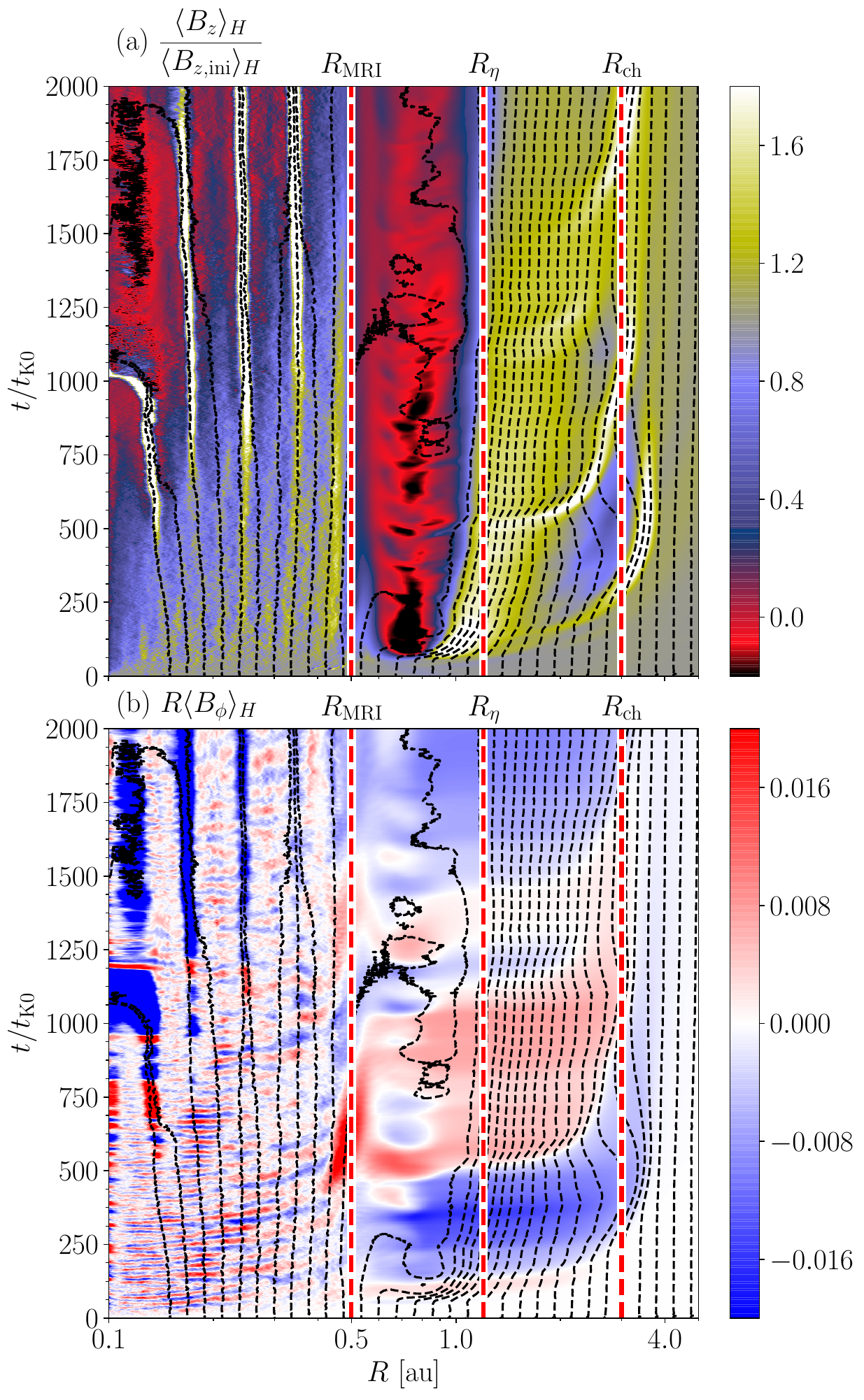}
    \end{center}
    \caption{
    Radius-time diagrams of 
    (a)$\avphi{B_z}_H/\avphi{B_{z,\mathrm{ini}}}_{H}$ and 
    (b)$R\avphi{B_\phi}_H$, 
    where $\avphi{B_{z,\mathrm{ini}}}_{H}$ 
    is $\avphi{B_z}_{H}$ at $t=0$.
    The vertical red dashed lines correspond to $\Rmri$, $R_\eta$, and $R_\mathrm{ch}$ from left to right.
    The black dashed lines show the contours of $\Psi_\mathrm{rad}$, which is defined 
    in Equation (\ref{psirad}).
    They correspond to the trajectory of the magnetic flux 
    originated at each equally-spaced position at $t=0$.
    All the quantities are shown in the code units.
    }
    \label{fig:tRdiagram}
\end{figure}

Figure \ref{fig:tRdiagram} shows the radius-time diagrams 
of $\avphi{B_z}_H$ and $\avphi{B_\phi}_H$.
As shown in Section \ref{sec:structureform_active}, 
in the active zone ($R\le \Rmri$), the ring and gap structures develop
in $\avphi{B_z}_H$ and $\avphi{\rho}_H$.

In the transition zone ($\Rmri < R < R_\eta$), 
as shown in Section \ref{sec:loopform}, 
the combination of OR and AD causes the magnetic flux 
to be transferred outward rapidly in the early evolution $\sim 100\tK$.
As a result, $\avphi{B_z}_H$ is almost zero in the transition zone.
The gap structure is maintained at least during the simulation.

Before showing the evolution of $\avphi{B_\phi}_H$, 
we recall the spatial structure of the 
magnetic fields found in 
Figures \ref{fig:overall_slice}c and \ref{fig:BvR_tr}d.
Just outside the inner edge of 
the transition zone, $B_\phi$ is amplified 
above and below the mid-plane, and 
the current sheet where the sign of $B_\phi$ is 
flipped is located around the mid-plane.
As shown in Figure \ref{fig:dBphidt_0.5}, 
the profile of $B_\phi$ is 
determined by the balance between dissipation of $B_\phi$ due to OR 
and $B_\phi$ amplification due to AD and the induction term.
For $R\gtrsim 0.7~\au$,
strong magnetic diffusion moves the current sheet to either upper or lower AD dead-zone boundaries
(Figure \ref{fig:overall_slice}c).

Figure \ref{fig:tRdiagram}b shows that 
$\avphi{B_\phi}_H$ just outside the 
inner edge of 
the transition zone changes their sign quasi-periodically.
Since the position where $\avphi{B_\phi}=0$ 
is around the mid-plane, 
the quasi-periodic variations 
of $\avphi{B_\phi}_H$ are not caused by 
change of the current sheet position.
The vertical profile of $\avphi{B_\phi}$ 
is not perfectly inversely symmetric
with respect to the mid-plane, and
the difference between 
$|\avphi{B_\phi}|$ for $z>0$ and 
for $z<0$ varies quasi-periodically.

Figure \ref{fig:tRdiagram}b shows that 
the variations of $\avphi{B_\phi}_H$ propagate outward rapidly in the radial direction while
they do not propagate beyond $R\sim \Rch$.
These 
$\avphi{B_\phi}_H$ variations occur quasi-periodically.
The flip of the sign of $\avphi{B_\phi}_H$ occurs 
almost synchronously in $\Rmri \lesssim R \lesssim \Rch$ 
around $t\sim 150~\tK=13.4~t_\mathrm{K}(\Rmri)$, $\sim 520~\tK = 46.5~t_\mathrm{K}(\Rmri)$, 
$\sim 1100~\tK = 98.4~t_\mathrm{K}(\Rmri)$, and $\sim 1450~\tK = 130~t_\mathrm{K}(\Rmri)$.

\begin{figure}
    \begin{center}
    \includegraphics[width=8cm]{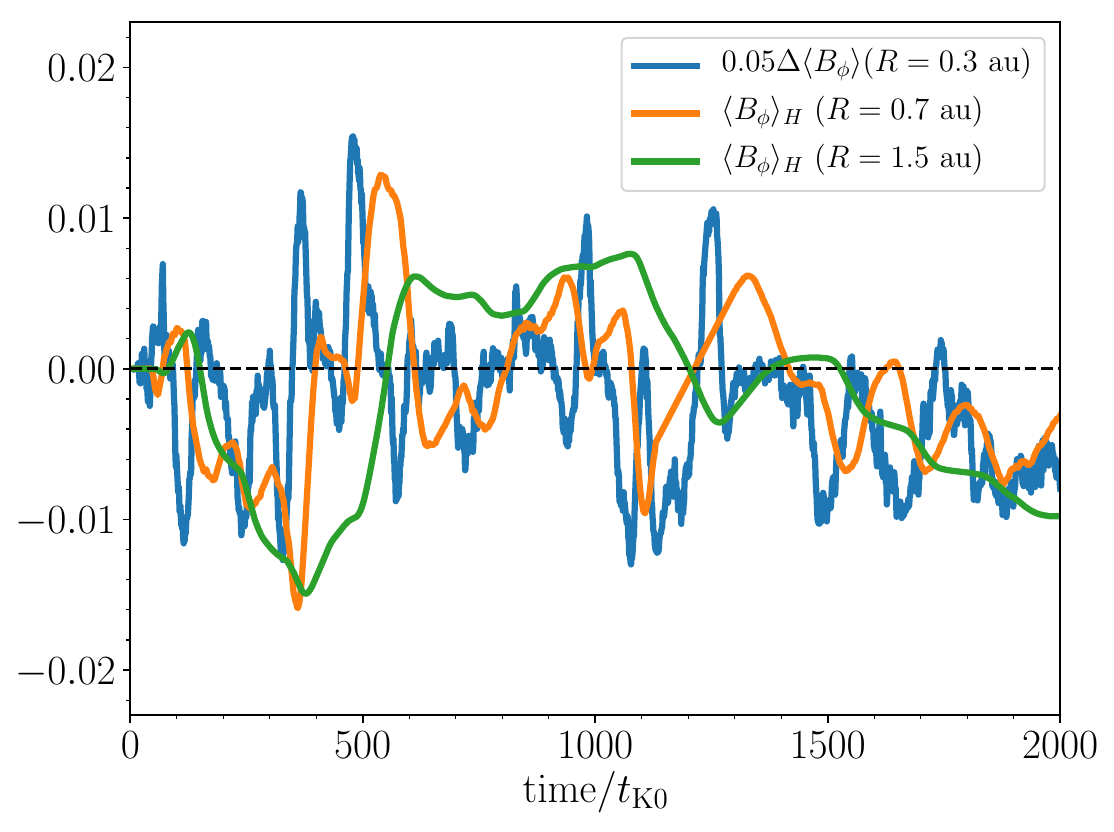}
    \end{center}
    \caption{
    Time variations of $B_\phi$ at three different radii.
    The blue line shows 
    $\Delta \avphi{B_\phi} \equiv \avphi{B_\phi}(z=-2.5H) + 
    \avphi{B_\phi}(z=2.5H)$ at $R=0.3~\mathrm{au}$, 
    and $\Delta \avphi{B_\phi}$ is multiplied by 0.05 
    to facilitate comparison with the other two lines.
    The orange and green lines correspond to 
    $\avphi{B_\phi}_H$ at $R=0.7$~au and $1.5$~au, respectively.
    All the quantities are shown in the code units.
    }
    \label{fig:Bphi_timeevo}
\end{figure}

In order to investigate the origin of the quasi-periodic time variation in $\avphi{B_\phi}_H$ in 
$\Rmri\lesssim R \lesssim \Rch$, 
we compare the time variations of 
$B_\phi$ at three different radii, 
$R=0.3$~au, 0.7~au, and 1.5~au.
In the active zone, net toroidal field 
$\avphi{B_\phi}$ is almost zero for $|z|\le 2H$ 
because of the MRI turbulence, and 
the quasi-periodic variations of 
$\avphi{B_\phi}$ are seen in higher latitudes as  the
butterfly structure in the time-$z$
diagram (see Figure \ref{fig:active_ver}a).
In order to quantify the time 
variations of $\avphi{B_\phi}$, we 
define 
$\Delta \avphi{B_\phi} \equiv \avphi{B_\phi}(z=-2.5H) + 
\avphi{B_\phi}(z=2.5H)$ at $R=0.3~\mathrm{au}$.
Since the signs of $\avphi{B}_\phi$ in $z>0$ 
and $z<0$ are opposite,
$\Delta \avphi{B_\phi}$ is a measure of the 
antisymmetry of $\avphi{B_\phi}$ with respect to the mid-plane.

Figure \ref{fig:Bphi_timeevo} shows that 
the time variations of $\Delta \avphi{B_\phi}$ at $R=0.3~\au$
are correlated with those of $\avphi{B_\phi}_H$ at $R=0.7~\au$. 
This means that the drift of the magnetic fields 
in the active zone disturbs 
the transition zone.
The variations of $\avphi{B_\phi}_H$ in the transition zone 
propagate rapidly in the radial direction owing to 
efficient magnetic diffusion.
This behavior can be seen in Figure \ref{fig:Bphi_timeevo} 
that shows a correlation between 
$\avphi{B_\phi}_H$ at $R=1.5$~au and $\avphi{B_\phi}_H$ at
$R=0.7~$au.

\begin{figure}
    \begin{center}
    \includegraphics[width=8cm]{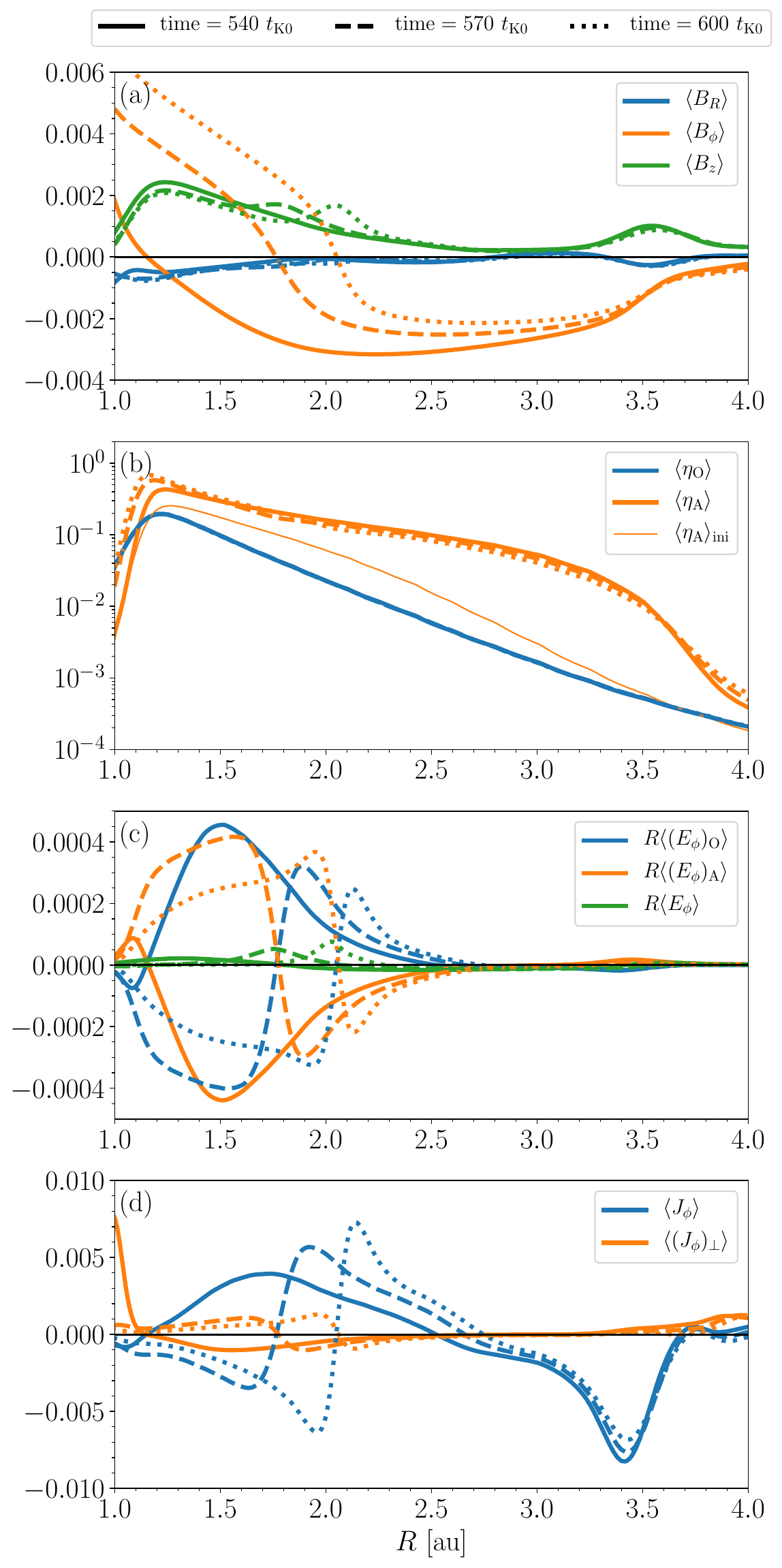}
    \end{center}
    \caption{
    Radial profiles of quantities related to diffusion of the magnetic fields in the coherent zone.
    In each panel, the solid, dashed, and dotted lines correspond to 
    the results at the mid-plane at $t=540~\tK$, $570~\tK$, and 
    $600~\tK$, respectively.
    (a) Radial profiles of  $\avphi{B_R}$, 
     $\avphi{B_\phi}$, and  $\avphi{B_z}$.
    (b) Radial profiles of 
     $\avphi{\etaO}$  and  $\avphi{\etaA}$.
    The thin solid orange line shows the radial profile 
    of $\avphi{\etaA}$ at the initial condition.
    (c) Radial profiles of $R\avphi{(E_\phi)_\mathrm{O}}$,
     $R\avphi{(E_\phi)_\mathrm{A}}$, and 
     $R\avphi{E_\phi} = 
    R\left[\avphi{(E_\phi)_\mathrm{I}} + \avphi{(E_\phi)_\mathrm{O}} +
    \avphi{(E_\phi)_\mathrm{A}}
    \right]$. 
    They determine the radial diffusion of $\avphi{B_z}$.
    (d) Radial profiles of 
     $\avphi{J_\phi}$ and  $\avphi{(J_\phi)_\perp}$.
    All the quantities are shown in the code units.
    }
    \label{fig:Eph_eta}
\end{figure}

Next, we investigate the radial diffusion of the magnetic fields 
in the coherent zone ($R>R_\eta$).
We focus on the second flip of $\avphi{B_\phi}_H$ starting around $t\sim 500~\tK$ (Figure \ref{fig:tRdiagram}b).

Figure \ref{fig:Eph_eta}a shows the time evolution of the radial profiles 
of $\avphi{B_R}$, $\avphi{B_\phi}$, and $\avphi{B_z}$ at the mid-plane.
When the time passes from $t=540~\tK$ to $600~\tK$, 
the radius where the sign of $\avphi{B_\phi}$ is reversed 
moves outward while $\avphi{B_R}$ and $\avphi{B_z}$ do not change significantly. 
At $t=600~\tK$, a concentration of $\avphi{B_z}$ appears around $R\sim 2.1~\au$ where 
the sign of $\avphi{B_\phi}$ is flipped.

The radial diffusion speed of $\avphi{B_\phi}$ is roughly estimated by 
$\avphi{\etaO}$ and $\avphi{\etaA}$, which are shown in Figure \ref{fig:Eph_eta}b.
$\avphi{\etaA}$ increases by increasing the field strength while 
$\avphi{\etaO}$ does not change in time significantly.
Around $R\sim 1.2~\au$, $\avphi{\etaO}$ and $\avphi{\etaA}$ are both roughly $\sim 0.3$.
The spatial scale of the spatial variation of $\avphi{B_\phi}$ is around $\Delta R \sim 0.5~\au$.
The typical speed of the radial diffusion of the magnetic fields is estimated 
to be $\avphi{\eta}/\Delta R \sim 0.1~\mathrm{au}~\tK^{-1}$, indicating 
that the structure of $\avphi{B_\phi}$ moves 1~au per $10~\tK$.
Although this speed is a few times larger than estimated from Figure \ref{fig:Eph_eta}a,
they are consistent based on the rough estimate. 

The $\avphi{B_z}$ concentration 
is determined by $R\avphi{E_\phi}$.
Figure \ref{fig:Eph_eta}c shows the radial profiles of $R\avphi{E_\phi}$ and 
the contributions of OR and AD to $R\avphi{E_\phi}$
which are denoted by
$R\avphi{(E_\phi)_\mathrm{O}}$ and $R\avphi{(E_\phi)_\mathrm{A}}$, respectively. 
The contribution from the ideal-MHD (induction) term 
$-\bm{v}\times \bm{B}$
is not plotted in Figure \ref{fig:Eph_eta}c because it is negligible.
Figure \ref{fig:Eph_eta}c shows that 
$R\avphi{(E_\phi)_\mathrm{O}}$ and $R\avphi{(E_\phi)_\mathrm{A}}$ have 
opposite signs and nearly cancel each other.
This indicates that AD behaves anti-diffusion 
since OR provides pure diffusion of magnetic fields. 
As shown in Figure \ref{fig:Eph_eta}d, 
the opposite sign of $R\avphi{(E_\phi)_\mathrm{O}}$ and 
$R\avphi{(E_\phi)_\mathrm{A}}$ 
comes from the fact that the sign of $\avphi{J_\phi}$ 
is opposite to $\avphi{(J_\phi)_\perp}$ 
since $\avphi{(E_\phi)_\mathrm{O}}\propto \avphi{J_\phi}$.
and $\avphi{(E_\phi)_\mathrm{A}}\propto \avphi{J_\phi}_\perp$.
This feature was previously pointed out by \citet{Bethune2017}. 
Anti-diffusion owing to AD is not 
completely cancelled out 
by the normal diffusion owing to OR,
and triggers the concentration of $\avphi{B_z}$.

Comparison between Figures \ref{fig:tRdiagram}a and \ref{fig:tRdiagram}b shows that 
the concentrations of $\avphi{B_z}_H$ are located at the radii where 
the sign of $\avphi{B_\phi}_H$ is flipped for $R\gtrsim R_\eta$.
The outward propagation of the concentrations of $\avphi{B_z}_H$ slows 
down as the radius increases, and 
almost stall around $R\sim \Rch$.
This is because $\etaA$ decreases
rapidly beyond $R=\Rch$, 
resulting in a rapid decrease of the propagation speed of the condensations of $\avphi{B_z}_H$.

\subsubsection{Radial Propagation of the Velocity Disturbances Originating from 
the MRI Turbulence}\label{sec:sound}

The MRI turbulence driven in the active zone generates velocity disturbances that 
propagate outward and disturb the transition and coherent zones as seen in Figure \ref{fig:overall_slice}b and 
\ref{fig:overall_slice_outer}b.

\begin{figure}
    \begin{center}
   \includegraphics[width=8cm]{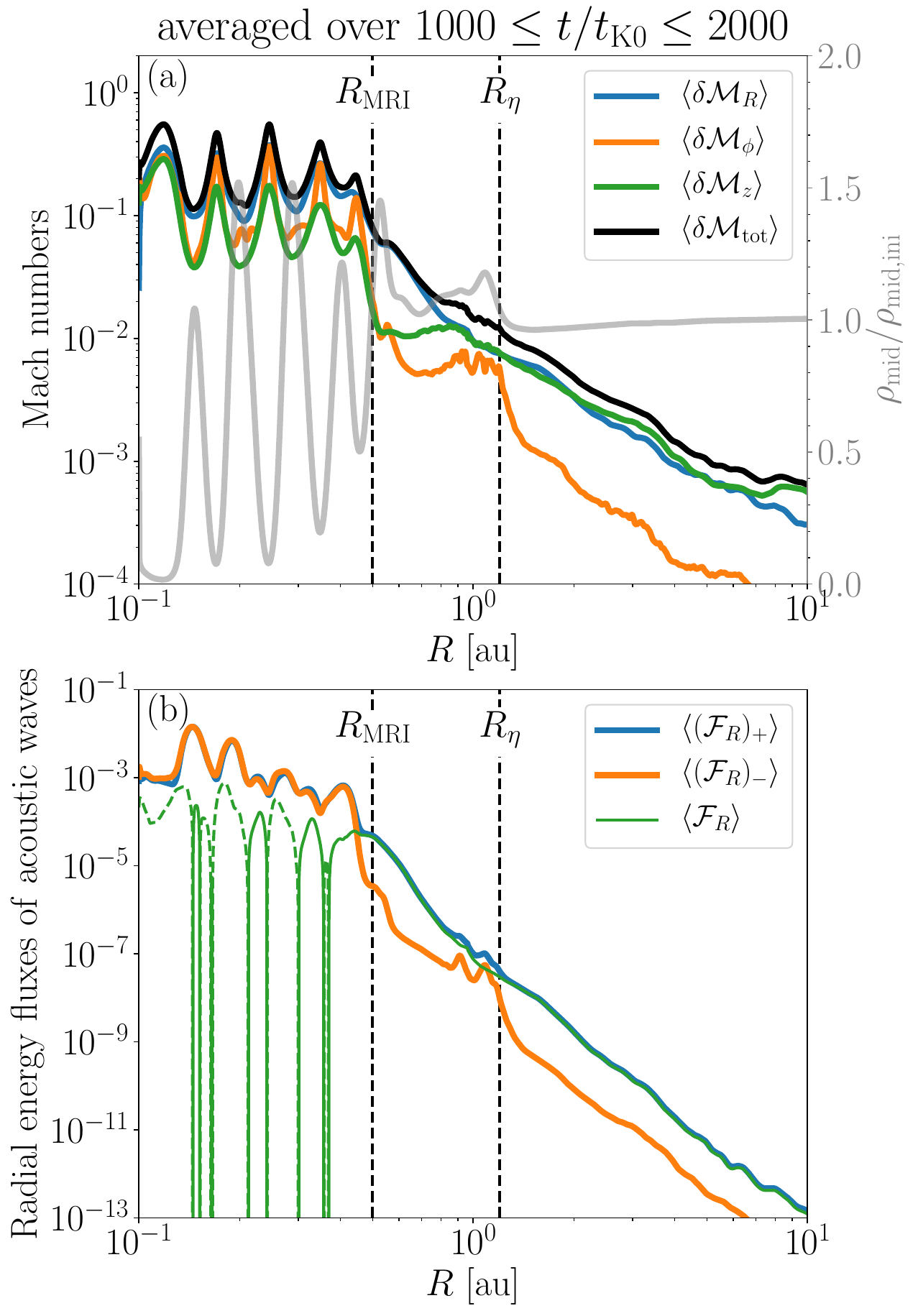}
    \end{center}
    \caption{
    Radial propagation of the sound waves in the whole disk.
    (a) The Mach numbers in the $R$, $\phi$, and $z$ directions at 
    the mid-plane as a function of $R$ (the left vertical axis).
    The radial profile of the mid-plane density is shown by the gray line (the right vertical axis).
    (b) The radial profiles of the energy fluxes of the sound waves.
    The blue, orange, and green lines correspond to 
    $\avphi{ ({\cal F}_R)_+}$,
    $\avphi{ ({\cal F}_R)_-}$, and $\avphi{ {\cal F}_R}$ at the mid-plane, respectively.
    In all the panels, all the data are averaged over $1000\le t/\tK\le 2000$, and 
    the vertical dashed lines denote $R=\Rmri$ and $R_\eta$ from left to right.
    All the quantities are shown in the code units.
    }
    \label{fig:soundwave}
\end{figure}

Figure \ref{fig:soundwave}a shows the Mach numbers of the velocity dispersion in 
the $R$, $\phi$, and $z$ directions at the mid-plane.
They are defined as
\begin{equation}
    \langle \delta {\cal M}_{R,\phi,z}\rangle 
    = \sqrt{
    \frac{\bigl \langle {\rho \delta v_{R,\phi,z}^2}\bigr \rangle}{\avphi{P}}}.
\end{equation}
Inside the active zone, $\avphi{ {\cal M}_{R,\phi,z}}$ are 
as large as $\sim 0.1$.
Comparison between the Mach numbers and density in Figure \ref{fig:soundwave}a 
shows that the velocity dispersion is anti-correlated with the density; 
it is larger in the gaps than in the rings.

Inside the transition zone ($\Rmri < R < R_\eta$), 
$\dmachR$, $\dmachphi$, and $\dmachz$ behave differently.
$\dmachR$ continues to decrease in the transition zone, 
and reaches $\sim 10^{-2}$ at $R=R_\eta$.
By contrast, 
after the rapid decrease at $R\approx \Rmri$, 
$\dmachphi$ and $\dmachz$ become almost constant in the outer part of the transition zone.
Around $R=R_\eta$, all the components of the Mach numbers 
have a similar value of $\sim 10^{-2}$.

In the coherent zone ($R\ge R_\eta$), 
$\dmachR$ and $\dmachz$ are of similar magnitude and decrease with increasing radius. 
$\dmachphi$ is smaller than the other two.

We investigate the radial propagation of the radial velocity fluctuations.
The energy flux of the sound waves is estimated as 
\begin{equation}
    \avphi{ {\cal F}_{R}} = \avphi{\delta \rho \delta v_R} c_\mathrm{s}^2,
\end{equation}
where $\delta \rho \equiv \rho - \avphi{\rho}$.
It can be divided into the energy flux propagating in the $+R$ direction 
\begin{equation}
   \avphi{\left( {\cal F}_{R}\right)_+} = c_\mathrm{s} 
   \biggl \langle \rho \left\{ 
   \frac{1}{2}\left(\delta v_R + 
   c_\mathrm{s}\frac{{\delta \rho}}{\avphi{\rho}}\right)\right\}^2\biggr \rangle,
\end{equation}
and that propagating in the $-R$ direction
\begin{equation}
    \avphi{\left({\cal F}_{R}\right)_-} = c_\mathrm{s} 
   \biggl \langle \rho \left\{ 
   \frac{1}{2}\left(\delta v_R - 
   c_\mathrm{s}\frac{{\delta \rho}}{\avphi{\rho}}\right)\right\}^2\biggr \rangle.
\end{equation}
They satisfy $\avphi{ {\cal F}_R} =  \avphi{({\cal F}_R)_+} - \avphi{({\cal F}_R)_-}$.

The radial distributions of 
$\avphi{({\cal F}_{R})_+}$, $\avphi{({\cal F}_R)_-}$, and 
$\avphi{{\cal F}_R}$ are shown in 
Figure \ref{fig:soundwave}b.
In the active zone, 
$\avphi{({\cal F}_R)_+}$ and $\avphi{({\cal F}_R)_-}$  
are comparable;
there are equal amounts of outward and inward propagating sound waves.

Outside the active zone, 
$\avphi{({\cal F}_R)_+}$ dominates over $\avphi{({\cal F}_R)_-}$.
This clearly shows that 
the sound waves transport more energy outward.
Around $R\sim R_\eta$, $\avphi{({\cal F}_R)_-}$ is enhanced and becomes comparable to 
$\avphi{({\cal F}_R)_+}$ although the outward flux is still larger than the inward flux.
This indicates that a part of the outgoing waves are reflected by the 
density bump around $R=R_\eta$ (Figure \ref{fig:soundwave}a).

\section{ Detailed Analyses on Transfer of Mass, Angular Momentum, and 
Magnetic Flux }\label{sec:mass_angmom_flux}

\subsection{Mass and Angular Momentum Transfer Throughout the Disk}\label{sec:angmomtrans}

In this section,
we investigate the radial dependence of the mass accretion rate throughout the disk. 
The mass transfer rate at a given $R$ is defined as
\begin{equation}
    \dot{M}^{z_\mathrm{b}}  = \int_{-z_\mathrm{b}}^{\zb} 2\pi R\avphi{\rho v_R}dz,
\end{equation}
where $\zb$ is a reference height, 
and it will take the value of either $H$ or $z_\mathrm{atm}$.

We derive some equations to analyze the angular momentum transfer.
We start from 
the continuity equation 
\begin{equation}
     \frac{\partial \avphi{\rho}}{\partial t} 
     + \frac{1}{R} \frac{\partial}{\partial R} \left( 
R \avphi{\rho v_R}\right)
    + \frac{\partial}{\partial z} \left( \avphi{\rho v_z}\right)=0,
    \label{eoc_phiav}
\end{equation}
and the momentum conservation equation in the $\phi$ direction 
\begin{eqnarray}
    \frac{\partial \avphi{\rho v_\phi}}{\partial t} 
    &+ &\frac{1}{R^2} \frac{\partial}{\partial R} \left[ 
R^2 \left( \avphi{\rho v_R v_\phi}
- \frac{\avphi{B_RB_\phi}}{4\pi} \right)\right] \nonumber \\
   & +& \frac{\partial}{\partial z} \left( 
    \avphi{\rho v_{\phi}v_z}- \frac{\avphi{B_z B_\phi}}{4\pi}\right)
    =0,
    \label{momphi_phiav}
\end{eqnarray}
where they are averaged over $0\le \phi < 2\pi$.

The $R\phi$ and $\phi z$ components of the Reynolds stress tensor are 
decomposed into the laminar and turbulent parts,
\begin{equation}
    \avphi{\rho v_i v_\phi}= \avphi{\rho v_i}\avphi{v_\phi}+ 
    \avphi{\rho \delta v_i \delta v_\phi},
    \label{rhovivphi}
\end{equation}
where $i=(R,z)$ (Equation (\ref{vBaveturb})).
Combining the Maxwell stress tensor and the turbulent Reynolds stress tensor, 
we define the $R\phi$ and $\phi z$ stresses as 
\begin{equation}
    \avphi{W_{R\phi}}= \avphi{\rho \delta v_R \delta v_\phi}
    - \frac{\avphi{B_R B_\phi}}{4\pi},
    \label{WRphi}
\end{equation}
and
\begin{equation}
    \avphi{W_{\phi z}}= \avphi{\rho \delta v_z \delta v_\phi}
    - \frac{\avphi{B_\phi B_z}}{4\pi},
    \label{Wphiz}
\end{equation}
respectively.
Substituting Equations (\ref{rhovivphi})-(\ref{Wphiz}) into 
Equations (\ref{eoc_phiav}) and (\ref{momphi_phiav}) and
averaging 
Equations (\ref{eoc_phiav}) and (\ref{momphi_phiav}) 
over $t_1\le t\le t_2$,
one obtains a prediction of the radial mass flux $\dot{M}_\mathrm{pred}^{\zb}$, which 
is given by integrating 
$2\pi R\avphi{\rho v_R}$ over $-\zb\le z\le \zb$ as a function of radius as follows:
\begin{equation}
    \dot{M}_\mathrm{pred}^{\zb}
    = \dot{M}_{R\phi}^{\zb}
    + \dot{M}_{\phi z}^{\zb}
    + \dot{M}_{\rho v_z}^{\zb}
    \label{mass_acc_pred}
\end{equation}
where 
\begin{equation}
    \dot{M}_{R\phi}^{\zb} = \frac{4\pi R}{\OmegaK} 
    \int_{-\zb}^{\zb}
    {\cal J}_{R\phi}
    dz,\;\;
    {\cal J}_{R\phi} = -\frac{1}{R^2} \frac{\partial}{\partial R}
    \left( R^2 \avphi{W_{R\phi}}_t \right),
    \label{rhovR_Rphi}
\end{equation}
\begin{equation}
    \dot{M}_{\phi z}^{\zb} =  \frac{4\pi R}{\OmegaK} 
    \int_{-\zb}^{\zb}
    {\cal J}_{\phi z}
    dz,
    \;\;
    {\cal J}_{\phi z} = 
    -\frac{\partial \avphi{W_{\phi z}}_t}{\partial z} 
    \label{rhovR_phiz}
\end{equation}
and 
\begin{equation}
    \dot{M}_{\rho v_z}^{\zb} = - \frac{4\pi R}{\OmegaK}
    \frac{1}{t_2-t_1}\int_{t_1}^{t_2}dt
    \int_{-\zb}^{\zb}\avphi{\rho v_z}\frac{\partial \avphi{v_\phi}}{\partial z}dz,
\end{equation}
where the terms with $\partial /\partial t$ are neglected.
In the derivation of the above equations, $\avphi{v_\phi}=R\OmegaK$ is assumed.
We note that this approximation is not very accurate in the ring and gap regions 
where the rotation velocity deviates from the Keplerian profile due to the gas pressure gradient.
The first and second terms correspond to the radial mass fluxes driven by the torques due to
$\avphi{W_{R\phi}}$ and $\avphi{W_{\phi z}}$, respectively. 
These two provide the dominant contribution to $\dot{M}_\mathrm{pred}^{z_\mathrm{b}}$.
The third term comes from the radial mass flux affected by the vertical mass flux.

\begin{figure*}
    \begin{center}
    \includegraphics[width=17cm]{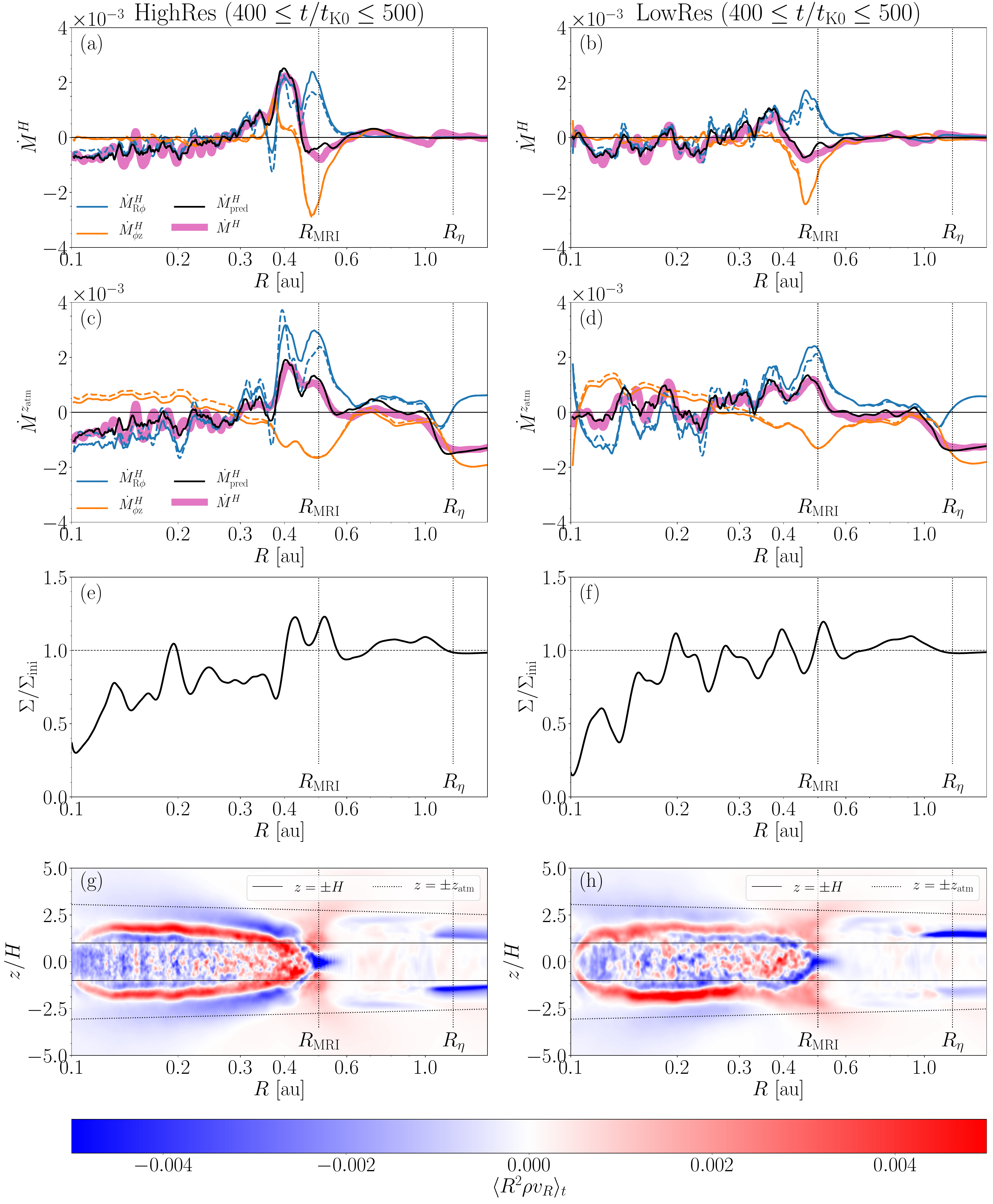}
    \end{center}
    \caption{
    Radial dependence of the mass transfer rate in the early stage for HighRes (left) and LowRes (right).
    All the quantities shown in this figure are averaged over $400\tK\le t\le 500\tK$.
    The first row compares  the mass transfer rate (pink) integrated over $|z|\le H$ and  their predictions (black) given 
    by Equation (\ref{mass_acc_pred}). The blue and orange  solid lines 
    correspond to $\dot{M}_{R\phi}^H$ and $\dot{M}_{\phi z}^H$, respectively.
    The blue and orange dashed lines represent $\dot{M}_{R\phi}^H$ and $\dot{M}_{\phi z}^H$
    evaluated without the Reynolds stress, respectively.
    The second row is the same as the first row but the mass transfer rate is integrated over $|z|\le z_\mathrm{atm}$.
    The radial profiles of the surface densities are shown in the third row.
    In the fourth row, the color maps of 
    $R^2 \avphi{\rho v_R}_t$ are displayed in the plane of ($\log_{10}R, z/H$), 
    and the black solid and dashed lines correspond to $z=\pm H$ and $z=\pm \zatm$.
    In each panel, 
    the two vertical dotted lines show $R=\Rmri$ and $R=\Reta$.
    In the first and second row, to reduce significant spatial fluctuations of 
    the mass  transfer rates $\dot{M}_{R\phi}^{H,\zatm}$, 
    $\dot{M}_{\phi z}^{H,\zatm},\dot{M}_\mathrm{pred}^{H,\zatm}$, 
    the centered moving averages over $\Delta R/R = 0.4$ are taken.
    All the quantities are shown in the code units.
    }
    \label{fig:angmom_early}
\end{figure*}

\subsubsection{Early Evolution}\label{sec:earlyevo_angmom}

First, we investigate the angular momentum transfer 
in the early evolution $400\tK \le t \le 500\tK$. The simulation time is long enough for 
the turbulence in the active zone to reach saturation (Figure \ref{fig:resolution}),
but not long enough to capture the secular evolution, 
such as the ring formation and flux concentration
(Figure \ref{fig:Ring}).
At $R=R_\eta$, the gas rotates only 14 
rotations at $t=450\tK$, indicating 
that both the transition and coherent zones have not reached quasi-steady states.
In this section, we investigate the angular 
momentum transport mechanism especially in the active zone and around the inner edge of the transition zone. 

\noindent
\\
{\bf Mass Transfer in the Active Zone}

The top and second rows of Figure \ref{fig:angmom_early} 
show comparison between the measured and predicted mass transfer rates
by integrating over $|z|\le H$ and $|z|\le \zatm$, respectively.
Equation (\ref{mass_acc_pred}) reproduces the actual mass  transfer rates reasonably well for all the panels.

Next, we investigate the mass transfer in the active zone around the mid-plane ($|z|\le H$).
Figures \ref{fig:angmom_early}a and \ref{fig:angmom_early}b  show that 
the mass transfer is mainly driven by 
the $R\phi$ torque due to the MRI turbulence while the contribution from the $\phi z$ torque is negligible.
 For both torques, the Maxwell stress plays a more dominant role than the Reynolds stress does. 
The active zone shows that the radial mass fluxes are 
negative at $R\lesssim 0.25~$au while 
they are positive in the outer region.
This is a typical feature of viscous evolution of an isolated accretion disk where 
the outer region expands outward by receiving 
the angular momentum from the inner region 
where the gas accretes inwards.
The resolution dependence of the radial mass flux is seen 
especially in the expanding outer region
($0.25~\mathrm{au}\lesssim R\lesssim \Rmri$).
The outward radial mass flux around $R\sim 0.4~\mathrm{au}$ is 
more significant in HighRes than in LowRes run.

When the upper layers of the disk is considered in the estimation of 
the mass  transfer rate, the resolution dependence becomes more significant.
For HighRes run, the mass transfer rate is not influenced by 
the upper layers significantly, or $\dot{M}^H\sim \dot{M}^{\zatm}$ 
(Figures \ref{fig:angmom_early}a and \ref{fig:angmom_early}c).
By contrast, LowRes run shows that 
$\avphi{W_{\phi z}}$ owing to large-scale magnetic fields in the upper layers 
makes $\dot{M}_{\zatm}$ positive in the inner active zone 
although $\dot{M}^H$ is negative there.
The corresponding structure in  Figure \ref{fig:angmom_early}h 
is the outgoing gas upper layers in the active zone.
This is because the magnetic fields are more coherent for LowRes run than HighRes run.

\begin{figure}
    \begin{center}
    \includegraphics[width=8cm]{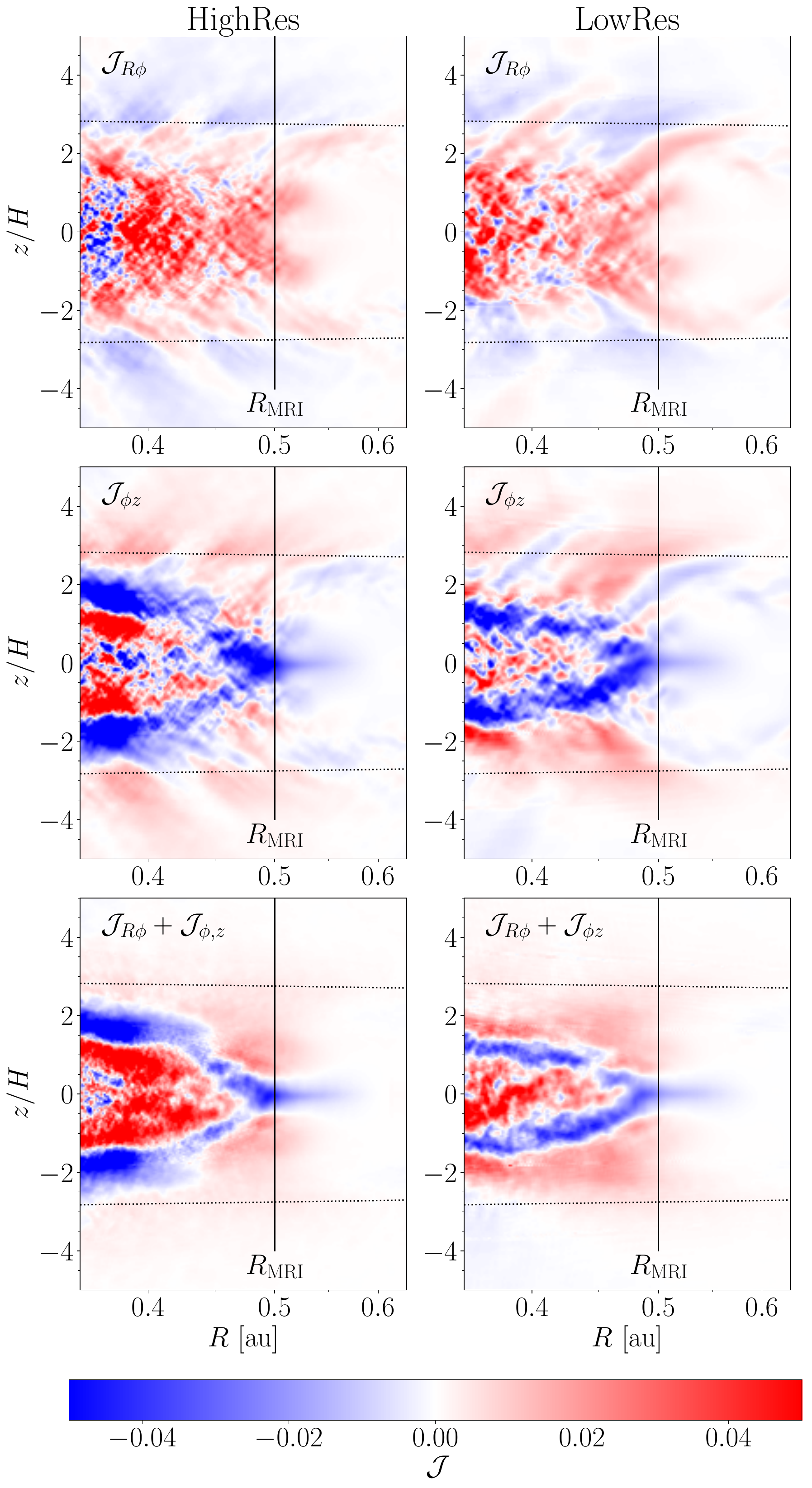}
    \end{center}
    \caption{
    Color maps of ${\cal J}_{R\phi}$, ${\cal J}_{\phi z}$, and ${\cal J}_{R\phi}+{\cal J}_{\phi z}$ averaged over 
    $400\tK\le t\le 500\tK$ for  HighRes ({\it left panels}) and LowRes runs ({\it right panels}).
    The dotted lines correspond to $z=\pm \zatm $, respectively.
    All the quantities are shown in the code units.
    }
    \label{fig:angmom_rph_phz}
\end{figure}

\noindent
\\
{\bf Mass Transfer Around the Inner Edge of the Transition Zone}

Next, we investigate the mass  transfer around the inner edge of the 
transition zone.
Since the angular momentum transfer mechanisms change significantly 
both radially and vertically, 
it is unclear which heights of the gas mainly contribute to 
$\dot{M}_{R\phi}$ and $\dot{M}_{\phi z}$ at a given radius only from 
the first and second rows of Figure \ref{fig:angmom_early}.
The color maps of the integrands of $\dot{M}_{R\phi}$ and $\dot{M}_{\phi z}$, or ${\cal J}_{R \phi}$ and 
${\cal J}_{\phi z}$ (see Equations (\ref{rhovR_Rphi}) and (\ref{rhovR_phiz})) are displayed 
in Figure \ref{fig:angmom_rph_phz}.

The bottom rows of Figures \ref{fig:angmom_early}
and \ref{fig:angmom_rph_phz}, or 
the distributions of
$\avphi{\rho v_R}$ and ${\cal J}_{R\phi} + {\cal J}_{\phi z}$,
look quite similar, showing that 
the radial mass flux is determined by the local torque 
acting on the gas.
Interestingly, Figure \ref{fig:angmom_rph_phz} shows that 
small-scale substructures in ${\cal J}_{R\phi}$ and ${\cal J}_{\phi z}$ 
have complementary spatial distributions to each other, and 
their sum has a much smoother distribution.

As explained in Section \ref{sec:structureform_transition}, 
the hourglass-shaped magnetic field with a steep gradient of $B_\phi$ 
at the mid-plane is generated mainly by AD (Figure \ref{fig:dBphidt_0.5}b).
The middle row of Figures \ref{fig:angmom_rph_phz} clearly shows that 
the $\phi z$ stress drives gas accretion at the mid-plane. 
By contrast, the outflows just above the mid-plane accreting layer identified 
in the bottom row of Figure \ref{fig:angmom_early}
are driven by the $R \phi$ stress.

From Figures \ref{fig:angmom_early}a and \ref{fig:angmom_early}b, 
it is found that the gas within $|z|\le H$ moves inward 
around $R=\Rmri$
because the gas accretion rate due to ${\cal J}_{\phi z}$ 
dominates over the outflow rate due to ${\cal J}_{R\phi}$.
When the contribution of the upper layers ($H\le |z|\le \zatm$) 
is taken into account, the net mass flux $\dot{M}^{\zatm}$ 
is directed outward (Figures \ref{fig:angmom_early}c and \ref{fig:angmom_early}d),
leading to the mass supply from the active zone to the transition zone (Section \ref{sec:masstrans}).

As one moves inward from $R=\Rmri$, the layer with ${\cal J}_{\phi z}<0$ 
is divided into two layers that sandwitch the 
active zone in between (the middle row of Figure \ref{fig:angmom_rph_phz}).
Inside the active zone, the $R\phi$ stress moves the gas outward.
Just above the two accreting layers, the outflows are driven mainly by 
the $\phi z$ stress, although the $R\phi$ stress also contributes.

\noindent
\\
{\bf Ring Structures Around the Inner Edge of the Transition Zone}

The surface density radial profiles are shown in Figures \ref{fig:angmom_early}e 
and \ref{fig:angmom_early}f. 
Roughly speaking, the gas is accumulated around the 
inner-edge of the transition zone by outward 
turbulent diffusion in the outer region of the active zone.
Looking more closely at the inner edge of the 
transition zone in Figure \ref{fig:angmom_early}e, 
there are two density peaks around 
$R\sim 0.43~\mathrm{au}$ and $R\sim 0.52~\mathrm{au}$ 
for HighRes run. The inner peak is formed by the viscous 
expansion of the active zone 
while the outer peak is formed by the wind launched 
around the inner-edge of the transition zone 
(Section \ref{sec:structureform_transition}).
The gap between the two peaks is located at the 
boundary between 
the expanding layer due to the MRI turbulence 
and the mid-plane accreting layer (Figure \ref{fig:angmom_early}h). 
For LowRes run, the inner density peak 
is slightly closer to the central star than for 
HighRes run since the outward radial mass flux around 
$R\sim 0.4~\au$ is smaller 
for LowRes run (Figures \ref{fig:angmom_early}c and \ref{fig:angmom_early}d).

\subsubsection{Later Evolution}\label{sec:angmom_later}

\begin{figure}
    \begin{center}
    \includegraphics[width=8cm]{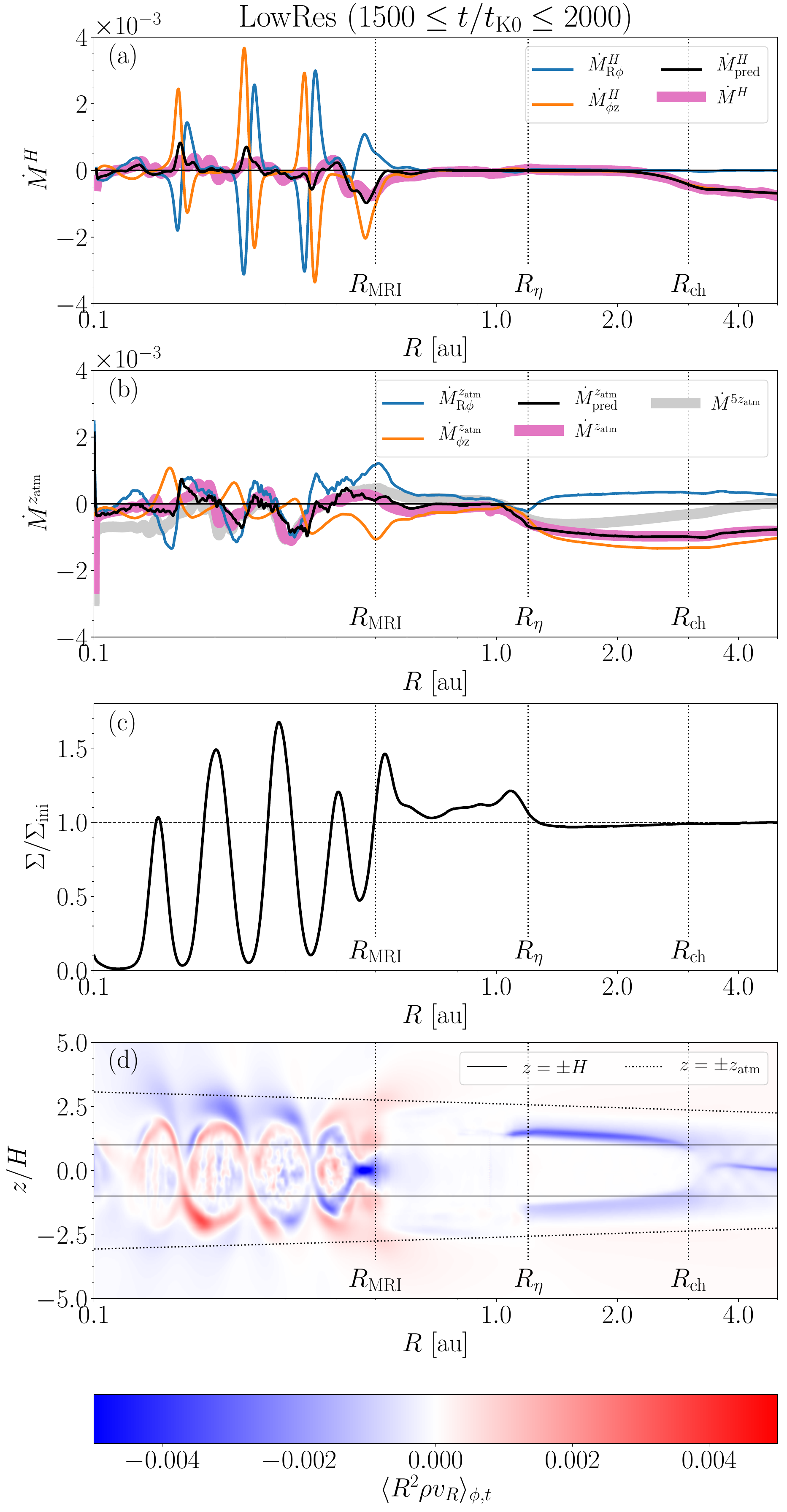}
    \end{center}
    \caption{
    The same as Figure \ref{fig:angmom_early} but the quantities are averaged over $1500\le t/t_\mathrm{K0}\le 2000$
    for LowRes run.
    In Panel (b), the {\color{red} mass transfer} rate integrated over $|z|\le 5\zatm$ is plotted.
    In Panels (a) and (b), to reduce significant spatial fluctuations of 
    the mass transfer rates $\dot{M}_{R\phi}^{H,\zatm}$, 
    $\dot{M}_{\phi z}^{H,\zatm},\dot{M}_\mathrm{pred}^{H,\zatm}$,  
    and $\dot{M}^{H,\zatm,5\zatm}$, 
    the centered moving averages over $\Delta R/R = 0.4$ are taken.
    All the quantities are shown in the code units.
    }
    \label{fig:angmom}
\end{figure}

Figure \ref{fig:angmom} is the same as Figure \ref{fig:angmom_early} 
but showing the late evolution for LowRes run.
In the active zone, 
as shown in Section \ref{sec:structureform_active}, 
the ring structures and flux concentrations develop.

Unlike in the early evolution, 
the outer regions of the active zone ($R\sim 0.4~\au$) do not show 
outward mass flux $\dot{M}^H>0$ and $\dot{M}^{\zatm}>0$
in the long-term evolution (Figures \ref{fig:angmom}a and \ref{fig:angmom}b).
The net mass transfer rates
through the disk with $|z|\le H$ and $|z|\le \zatm$ are almost zero.
One can identify the gas flow converging to the rings both in Figures \ref{fig:angmom}a and 
\ref{fig:angmom}b.

The fact that 
gas accretion does not occur around the disk mid-plane
may be due to the ring structures.
\citet{Jacquemin-Ide2021} investigated an MRI disk (model SEp) similar to 
our active zone and found that the radial migration of the ring structures is not seen.
However, the physical mechanism is unclear and needs to be investigated.
We note that the insufficient resolution of LowRes run may 
cause the almost zero accretion rate.
The gas accretion continues in the higher latitudes although 
it halts inside the disk.
This is evident that 
the mass transfer rate integrated over $|z|\le 5\zatm$ ($\dot{M}^{5\zatm}$) 
takes negative values except around the rings
(also see Figure \ref{fig:massactive}).

The behaviors around the inner edge of the transition zone 
shown in Figure \ref{fig:angmom} are 
similar to those shown in Figure \ref{fig:angmom_early}.
Inside the disk ($R\sim \Rmri$, $|z|\le H$), 
the mass is transferred inward by the $\phi z$ torque due 
to the hour-glass-shaped magnetic fields. 
When the upper layers of the disk ($|z|\le \zatm$) is considered,
the inward angular momentum transfer is almost compensated 
by the outward angular momentum transfer owing to the $R\phi$ torque.
In Section \ref{sec:masstrans}, we will see that 
the absence of net angular momentum transport around the inner edge of the transition zone 
results in almost zero radial mass transfer from the active zone to the transition zone
(Figures \ref{fig:masstrans}a and \ref{fig:masstrans}c).

For $R_\eta\le R \lesssim \Rch$, surface gas accretion flows are seen in 
$H<|z|<\zatm$ in Figure \ref{fig:angmom}d.
The surface gas accretion flows are anti-symmetric with respect to the mid-plane 
because the current sheet is not 
at the mid-plane but at the lower AD dead-zone boundary (see Figure \ref{fig:overall_slice}c).
Figures \ref{fig:angmom}a and \ref{fig:angmom}b show that
this accretion flow is driven by the magnetic braking owing 
to the coherent magnetic field ($\avphi{W}_{\phi z}$).
Since the outer region of the transition zone does not have gas accretion, 
the gas is accumulated around $R=R_\eta$ (Figure \ref{fig:angmom}c).

The mass transfer rate in the coherent zone is determined by $\avphi{W_{\phi z}}$.
Using the ratio $f_{\phi z}=B_\phi/B_z$ and Equation (\ref{rhovR_phiz}), the mass transfer rate driven by magnetic braking is estimated to be 
\begin{equation}
    \dot{M} = -\frac{8\pi R}{\Omega_\mathrm{K}} \frac{f_{\phi z} B_z^2}{4\pi} = 
    -2\pi R \Sigma c_\mathrm{s,mid}
    \times \left(4\sqrt{2/\pi} f_{\phi z} \beta_z^{-1} \right),
    \label{dM_dead}
\end{equation}
where $\beta_z = 8\pi \rho_\mathrm{mid} c_\mathrm{s,mid}^2/B_{z}^2$.
Equation (\ref{dM_dead}) predicts that 
$\dot{M} \sim -10^{-3}$ in the code units at $R=1~\au$ and 
$|\dot{M}|$ decreases in proportion to 
$\propto R^{-1/4}$ since $\Sigma \propto R^{-1}$ (Equation (\ref{sigma})),
where we use the fact that the radial magnetic flux transport from the transition zone decreases $\beta_z$ in the coherent zone 
by a factor of two from the initial plasma beta $10^4$ 
(Section \ref{sec:loopform})
and $f_{\phi z}$ is estimated to $\sim 5$ from the simulation result.

Beyond $R\sim \Rch$, the surface gas accretion flows disappear 
and mid-plane gas accretion is driven by the magnetic torque exerted 
at the mid-plane (Figures \ref{fig:angmom}a and \ref{fig:angmom}d)
since $\etaO$ becomes small enough 
for the toroidal field to be amplified inside the disk (Figure \ref{fig:overall_slice_outer}c).

\subsection{Time Evolution of the Masses of 
the Three Zones}\label{sec:masstrans}

\begin{figure*}
    \begin{center}
   \includegraphics[width=17cm]{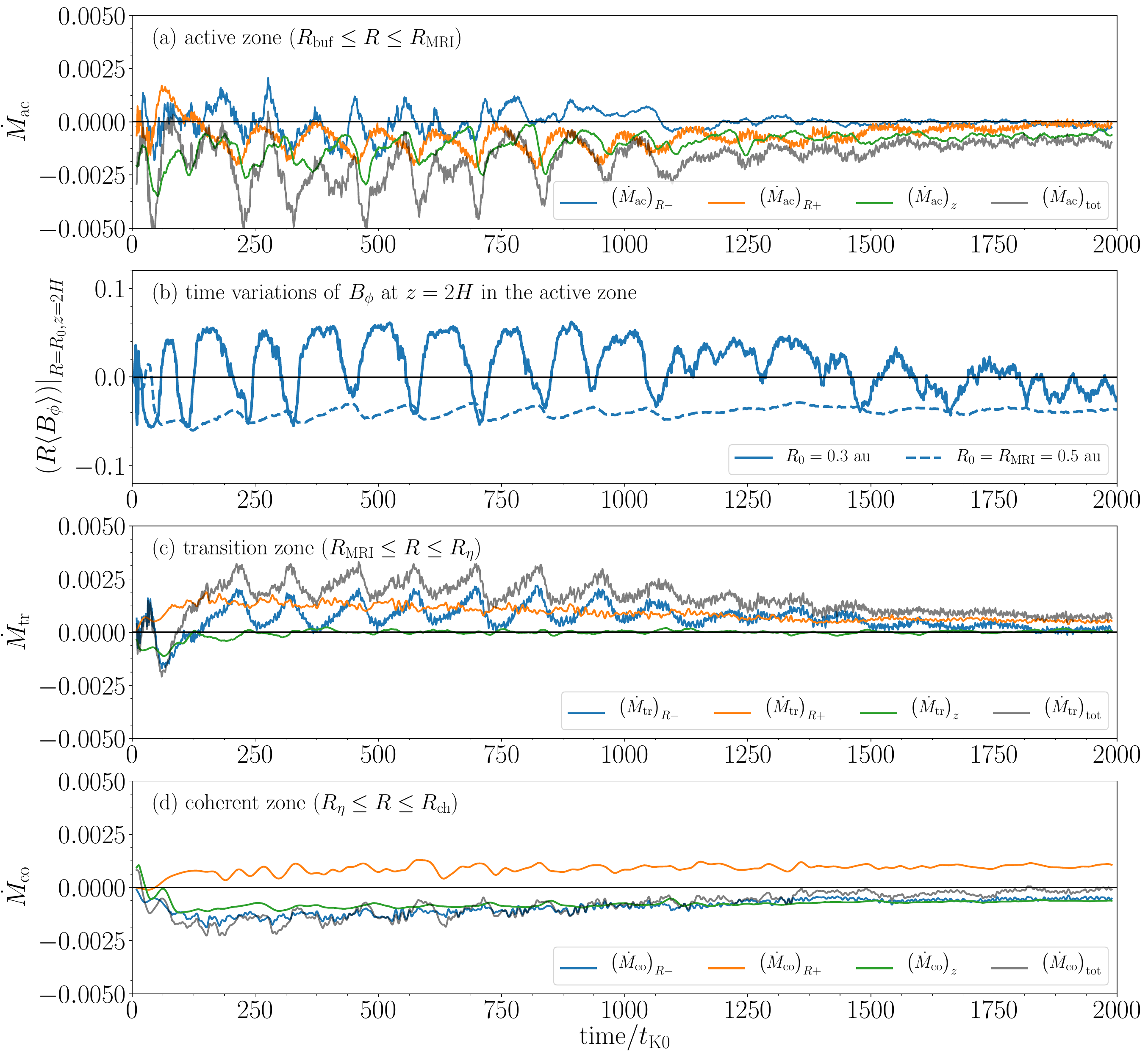}
    \end{center}
    \caption{
    Mass transfer rate through the four surfaces 
    of (a) the active, (c) 
    transition, and (d) coherent zones for LowRes run.
    The blue, orange, and green lines correspond to 
    $\left(\dot{M}_\mathrm{ac,tr,co}\right)_{R+}$, 
    $\left(\dot{M}_\mathrm{ac,tr,co}\right)_{R-}$, and
    $\left(\dot{M}_\mathrm{ac,tr,co}\right)_{z}$, respectively. 
    The gray lines show $\left(\dot{M}_\mathrm{ac,tr,co}\right)_\mathrm{tot}$.
    To reduce significant temporal fluctuations of the mass transfer rates, 
    we take the $20\tK$ centered moving average for them.
    Panel (b) shows the time variations of 
    $R\avphi{B_\phi}$ at $z=2H$ and 
    two different radii ($R=0.3~$au and $\Rmri=0.5$~au).
    All the quantities are shown in the code units.
    }
    \label{fig:masstrans}
\end{figure*}

In this section, we investigate how much mass is transferred between the active, transition, and 
coherent zones and is ejected from or falls onto the disk in the vertical directions.

We measure the mass fluxes passing through 
surfaces enclosing each of
the active, transition, and coherent zones.
In each zone, the inner and outer radii are denoted by 
$R_-$ and $R_+$, 
respectively ($R_-=\Rbuf$ and 
$R_+=\Rmri$ for the active zone, 
$R_-=\Rmri$ and 
$R_+=\Reta$ for the transition zone, and 
$R_-=\Reta$ and $R_+=\Rch$
for the coherent zone).
The upper and lower boundaries 
in the vertical directions are 
$z=+\zatm(R)$ and $z=-\zatm(R)$, respectively.

The time evolution of the total masses 
of the active, transition, and dead zones is given by 
\begin{equation}
(\dot{M}_\mathrm{ac,tr,co})_\mathrm{tot} = 
(\dot{M}_\mathrm{ac,tr,co})_{R+} + 
(\dot{M}_\mathrm{ac,tr,co})_{R-} + (\dot{M}_\mathrm{ac,tr,co})_z,
\end{equation}
where the subscripts "ac", "tr", and "co" stand for 
the active, transition, and coherent zones, respectively, 
the mass increasing rates passing through 
the surfaces $R=R_-$ and $R_+$ 
are denoted by 
$(\dot{M}_\mathrm{ac,tr,co})_{R-}$ and
$(\dot{M}_\mathrm{ac,tr,co})_{R+}$, 
respectively, and 
the mass increasing rates from the surfaces 
$z=-\zatm$ and $z=+\zatm$ 
are added and denoted as $(\dot{M}_\mathrm{ac,tr,co})_z$.
The signs of 
$(\dot{M}_\mathrm{ac,tr,co})_{R-}$, 
$(\dot{M}_\mathrm{ac,tr,co})_{R+}$, and  
$(\dot{M}_\mathrm{ac,tr,co})_{z}$  
are determined so that they are positive when the total mass inside the enclosed region
increases owing to its contribution.
Since the surface at $R=\Rmri$ ($R=\Reta$) is shared between
the active and transition zones (the transition and coherent zones), 
$(\dot{M}_\mathrm{ac})_{R+} =
-(\dot{M}_\mathrm{tr})_{R-} 
$
($(\dot{M}_\mathrm{tr})_{R+} =
-(\dot{M}_\mathrm{co})_{R-} 
$) 
is satisfied.
Figures \ref{fig:masstrans}a, \ref{fig:masstrans}c, and 
\ref{fig:masstrans}d show the time evolution of 
$(\dot{M}_\mathrm{ac,tr,co})_{R+}$,
$(\dot{M}_\mathrm{ac,tr,co})_{R-}$ ,
$(\dot{M}_\mathrm{ac,tr,co})_z$,
and 
$(\dot{M}_\mathrm{ac,tr,co})_\mathrm{tot}$.

\subsubsection{The Active Zone}\label{sec:massactive}

Figure \ref{fig:masstrans}a shows that 
$(\dot{M}_\mathrm{ac})_\mathrm{tot}$ is always negative, indicating that 
the total mass of the active zone decreases with time.
The time evolution of $(\dot{M}_\mathrm{ac})_{R-}$, 
$(\dot{M}_\mathrm{ac})_{R+}$, and 
$(\dot{M}_\mathrm{ac})_z$ can be divided into the early and late phases.
The former shows quasi-periodic oscillations and the latter is characterized by 
quasi-steady state (Figure \ref{fig:active_ver}).
The late phase corresponds to the development of the ring structures.

\noindent
\\
{\bf Early Evolution ($t\lesssim 1000~\tK$)}

In the early phase ($t\lesssim 1000~\tK$), 
the mass of the active zone is lost both 
vertically and radially with significant 
time variations. 
The fact that the mass of the active zone is lost through
$R=\Rmri$ ($(\dot{M}_\mathrm{ac})_{R+}<0$)
is consistent with the positive mass transfer rate at $R=\Rmri$ in 
Figures \ref{fig:angmom_early}c and \ref{fig:angmom_early}d.
 
In the radial directions, 
the mass of the active zone flows out more from $R=R_+=\Rmri$ than 
from $R=R_-=R_\mathrm{buf}$. 
The mass ejection rate by the disk wind $(\dot{M}_\mathrm{ac})_z$ is 
comparable to $(\dot{M}_\mathrm{ac})_{R+}$.
This indicates that the vertical mass 
loss is as important as the radial mass transport.

Why does $(\dot{M}_\mathrm{ac})_{R+}$ exhibit quasi-periodic variation
even though the MRI disappears around $R=\Rmri$ (Figure \ref{fig:angmom_rph_phz})?
The quasi-periodic variations generated in the atmospheres of the active zone
disturb the magnetic field around $R=\Rmri$.
Figure \ref{fig:masstrans}b shows that 
the time variations of $\avphi{B_\phi}$ at $z=2H$ at two different radii.
As a representative radius in the active zone, $R=0.3~\au$ is chosen because 
the quasi-periodic variations caused by the MRI are seen 
in $0.2~\au\lesssim R\lesssim 0.4~\au$.
At $R=0.3~\au$, 
$\avphi{B_\phi}$ at $z=2H$ is positive for a longer time than it is negative 
because it is amplified from negative $B_R$
that is generated by dragging of the magnetic fields
toward the central star \citep{Suzuki2014,Takasao2018}.
The amplified positive $\avphi{B_\phi}$ suddenly 
breaks and the sign of $\avphi{B_\phi}$ is flipped.
Afterwards, the positive $\avphi{B_\phi}$ is amplified again.
This cycle is repeated in a quasi-periodic manner in the active zone
(Figure \ref{fig:active_ver}).
When the sign of $\avphi{B_\phi}$ reverses and reverts in 
the active-zone atmospheres,
the disturbances propagate outwards and affect the magnetic fields around $R=\Rmri$.
This can be seen in the fact that $\avphi{B_\phi}(R=\Rmri,z=2H)$ suddenly drops slightly after
$\avphi{B_\phi}(R=0.3~\au,z=2H)$ becomes negative in Figure \ref{fig:masstrans}b.
The quasi-periodic variations in the magnetic fields at $R = \Rmri$ produce the fluctuations of $(\dot{M}_\mathrm{ac})_{R+}$.

Comparison between Figures \ref{fig:masstrans}a and \ref{fig:masstrans}b
indicates that the time variation of 
the vertical mass transfer rate $(\dot{M}_\mathrm{ac})_z$ is also correlated with 
that of $\avphi{B_\phi}$ at $R=0.3~\au$.
This is because the gas is ejected radially obliquely upward when 
the sign of $\avphi{B_\phi}$ returns to positive 
in the northern hemisphere.

The variations of $(\dot{M}_\mathrm{ac})_{R-}$ are not
quasi-periodic and 
correlate little with $(\dot{M}_\mathrm{ac})_{R+}$ and 
$(\dot{M}_\mathrm{ac})_{R+}$.
The variations of $(\dot{M}_\mathrm{ac})_{R-}$ 
are also likely to be produced by the MRI activity, but it
may be affected by the buffer region and the 
inner boundary conditions (Section \ref{sec:buffer}).

\noindent
\\
{ \bf Later Evolution ($t\gtrsim 1000~\tK$)}

The time evolution of $(\dot{M}_\mathrm{ac})_{R+}$ 
transitions to a stationary state because
the magnetic field around $R=\Rmri$ 
is no longer subject to quasi-periodic disturbances 
from the atmospheres of the active zone (Figure \ref{fig:masstrans}b).
As mentioned in Section \ref{sec:angmom_later}, $(\dot{M}_\mathrm{ac})_{R+}$
approaches zero because the mid-plane gas accretion is almost compensated 
by the outflow in higher latitudes.

\begin{figure}
    \begin{center}
    \includegraphics[width=9cm]{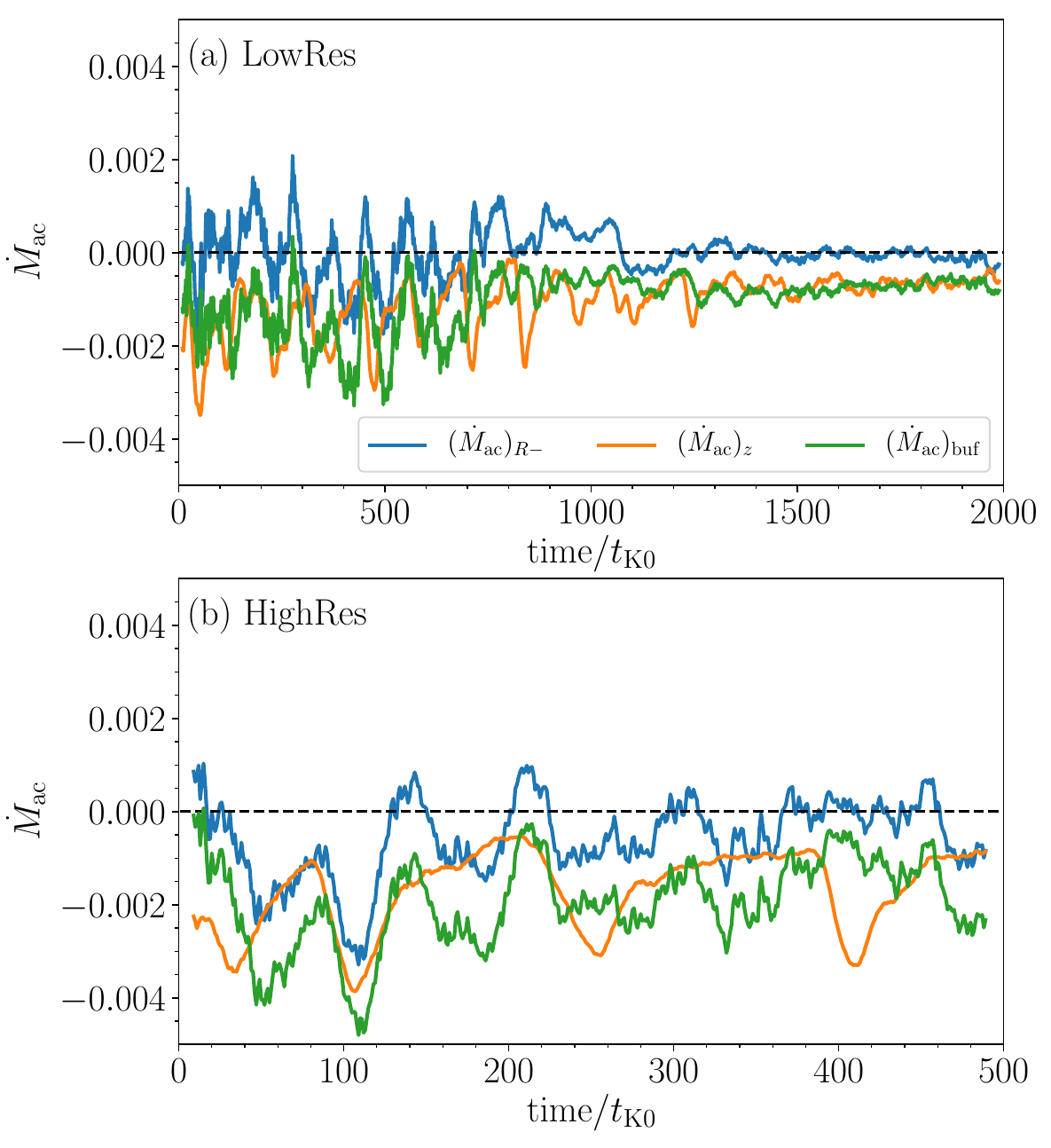}
    \end{center}
    \caption{
    Mass transfer in the active zone.
    Time evolution of $(\dot{M}_\mathrm{ac})_{R-}$, $(\dot{M}_\mathrm{ac})_z$, and $(\dot{M}_\mathrm{ac})_\mathrm{buf}$ 
    for (a) LowRes and (b) HighRes runs.
    To reduce significant temporal fluctuations of the mass transfer rates, 
    we take the $20\tK$ centered moving average for them.
    All the quantities are shown in the code units.
    }
    \label{fig:massactive}
\end{figure}

{
The vertical mass flow $(\dot{M}_\mathrm{ac})_z$ mainly determines the time evolution of the mass of the active zone because 
$(\dot{M}_\mathrm{ac})_{R-}$ and $(\dot{M}_\mathrm{ac})_{R+}$ approach zero 
(Figure \ref{fig:masstrans}a).
}
Where does the mass ejected vertically from $z=\pm \zatm(R)$ go?
Since the radial velocity in $|z|>\zatm$ 
above the active zone is negative 
(the bottom panels of Figure 
\ref{fig:angmom_early}
and Figure \ref{fig:angmom}), 
the disk wind falls onto the center.
In order to take into account mass accretion 
at high latitudes at $R=R_\mathrm{buf}$,
we measure the mass transfer rate $(\dot{M}_\mathrm{ac})_\mathrm{buf}$ through
the sphere $r=R_\mathrm{buf}$,
where we take into account only the regions where 
$\rho > 10^{-2}$ to remove the funnel regions.
Figure \ref{fig:massactive} 
shows that $(\dot{M}_\mathrm{ac})_\mathrm{buf}$ is comparable to 
$(\dot{M}_\mathrm{ac})_z$ in all the times for LowRes and HighRes runs.
This indicates that most mass ejected vertically 
falls onto the center \citep{Takasao2018}.

\subsubsection{The Transition Zone}\label{sec:masstransition}


Figure \ref{fig:masstrans}c shows that 
the mass transfer rate $(\dot{M}_\mathrm{tr})_{R+}$ is 
positive and does not show a significant 
time variation.
This is consistent with the fact 
that the gas is transferred 
from the coherent zone to the transition zone 
by the quasi-steady surface accretion flow. 

\begin{figure}
    \begin{center}
    \includegraphics[width=8cm]{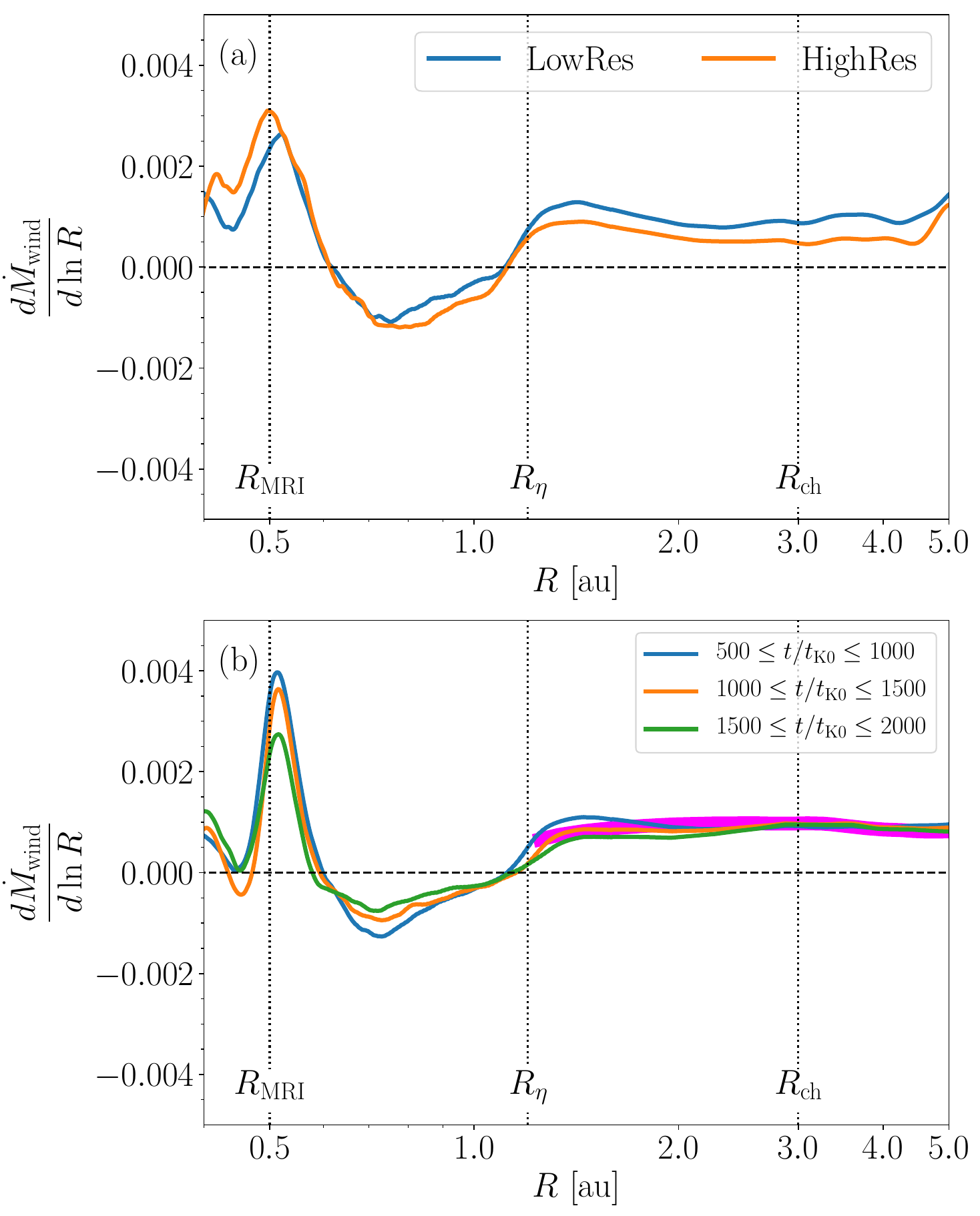}
    \end{center}
    \caption{
    Vertical mass outflow rate from $R$ to $R+dR$ as a function of radius.
    Panel (a) compares the results for LowRes and HighRes, and 
    they are averaged over $400\le t/\tK\le 500$.
    Panel (b) displays the time evolution of $d\dot{M}_\mathrm{wind}/d\ln R$ by 
    showing the result averaged over three different time intervals, 
    $500\le t/\tK\le 1000$,
    $1000\le t/\tK\le 1500$, and $1500\le t/\tK\le 2000$.
    The mass transfer rate in the coherent zone is shown by 
    the magenta line.
    All the quantities are shown in the code units.
    }
    \label{fig:wind}
\end{figure}

Figure \ref{fig:masstrans}c shows 
that the vertical mass supply 
$(\dot{M}_\mathrm{tr})_z$ stays almost zero.
This does not mean that there are no mass 
transfer at each radius.
To investigate the radial profile of the mass ejection rate by 
the disk wind, in Figure \ref{fig:wind} we plot 
the mass ejection rate from the surfaces of $z=\pm \zatm$ from $R$ to $R+dR$ that is given by 
\begin{equation}
    \frac{d \dot{M}_\mathrm{wind}}{d \ln R} 
    = 2\pi R^2 \left[ (\avphi{\rho v_n})_{\zatm} + (-\avphi{\rho v_n})_{-\zatm}\right],
    \label{wind}
\end{equation}
where the gas is ejected from the surfaces of $z=\pm \zatm$ when 
$d\dot{M}_\mathrm{wind}/d\ln R > 0$.
Figure \ref{fig:wind} shows that 
the mass is lost in $\Rmri\le R \le 0.6~\au$ 
and is supplied in the outer region.
This profile can be understood by the velocity 
field above the transition zone.
Most amount of the gas outflowing 
from the inner part falls back within the 
transition zone owing to the failed disk wind 
(Figure \ref{fig:BvR_tr}f).

\subsubsection{The Coherent Zone}

Unlike the active and transition zones, 
the change rate of the coherent zone mass 
($R_\eta \le R\le \Rch$) appears to 
approach zero in the long-term evolution
(Figure \ref{fig:masstrans}d).
The mass supply through $R=R_+=\Rch$ almost 
balances with the mass loss through 
$R=R_-=R_\eta$ and disk wind.
We note that our calculation time 
may not be long enough for $(\dot{M}_\mathrm{co})_{R+}$ 
to reach a quasi-steady state.
About half of the mass supplied through $R=R_+=\Rch$ is ejected by the disk wind, and 
the remaining half is transferred to the transition zone since 
$(\dot{M}_\mathrm{co})_z\sim (\dot{M}_\mathrm{co})_{R-}$ 
(Figure \ref{fig:masstrans}d).

\citet{Ferreira1997} defines "the ejection index" as 
\begin{equation}
    \xi \equiv  \frac{d\dot{M}_\mathrm{wind}/d\ln R}{\dot{M}^{\zatm}},
    \label{ejectionindex}
\end{equation}
which indicates a mass ejection efficiency 
because $\dot{M}_\mathrm{wind}$
obtained from the radial integration 
of $d\dot{M}_\mathrm{wind}/d\ln R$  
increases more rapidly compared with 
the mass transfer rate $\dot{M}^{\zatm}$ 
as the radial extent considered in the integration for larger $\xi$.

Figure \ref{fig:wind}b shows that 
$\dot{M}^{\zatm} \sim d\dot{M}_\mathrm{wind}/d\ln R$, 
indicating 
that significant mass loss occurs in our simulations.
The ejection index is about unity, or $\xi \sim 1$ 
(Equation (\ref{ejectionindex})).
By the angular momentum conservation and 
the induction equation, $\xi$ and the 
magnetic lever arm $\lambda \equiv 
(R_\mathrm{A}/R_0)^2$ are related, 
where $R_0$ is the radius at a radius where 
the disk wind is launched 
and $R_\mathrm{A}$ is the Alfv\'en radius 
of the radius, or 
$\xi = \left[
2 \left(
\lambda -1
\right)\right]^{-1}$
\citep{PelletierPudritz1992}.
From the fact that $\xi \sim 1$, 
the magnetic lever arm is 
estimated to $\lambda\sim 1.5$.
The lever arm is much shorter than 
expected in a typical magneto-centrifugal wind 
\citep{Ferreira1997}.
Disk winds with short lever arms were
reported; $\lambda\sim 1.4$ 
\citep{Bethune2017} and $\lambda\sim 1.15$ \citep{Bai2017}.

\citet{Baietal2016} pointed out the wind kinematics is mainly controlled by 
the plasma beta around the wind base.
Supposing that the wind base is at $z=\zatm$, the plasma beta is 
$\beta_0 \times \exp(-(\zatm/H)^2) \sim 20$.
For such a high $\beta$, the 
magneto-centrifugal winds 
\citep{BlandfordPayne1982} do not occur
since the magnetic fields are too weak 
to corotate with the Kepler rotation at the wind base within the Alfv\'en radius.
Instead, the thermal and magnetic pressure gradient forces 
accelerate the winds with the 
short lever arm $\lambda\sim 1.5$.

\subsection{Magnetic Flux Transport}\label{sec:flux}

We define the following two kinds of the magnetic fluxes.
One is the magnetic flux threading the northern hemisphere at a given radius $r$ that is given by 
\begin{equation}
    \Psi_\mathrm{rad}(r) = \int_{0}^{\pi/2} \sin\theta d\theta \int_0^{2\pi} d\phi r^2 B_r(r,\theta,\phi).
    \label{psirad}
\end{equation}
The other is the magnetic flux threading the mid-plane from $R=r_\mathrm{in}$ to $R$,
\begin{equation}
    \Psi_\mathrm{mid}(R) = \int_{r_\mathrm{in}}^R R dR \int_0^{2\pi} d\phi B_z(r,\theta,\phi).
\end{equation}
The divergence-free condition in the northern hemisphere gives 
\begin{equation}
    \Psi_\mathrm{rad}(r) = \Psi_\mathrm{rad}(r_\mathrm{in}) + \Psi_\mathrm{mid}(
    R). 
    \label{Bflux}
\end{equation}
where $R$ is identical to $r$ at $\theta=\pi/2$,
because all the magnetic field lines passing through 
the northern hemisphere at the inner edge and the mid-plane 
from $R=r_\mathrm{in}$ to $R$ penetrate through 
the northern hemisphere at $r$.

From Equation (\ref{Bflux}), one obtains 
\begin{equation}
 \Psi_\mathrm{rad}(r_\mathrm{out}) = 
 \Psi_\mathrm{rad}(r_\mathrm{in}) + \Psi_\mathrm{mid}(R_\mathrm{out}),
 \label{bflux_cons}
\end{equation}
where $R_\mathrm{out}=r_\mathrm{out}$ 
is the outermost cylindrical  radius of the simulation box.
Equation (\ref{bflux_cons}) shows 
how the total magnetic flux $\Psi_\mathrm{rad}(r_\mathrm{out})$ 
is distributed between $\Psi_\mathrm{rad}(r_\mathrm{in})$ and 
$\Psi_\mathrm{mid}(R_\mathrm{out})$.

\begin{figure}
    \begin{center}
    \includegraphics[width=8cm]{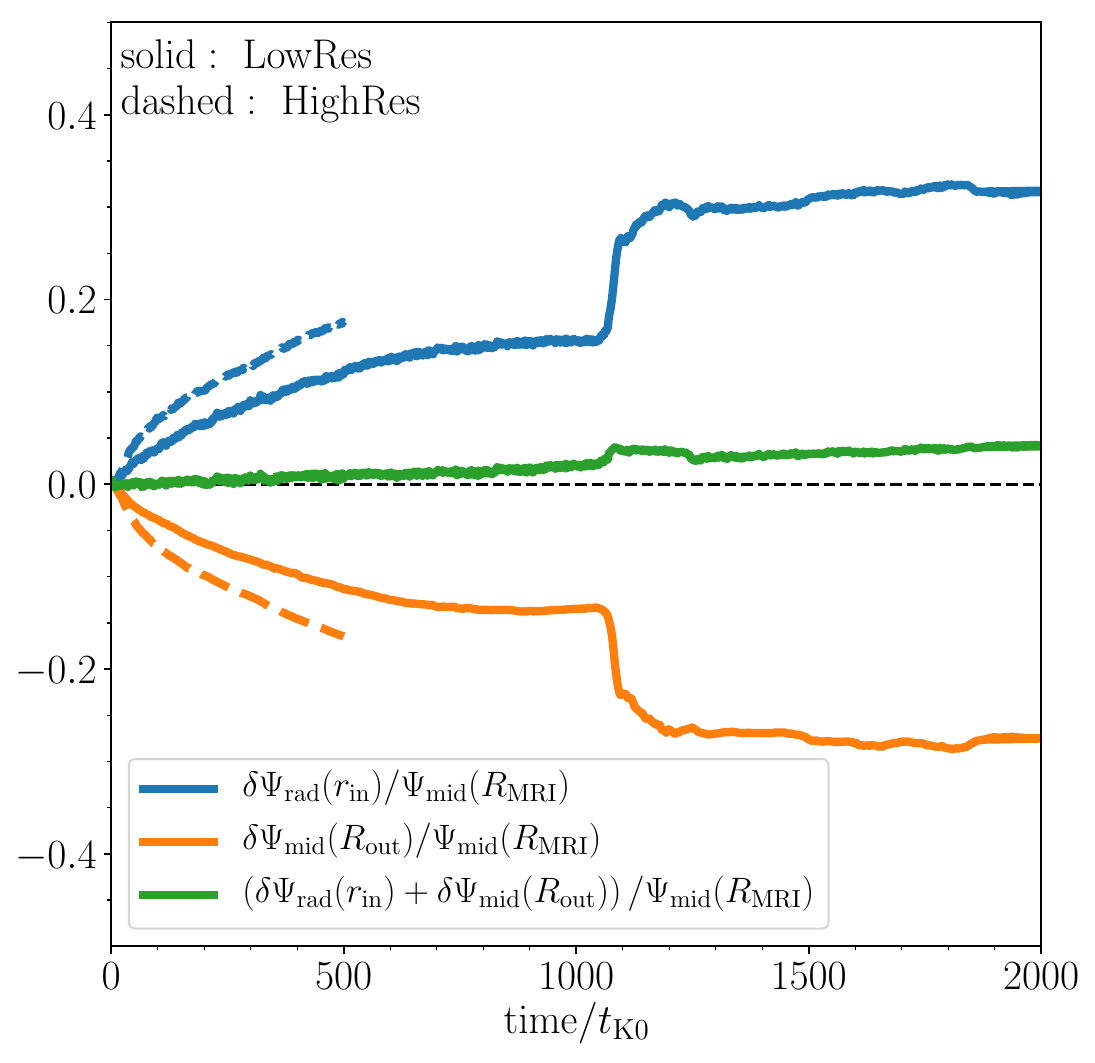}
    \end{center}
    \caption{
    Time Evolution of $\Psi_\mathrm{rad}(r_\mathrm{in})$, 
    $\Psi_\mathrm{mid}(R_\mathrm{out})$, and 
    $\Psi_\mathrm{rad}(r_\mathrm{out})=\Psi_\mathrm{rad}(r_\mathrm{in})+\Psi_\mathrm{mid}(R_\mathrm{out})$ for 
    LowRes (solid) and  HighRes (dashed).
The deviations from the respective initial values (
$\delta \Psi_\mathrm{rad}(r_\mathrm{in})$ and
$\delta \Psi_\mathrm{mid}(R_\mathrm{out})$) are displayed.
In addition, their values are normalized by the initial mid-plane magnetic flux within $R=\Rmri$.
    }
    \label{fig:bflux_cons}
\end{figure}

\begin{figure*}
    \begin{center}
    \includegraphics[width=17cm]{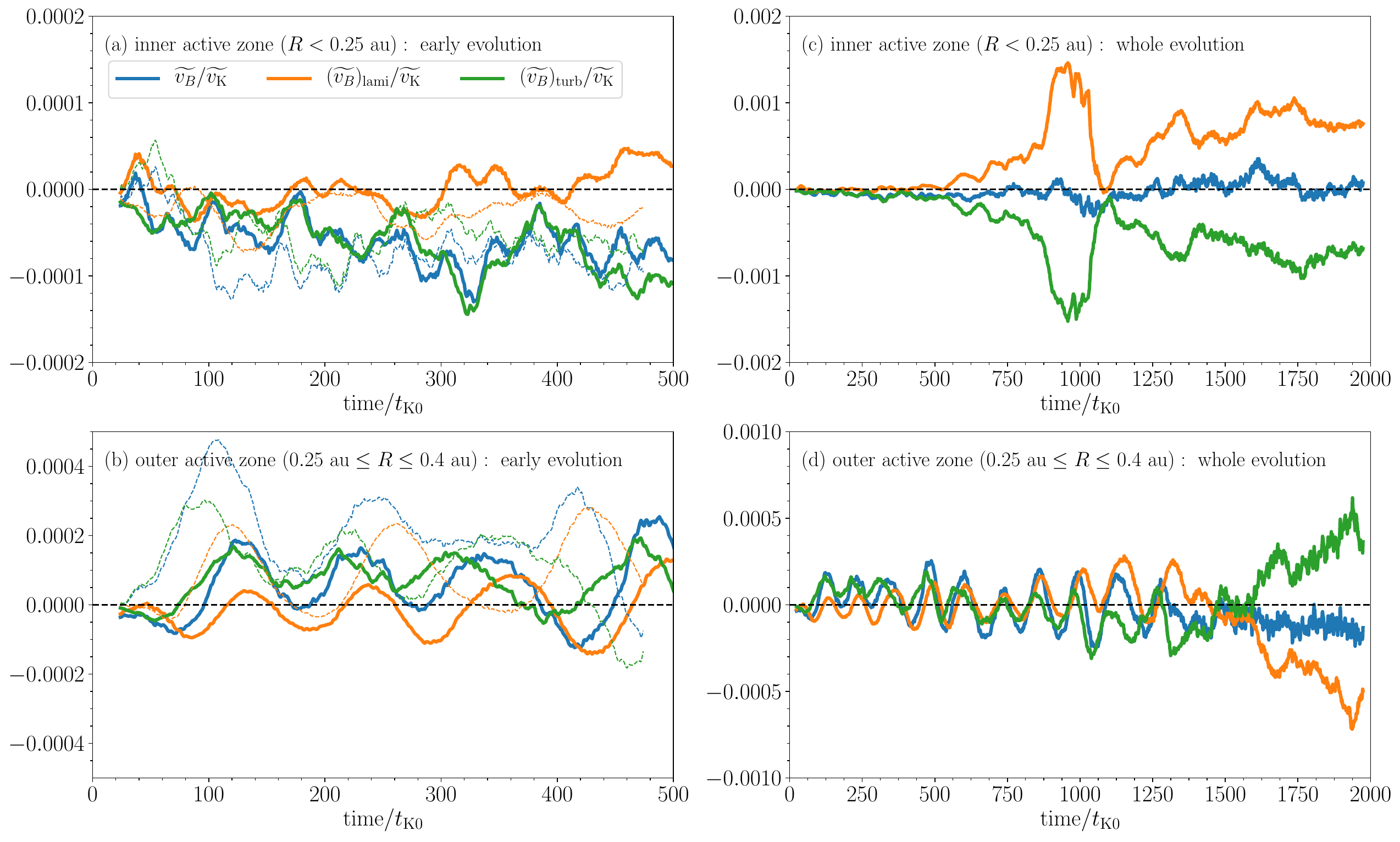}
    \end{center}
    \caption{
    Time evolution of the drift velocities 
    averaged over  $R<0.25~$au (top) and 
     $0.25~\au\le R\le 0.4~\au$ (bottom) for LowRes.
    The left and right panels show the early and whole evolution,
    respectively.
    For comparison, the results for HighRes are shown by the thin dashed line in the left panels.
    In each panel, 
    $\left(\widetilde{v_\mathrm{B}}\right)_\mathrm{lami}$, $\left(\widetilde{v_\mathrm{B}}\right)_\mathrm{turb}$, and 
    $\left(\widetilde{v_\mathrm{B}}\right)$ are shown, 
    and are normalized by the Kepler speed $\widetilde{v_\mathrm{K}}$ averaged over the corresponding radius range.
    To reduce significant temporal fluctuations of the drift velocities, 
    we take the $20\tK$ centered moving average for them.
    }
    \label{fig:vB_active}
\end{figure*}

The time evolution of $\Psi_\mathrm{rad}(r_\mathrm{in})$
and $\Psi_\mathrm{mid}(R_\mathrm{out})$ is shown in
Figure \ref{fig:bflux_cons}. 
Because $\Psi_\mathrm{mid}(R_\mathrm{out})$ is much larger than $\Psi_\mathrm{rad}(r_\mathrm{in})$
at the initial condition, 
Figure \ref{fig:bflux_cons} displays the deviations from the respective initial values,
$\delta \Psi_\mathrm{rad}(r_\mathrm{in})$ and
$\delta \Psi_\mathrm{mid}(R_\mathrm{out})$.
In addition, their values are normalized by the initial mid-plane magnetic flux within $R=\Rmri$.
Figure \ref{fig:bflux_cons} shows that 
$\Psi_\mathrm{rad}(r_\mathrm{in})$ 
increases with time 
while $\Psi_\mathrm{mid}(R_\mathrm{out})$ decreases with time,
keeping $\Psi_\mathrm{rad}(r_\mathrm{in})+\Psi_\mathrm{mid}(R_\mathrm{out})$ almost constant\footnote{
We note that in Figure \ref{fig:bflux_cons}, the total magnetic flux $\Psi_\mathrm{rad}(r_\mathrm{out})$
gradually increases with time although it remains small.
This is because
the boundary conditions adopted in this paper do not 
strictly prohibit inflow of the magnetic flux into the simulation box.
}. 
This clearly shows that the magnetic fluxes passing through 
the mid-plane accrete onto the inner edge.
Comparison between the results of LowRes and HighRes shows that 
the magnetic flux accretes onto the inner boundary more rapidly for HighRes than for LowRes run.

The increasing rate of $\Psi_\mathrm{rad}(r_\mathrm{in})$ decreases with time except for 
a sudden increase around $t=1000\tK$, which corresponds to 
the accretion of 
the innermost $\avphi{B_z}$ concentration onto the inner boundary of the simulation box
around $t=1000\tK$ (Figure \ref{fig:tRdiagram}).
In the late evolution ($1500\tK\le t\le 2000\tK$), the magnetic flux at the inner boundary hardly changes while
the gas accretion onto the inner boundary continues
(Figure \ref{fig:massactive}).
This decrease in the increasing rate of $\Psi_\mathrm{rad}(r_\mathrm{in})$ has been 
reported in \citet{Beckwith2009} and \citet{Takasao2018}.

In Sections \ref{sec:mag_active}, \ref{sec:fluxtransport_transition}, and \ref{sec:mag_dead},
we will investigate the radius dependence of the magnetic flux transport 
in the active, transition, and coherent zones, respectively. 
The trajectories of the magnetic flux are shown by 
the contours of $\Psi_\mathrm{rad}$ shown in Figure \ref{fig:tRdiagram}.
It is clearly seen that the behaviors of the magnetic field transport 
are different between the three zones.
We define a drift speed of the magnetic flux $v_\mathrm{B}$ as follows:
\begin{equation}
\avphi{v_{B}} \equiv {\avphi{E_\phi}}/{\avphi{B_z}} 
\end{equation}
\citep{GuiletOgilvie2012,BaiStone2017}.

\subsubsection{The Active Zone}\label{sec:mag_active}

{
In this section, we will show that 
the magnetic flux drifts inward (outward) in inner (outer) regions of the active zone in the early 
phase evolution.
In the later phase, the drift velocities oscillate between positive and negative with time in 
the inner active zone while the magnetic flux drifts inward in the outer active zone.
}

The trajectories of the magnetic flux and the color map of $\avphi{B_z}_H$ in Figure \ref{fig:tRdiagram}a 
show that the flux concentrations become prominent for $t\gtrsim 500\tK$, and 
the contours of $\Psi_\mathrm{rad}$ converges to the multiple $\avphi{B_z}$ concentrations.

First we investigate the magnetic flux transport 
before developing the flux concentrations. 
We divide the active zone into two regions, the inner region with $R<0.25~\au$ and 
the outer region with $0.25~\mathrm{au}\le R\le 0.4~\au$ because 
the magnetic flux moves inward at $R<0.25~\au$.
The outermost radius is set to $R=0.4~\au$ because 
beyond $R=0.4~\au$ the MRI turbulence is affected by OR (Figure \ref{fig:QC}).
In each of the inner and outer regions, 
we average $\avphi{v_B}$ in the radial direction as follows:
\begin{equation}
    \widetilde{v_B}
    \equiv \int 2\pi R \avphi{(E_\phi)_\mathrm{I}}_{H}dR
    \times
    \left(\int 2\pi R \avphi{B_z}_H dR\right)^{-1}.
    \label{vB}
\end{equation}

In order to investigate what determines the drift speed,
we decompose $\avphi{({E}_\phi)_\mathrm{I}}_H=
\avphi{v_R B_z}_H - \avphi{v_z B_R}_H$ into the laminar component 
\begin{equation}
    \left(\avphi{({E}_\phi)_\mathrm{I}}_{H}\right)_\mathrm{lami} = 
    \avphi{v_R}_{H,v}\avphi{B_z}_H - \avphi{v_z}_{H,v}\avphi{B_R}_H,
\end{equation}
and the turbulent component
\begin{eqnarray}
    \left(\avphi{ ({E}_\phi)_\mathrm{I}}_{H}\right)_\mathrm{turb} &= &
    \avphi{(v_R - \avphi{v_R}_{H,v})\delta B_z}_H \nonumber \\
    &&- \avphi{(v_z - \avphi{v_z}_{H,v})\delta B_R}_H,
\end{eqnarray}
where 
$\avphi{(E_\phi)_\mathrm{I}}_H = 
\left(\avphi{(E_\phi)_\mathrm{I}}_H\right)_\mathrm{lami} + 
\left(\avphi{(E_\phi)_\mathrm{I}}_H\right)_\mathrm{turb}  
$ is satisfied.
Although the mass-weighted average is taken when 
we calculate the mean values of the velocity 
in Section \ref{sec:average},
the volume-weighted average should be applied 
for the velocity when 
the flux transport is discussed.
To distinguish the volume-weighted mean velocity, the 
subscript "$v$" is used, or $\avphi{v_{R,z}}_{H,v}$.
The radially-averaged drift velocities originating from the laminar and turbulent components 
of $\avphi{(E_\phi)_\mathrm{I}}$
are defined by replacing 
$\avphi{(E_\phi)_\mathrm{I}}_{H}$ with
$\left(\avphi{(E_\phi)_\mathrm{I}}_{H}\right)_\mathrm{lami}$ and
$\left(\avphi{(E_\phi)_\mathrm{I}}_{H}\right)_\mathrm{turb}$ 
in  Equation (\ref{vB}), 
and 
are denoted by 
$(\widetilde{v_\mathrm{B}})_\mathrm{lami}$
and $(\widetilde{v_\mathrm{B}})_\mathrm{turb}$, respectively. 

In the early evolution ($t\le 500~\tK$), 
the magnetic flux moves inward in the inner active zone $(R<0.25~\au)$ and 
moves outward in the outer active zone  $(0.25~\au\ge R \le 0.4~\au)$ as shown in 
Figures \ref{fig:vB_active}a and \ref{fig:vB_active}b which display the 
time evolution of $\widetilde{v_\mathrm{B}}$ and 
the contributions from the laminar and turbulent components. 
In the inner active zone, the turbulent $(E_\phi)_\mathrm{I}$ mainly drives 
inward drift of the magnetic flux at a speed of $\sim 5\times 10^{-5}v_\mathrm{K}$.
Although the resolution dependence of $\widetilde{v_\mathrm{B}}$ is found in $t\le 300~\tK$, 
it disappears in $300\lesssim t/\tK\lesssim 500$.

\begin{figure}
    \begin{center}
    \includegraphics[width=8cm]{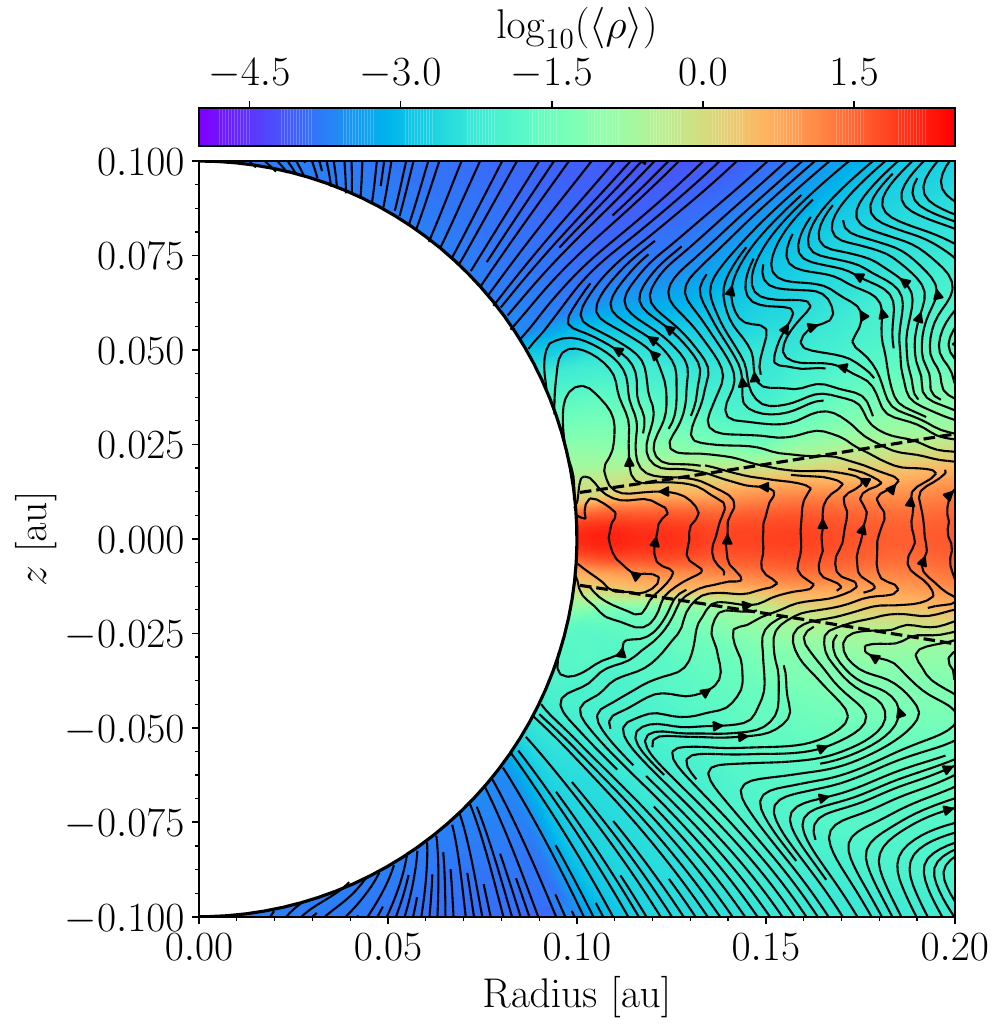}
    \end{center}
    \caption{
     Poloidal magnetic fields structure near the inner boundary at $t=500\tK$.
     The color map shows $\log_{10}\avphi{\rho}$, and the streaming lines show 
     the poloidal magnetic fields averaged over $0\le \phi <2 \pi$.
     The two dashed lines correspond to $z=\pm \zatm$.
    The densities are shown in the code units.
    }
    \label{fig:ringmag}
\end{figure}

The magnetic flux transport is not determined only by 
the dynamics inside the disk with $|z|<H$, but is affected by 
the upper atmospheres.
The poloidal magnetic field structures are shown in 
Figure \ref{fig:ringmag}. 
Just below the boundaries ($z\sim \pm 0.05~\au$ at $r=0.1~\au$) 
between the funnel regions and the atmospheres,
the poloidal magnetic fields are dragged toward the inner boundary. 
After the dragged poloidal fields accrete onto the inner boundary, 
loop-like poloidal fields penetrating the disk form \citep{Beckwith2009}.
Their edges are at the inner boundaries in the southern and northern hemispheres.
When the loop poloidal fields disappear through accretion and/or 
magnetic reconnection, the net flux at the inner boundary 
$\Psi_\mathrm{rad}(r_\mathrm{in})$ increases.

In the long-term evolution in the active zone, 
the turbulent and laminar contributions nearly cancel each other out and 
their summation, or $\widetilde{v_\mathrm{B}}$ remains low (Figure \ref{fig:vB_active}c).
The laminar and turbulent toroidal electric fields tend to move the magnetic flux 
outward and inward, respectively.
The drift velocities oscillate with time and takes both positive and negative values 
for $t\ge 1000~\tK$, indicating that there is almost no net magnetic transport.
Such a quasi-steady state has been obtained in previous global simulations \citep{Beckwith2009,Takasao2018}.

In the early evolution of the the outer active zone, outward drift of the magnetic flux 
is driven mainly by 
$\avphi{(E_\phi)_\mathrm{I}}_\mathrm{turb}$.
Both 
$\avphi{(E_\phi)_\mathrm{I}}_\mathrm{lami}$
and 
$\avphi{(E_\phi)_\mathrm{I}}_\mathrm{turb}$
exhibit quasi-periodic oscillations whose 
phases are correlated with those of $\avphi{B_\phi}$ oscillations in the upper atmospheres.
Unlike LowRes run, HighRes run shows that 
the laminar component plays an important role in the outward flux transport 
because the outward velocity in the outer active zone is larger for HighRes than for 
LowRes (Figure \ref{fig:angmom_early}a).

Figure \ref{fig:vB_active}d
shows that the quasi-periodic oscillations continue until $t\sim 1500~\tK$ beyond which 
the quasi-periodic MRI activities disappear (Figure \ref{fig:active_ver}).
Although $\widetilde{v_\mathrm{B}}$ is positive on average in the early phase, 
the net drift velocities become around zero in $500\le t/\tK\le 1000$.
In addition, unlike in the inner active zone, 
$(\widetilde{v_\mathrm{B}})_\mathrm{lami}$
and  $(\widetilde{v_\mathrm{B}})_\mathrm{turb}$ do not cancel each other out, and 
they oscillate in the same phase in the outer active zone.
In the later phase with $t\gtrsim 1500~\tK$, 
$(\widetilde{v_\mathrm{B}})_\mathrm{lami}$ and $(\widetilde{v_\mathrm{B}})_\mathrm{turb}$ do not 
completely  cancel each other out, 
and the magnetic flux drifts inward at $\widetilde{v_\mathrm{B}}\sim 10^{-4}v_\mathrm{K}$.
The magnetic flux drifts inward at a faster speed than the gas 
that accretes at a speed of $\sim 10^{-6}v_\mathrm{K}$.
\citet{Jacquemin-Ide2021} conducted 
a simulation (model SEp) similar to 
ours and obtained a drift speed consistent with our results.

\subsubsection{The Transition Zone}\label{sec:fluxtransport_transition}

\begin{figure}
\vspace{5mm}
    \begin{center}
    \includegraphics[width=8cm]{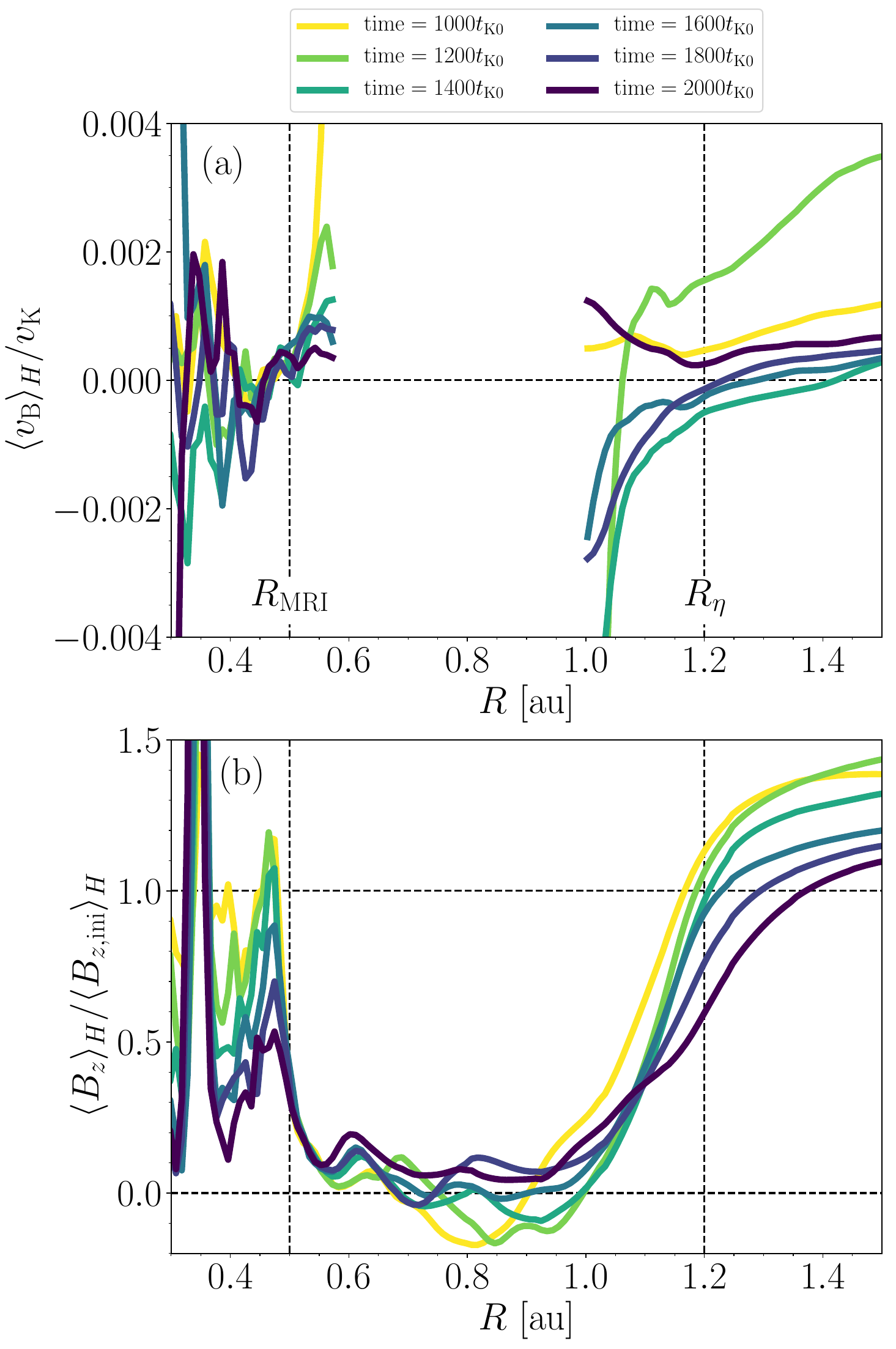}
    \end{center}
    \caption{
     Time Evolution of the radial profiles of $\avphi{v_\mathrm{B}}_H/v_\mathrm{K}$ 
     and $\avphi{B_z}_H/\avphi{B_z}_{H,\mathrm{ini}}$ from $t=1000\tK$ to $t=2000\tK$.
     In Panel (a), we do not plot $\avphi{v_\mathrm{B}}_H/v_\mathrm{K}$ in 
     $0.6\mathrm{au}\le R\le 1~\mathrm{au}$ because scatters of the data are significant owing 
     to almost zero $\avphi{B_z}_H$.
    All the quantities are shown in the code units.
    }
    \label{fig:vB_trans_long}
\end{figure}

In Section \ref{sec:loopform}, we found that the rapid outward transport of the magnetic flux 
is caused by OR and AD.
In this section, we investigate the long term evolution of the magnetic flux around the 
transition zone.
Figure \ref{fig:vB_trans_long}b shows that the outer edge 
of the region where $\avphi{B_z}_H\sim 0$ expands with time
because the magnetic flux in the 
coherent zone drift outward as will be 
shown in Section \ref{sec:mag_dead}.
A typical drift speed of the outer edge that is 
identified at the radius where
$\avphi{B_z}_H=\avphi{B_{z,\mathrm{ini}}}_{H}$ 
is estimated to about $10^{-3}~v_\mathrm{K}$ 
from Figures \ref{fig:vB_trans_long}a and \ref{fig:vB_trans_long}b.  
This drift speed 
is almost identical to that 
in the coherent zone (Figure \ref{fig:vB_dead}).

The flux concentration around $R\sim 0.5~\mathrm{au}$, where a density gap exists 
(Figure \ref{fig:angmom}c), weakens with time because $\avphi{v_\mathrm{B}}_H$ is positive 
around $R\sim 0.5~\au$ (Figure \ref{fig:vB_trans_long}a).
Further time evolution would eliminate the flux concentration around $R\sim 0.5~\mathrm{au}$, 
resulting in inward expansion of the region where $\avphi{B_z}_H\sim 0$.
However, the inward expansion would not proceed furthermore because of rapid 
decreases in $\etaO$ and $\etaA$ (Figure \ref{fig:profinimid}).

\subsubsection{The Coherent Zone}\label{sec:mag_dead}

\begin{figure}
    \begin{center}
    \includegraphics[width=8cm]{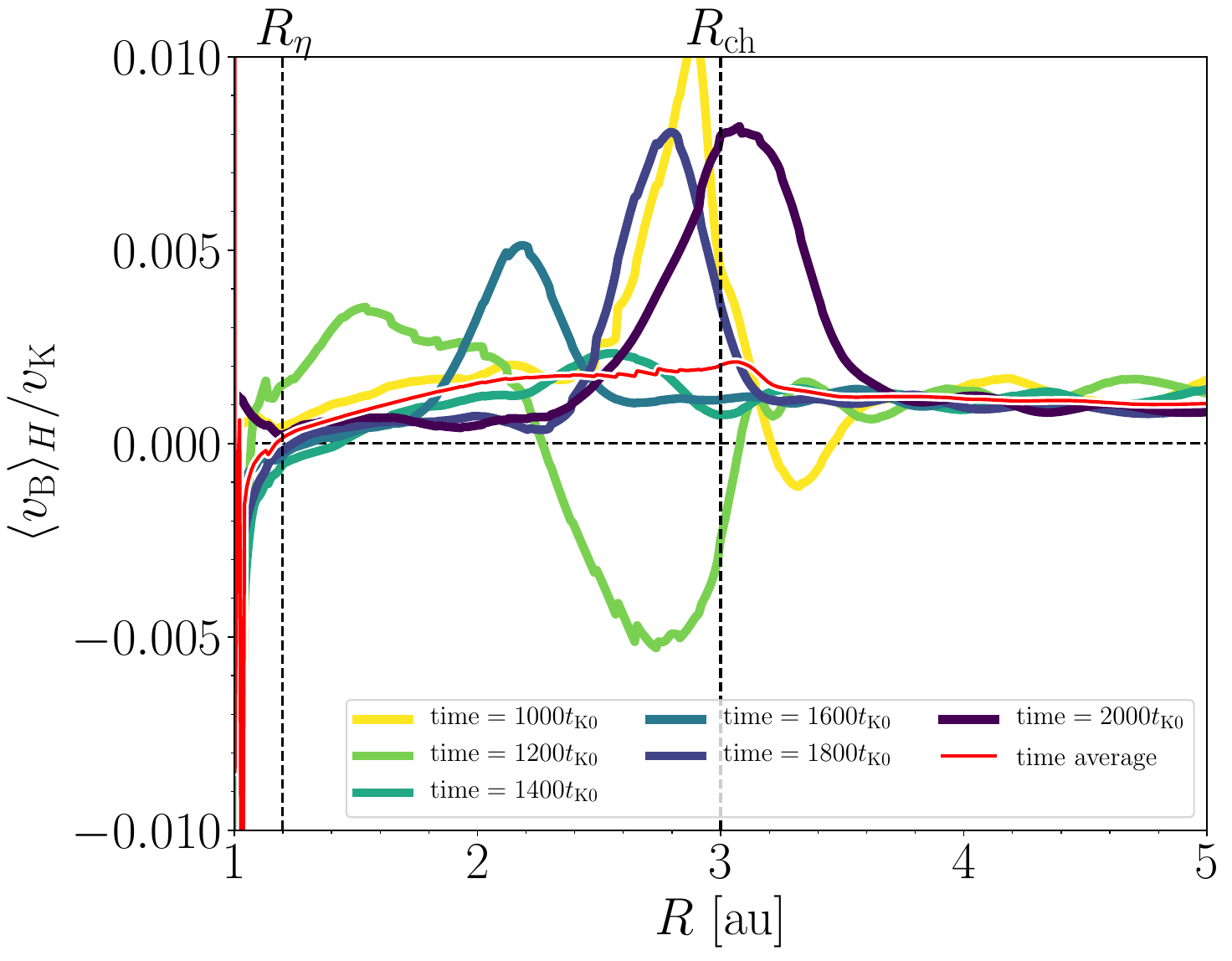}
    \end{center}
    \caption{
    Time sequence of the radial profiles of the drift velocity of the magnetic flux 
    $\avphi{v_\mathrm{B}}_H$ normalized by the Kepler speed from $t=1000\tK$ to $t=2000\tK$.
    The radial profiles of $\avphi{v_\mathrm{B}}_H$ averaged over $1000\tK\le t\le 2000\tK$
    are shown by the red line.
    The vertical dashed lines correspond to $R=\Reta$ and $\Rch$ from left to right.
    }
    \label{fig:vB_dead}
\end{figure}

\begin{figure}
    \begin{center}
   \includegraphics[width=8cm]{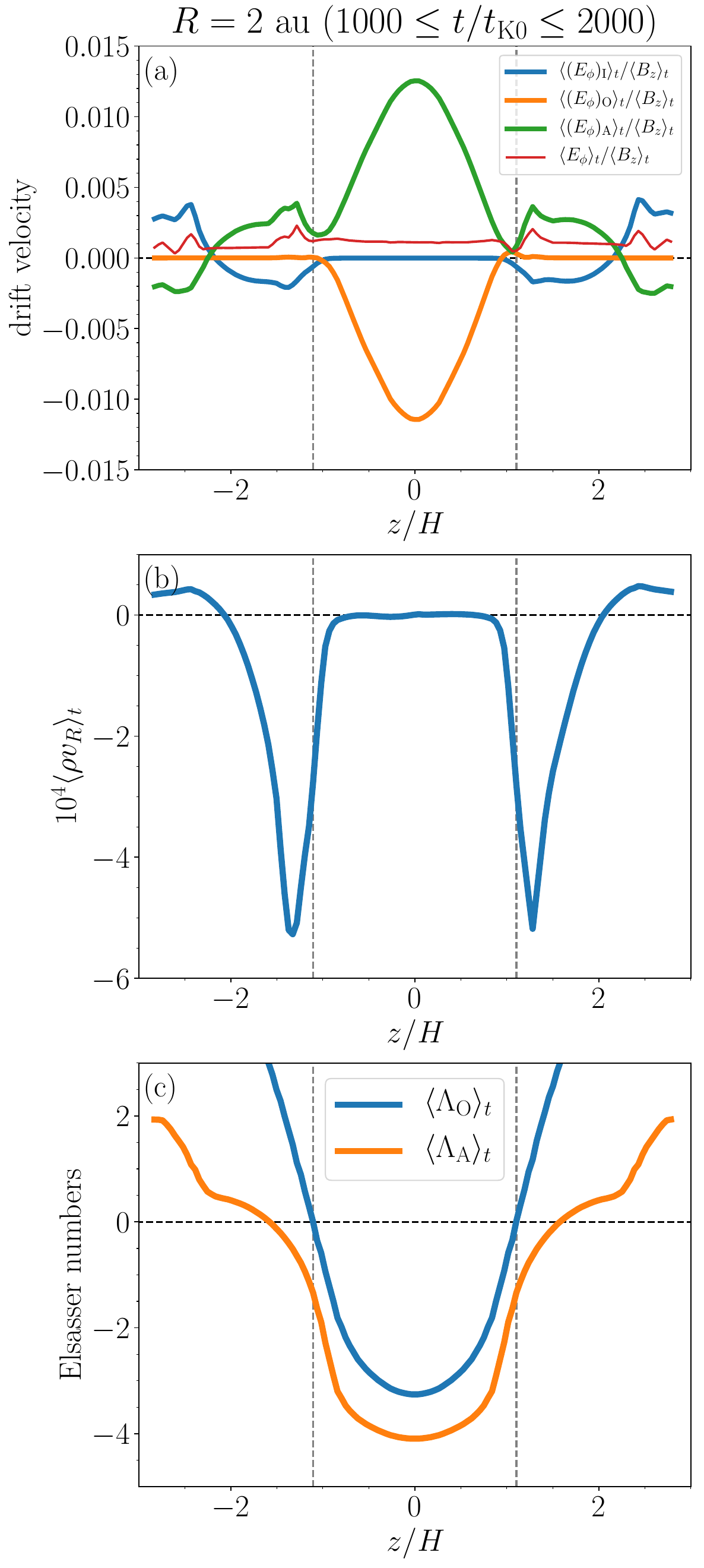}
    \end{center}
    \caption{
Vertical profiles of the quantities related to the magnetic field transport at 
$R=2~\au$.
All the data are averaged over $1000 \tK\le t \le 2000 \tK$.
(a) The vertical profiles of 
$\avphi{(E_\phi)_\mathrm{I}}_t/\avphi{B_z}_t$,
$\avphi{(E_\phi)_\mathrm{O}}_t/\avphi{B_z}_t$, 
$\avphi{(E_\phi)_\mathrm{A}}_t/\avphi{B_z}_t$, and 
$\avphi{(E_\phi)}_t/\avphi{B_z}_t$.
(b) The vertical profile of $\avphi{\rho v_R}_t$. 
(c) The vertical profiles of 
$\avphi{\LambdaO}_t$ and 
$\avphi{\LambdaA}_t$. 
In each panel, the vertical gray dashed lines correspond to 
the positions where $\avphi{\LambdaO}_t=1$.
    All the quantities are shown in the code units.
}
    \label{fig:Eph_z}
\end{figure}

Figure \ref{fig:vB_dead} shows the time evolution of 
the radial profiles of $\avphi{v_\mathrm{B}}_H/v_\mathrm{K}$.
For $R\lesssim 3.5~\au$, although $\avphi{v_\mathrm{B}}_H/v_\mathrm{K}$ is affected by 
the quasi-periodic disturbances from the active zone,
the drift velocity is positive on average. 
Beyond $R\gtrsim 3.5~\au$, $\avphi{v_\mathrm{B}}_H$ is steady. 
When the time average over $1000\tK\le t\le 2000\tK$ is taken,
$\avphi{v_\mathrm{B}}_{H,t}$ becomes almost constant, and 
is about $\avphi{v_\mathrm{B}}_{H,t}\sim 10^{-3}v_\mathrm{K}$.
This value is comparable to those in \citet{BaiStone2017} and \citet{Bai2017}.
Figure \ref{fig:Eph_z}a shows that
the drift velocity $v_\mathrm{B}=\avphi{E_\phi}_t/\avphi{B_z}_t$ does not depend on 
$z$ significantly, indicating that 
the magnetic fields drift outward keeping their shapes.
This feature is also consistent with \citet{BaiStone2017} and \citet{Bai2017}.

Since the magnetic field is almost completely decoupled with 
the gas inside the coherent zone where $\LambdaO\ll1$ and $\LambdaA\ll 1$,  
the magnetic flux transport is determined above the coherent zone.
We focus on the regions just above the 
OR dead zone where the surface gas accretion 
occurs (Figure \ref{fig:Eph_z}b in 
$H\lesssim |z| \lesssim 2H$).
The surface gas accretion tries to carry the magnetic flux inward since
$\avphi{(E_\phi)_I}_t \sim \avphi{v_R B_z}_t$ is negative.
Diffusion due to AD, which plays an important role 
in the surface accretion layers ($\LambdaA\sim 1$), tries to drift the magnetic flux outward.
Figure \ref{fig:Eph_z}a shows that the diffusion due to AD 
slightly dominates over the accretion, resulting in the outward transport of 
the magnetic flux.

Above the surface accretion layers, 
the wind drags the vertical field outward ($\avphi{(E_\phi)_I}>0$).
Unlike in the surface accretion layers, 
the diffusion due to AD is not sufficient to dominate over the wind drag 
because $\LambdaA$ increases with $z$ (Figure \ref{fig:Eph_z}c).
The drift velocity around the base of the wind ($|z|\sim 2.5H$) is close to 
that in the surface accretion layers.

Inside the OR dead-zone boundary ($\avphi{\LambdaO}_t<1$),
$(\avphi{E_\phi})_\mathrm{O}$ and 
$(\avphi{E_\phi})_\mathrm{A}$ almost cancel each other out, 
and the magnetic field structure inside the disk are adjusted so that 
$\left\{(\avphi{E_\phi})_\mathrm{O}+ 
(\avphi{E_\phi})_\mathrm{A}\right\}/\avphi{B_z}$ is comparable to
the drift velocity in the region with $\LambdaO>1$.

\section{Discussion}\label{sec:discuss}

\subsection{Structure Formation}

\subsubsection{The Active Zone}\label{sec:ring_discuss}

As shown in Figure \ref{fig:Ring}, 
the prominent multiple rings and flux concentrations are 
found in the active zone.
They are not transient structures but are maintained at least until $t/\tK =2000$.
It should be noted that not all of the simulations conducted in 
previous studies have shown ring formation;
some simulations 
\citep{BaiStone2014,Hawley2001,Steinacker2002,Dzyurkevich2010,Jacquemin-Ide2021} 
form rings and some \citep{Suzuki2014,Takasao2018,ZhuStone2018} do not.

In Section \ref{sec:structureform_active}, 
we examined two ring formation mechanisms proposed before, 
the viscous instability \citep{Suzuki2023} and the wind-driven instability \citep{RoilsLesur2019}. 
The viscous instability 
may be a possible ring structure formation mechanism in our simulations.
However, our analyses are insufficient to fully understand the ring formation in the ideal MHD.
The viscous instability found in \citet{Suzuki2023} forms 
transient ring structures, unlike our results.
The possibility of the wind-driven instability is rejected for our simulation, 
but we note that our result could be affected by the inner boundary condition.

We still do not know detailed mechanisms and what determines whether multiple-rings form or not.
We will address the formation of ring structures in MRI active disks by performing 
global simulations with a wide range of parameters in the future.

\subsubsection{The Transition and Coherent Zones}\label{sec:tr_co_discuss}

Previous studies investigating 
the inner dead-zone edge with 
global non-ideal MHD simulations 
consider only OR
\citep{Lyra2012,Dzyurkevich2010,Flock2017}.
Our simulation shows that AD significantly changes the structure around the inner dead-zone edge.

Without AD, no rapid magnetic flux transport
in the transition zone occurs.
As shown in Section \ref{sec:loopform}, 
if a finite $B_z$ exists in the initial condition, 
AD amplifies $\avphi{B_R}$, 
leading to the enhancement of $\avphi{E_\phi}_\mathrm{O}$.
Since $\etaO$ increases rapidly with $R$ in the transition zone (Figure \ref{fig:profinimid}), 
the magnetic flux is efficiently transported outward 
within a few rotations at $R=\Rmri$.

The previous studies focusing on outer disks show that
AD triggers spontaneous magnetic flux concentration and 
form ring-like substructures in disks
\citep{BaiStone2014,Bethune2017,Suriano2018,Suriano2019,CuiBai2021}.
Although the formation mechanism of 
the spontaneous flux concentration 
is still unclear, 
the previous papers suggest that  a driving mechanism of the flux concentration 
is related with the fact that AD steepens the gradients of magnetic fields around magnetic 
nulls \citep{BrandenburgZweibel1994}.
\citet{Suriano2018} found that AD generates a pinched magnetic 
field ($|B_R|\gg |B_z|$) at 
the mid-plane where the toroidal 
magnetic field is reversed. 
Then, magnetic reconnection is triggered,
resulting in magnetic flux transport.
\citet{CuiBai2021} identified that
the direction of $B_\phi$ is reversed in 
the radial direction where the 
magnetic flux is concentrated.
Another property of AD that may form substructures in disks is 
that ${J_{\phi,\perp}}$ and ${J_\phi}$ can have opposite sign when $B_\phi \gg B_R, B_z$.
\citet{Bethune2017} pointed out AD works as anti-diffusion.
This behavior is confirmed also in our simulations in Figures \ref{fig:Eph_eta} and \ref{fig:Eph_z}.

Although our simulations show  the $B_z$ concentrations around 
the regions where  the sign of $B_\phi$ is reversed, 
prominent gap structures do not form  in the coherent zone. 
This may be caused by the $B$ dependence of $\etaA$. 
Significant flux concentrations require that $\etaA$ increases as $\propto B^2$.
As shown in Figure \ref{fig:etadep}, 
$\etaA$ is almost independent of $B$ for $\Reta\lesssim R\lesssim \Rch$ where 
the dominant negative charge carrier is dust grains.
In addition to the Ohmic diffusion, 
this may be one of the reasons why flux concentration does not occur there.

For $R>\Rch$, since the physical conditions are similar to those in \citet{Suriano2019},
AD should generate the flux concentrations.
The reason why our results do not show the flux concentrations is that  
the simulation time is too short for them to develop;
2000 rotations at the inner boundary corresponds to $\sim8$ rotations at $R=4~\mathrm{au}$.

\subsubsection{ Vertical Shear Instability}\label{sec:VSI}

{
Since the locally isothermal equation of state is 
used in our simulations (Section \ref{sec:basiceq}),  
the VSI may occur \citep[e.g.,][]{UrpinBrandenburg1998,Nelson2013}.
In this section, we will show that 
our simulation time is too short to see non-linear growth of the VSI 
in the transition and coherent zones.

Vertically global linear analyses show that 
the VSI modes can be divided into 
the two body modes (corrugation and breathing modes) 
that have large amplitudes near the mid-plane and 
the surface modes that emerge when the boundaries are set in the vertical direction
\citep{Nelson2013,BarkerLatter2015,UmurhanNelsonGressel2016}.
In the nonlinear stage, the fundamental corrugation mode dominates 
over the other modes \citep[e.g.,][]{Nelson2013}.

In all the regions, the VSI does not appear to grow in our simulations.
In the active zone, the VSI is 
suppressed  by magnetic fields  \citep{LatterPapaloizou2018,CuiLin2021,LatterKunz2022}.
Also in the transition and coherent zones, 
our simulations do not show the non-linear development of the corrugation modes.

To estimate the growth rate of the VSI in the transition and coherent zones,
we used the reduced model proposed by \citet{Nelson2013} with $T\propto R^{-1/2}$ 
\citep[also see][]{BarkerLatter2015,UmurhanNelsonGressel2016}.
A significant difference from the previous studies \citep{Nelson2013,BarkerLatter2015,UmurhanNelsonGressel2016} 
lies in the vertical extent where the VSI occurs.
In our simulations, for $|z|\gtrsim H$, $\LambdaO$ is greater than unity 
and the gas is marginally coupled with the magnetic field (Figure \ref{fig:Eph_z}c).
It is reasonable to set the boundaries at $|z|=H$ because 
the VSI is suppressed by the Lorentz force in $|z|\gtrsim H$.
In addition, in the coherent zone, the surface gas accretion driven just above 
the OR dead-zone ($|z|\sim H$) is likely to suppress the VSI (Figure \ref{fig:Eph_z}b).
By contrast, the previous studies consider a 
larger vertical ranges ($z>~5\cs/\Omega$).
Unlike the previous studies   
\citep[see Figure 18 in][]{Nelson2013},
the limited vertical range in our setup ($|z|\le H$) 
makes the fundamental corrugation mode the most unstable in the body modes and 
reduces the growth rates of the higher-order body modes.  
This may be consistent with the fact that 
the velocity dispersion driven by the VSI is strongly suppressed when 
the vertical extent of the VSI unstable region is limited to $|z| \lesssim 2\cs/\Omega$ \citep{Fukuhara2023}.
The growth rate $\sigma$ of the fundamental corrugation mode 
is insensitive to the radial wavelength when the radial wavelength is larger than 
$O(\epsilon^2 R)$. The growth rate is about $\sim 0.1~\epsilon \Omega$.
The ratio of the growth timescale to the simulation time 
is about $\sim \sigma^{-1}/t_\mathrm{end} \sim 0.5 (R/1~\mathrm{au})^{1.25}$ (Table \ref{tab:model}).
This indicates that the simulation time is not long enough for the VSI to grow sufficiently nonlinear.

The surface modes, 
which develops near the vertical boundaries, 
is suppressed due to the limited vertical extent 
($|z|\le H$) when the radial wavelength is less than $\sim 40 (\epsilon^2 R)^{-1}$ in the reduced model.
Although the surface modes would emerge 
for higher $k_\mathrm{R}$ at higher growth rates,
the radial resolution of our simulations 
is not sufficient to resolve them.

}

\subsection{Implications for Long-term Evolution}\label{sec:longterm}

Although we conducted LowRes run until $t=2000\tK = 40~\mathrm{yr}$, 
it is much shorter than a typical disk lifetime.
In this section, we discuss a lone-term evolution of the disk inferred 
from our results.

\subsubsection{Gas Evolution}\label{sec:longterm_gas}

First, we consider a long-term evolution by assuming 
the magnetic flux is fixed.
If the gas accretion rate shown in Figure \ref{fig:massactive} is maintained
($\dot{M}\sim 5\times 10^{-4}$ in the simulation unit corresponding to 
$\dot{M}=2.8\times 10^{-7}M_\odot~\mathrm{yr}^{-1}$), 
the depletion time scale of the active zone is 
$\sim 4\times 10^5~\mathrm{yr}$ if there is no mass supply.
This may be a formation mechansim of transition disks
\citep[e.g.,][]{vanderMarel2023}.

What would be the mass transport between the transition and active zones?
As discussed in Sections \ref{sec:structureform_transition} and \ref{sec:angmomtrans}, 
the direction of the mass transfer is determined by the two competing mass flows; 
the mid-plane accretion and the outflow in the high latitudes.
If the mass of the active zone becomes sufficiently small, 
the mid-plane accretion from the transition to active zone will dominate over 
the outflows from the active zone to the transition zone.
This leads to inward mass transfer from the transition zone to the active zone.

In the coherent zone, the surface gas accretion continues to accumulate the gas 
in the outer edge of the transition zone.
If the gas accretion rate is kept to $\dot{M} = 10^{-3}$ in the simulation units (Figure \ref{fig:angmom}), 
the mass corresponding to the initial mass of the active zone is transferred to 
the transition zone in $2\times 10^5$~yr.
The accumulated gas is not directly transferred to the active zone
since gas accretion inside the transition zone does not occur because of no net magnetic flux. 
Thus, the transferred mass forms a ring structure whose surface density contrast 
is about $10^3$ if the radial width is $\sim 0.2~\au$ at $2\times 10^{5}~\mathrm{yr}$.
In reality, the ring will become vorticies by the Rossby-wave instability \citep{Lovelace1999,Ono2016,Ono2018}.

Gas accretion from the inner edge of the transition zone to the active zone 
may increase $\Rmri$ when the surface density becomes low enough for cosmic rays and/or ionizing radiation to reach 
the mid-plane. If $\Rmri$ gets close enough to $\Reta$,
the gas accumulated at the outer-edge of the transition zone may be transferred 
to the active zone.

\subsubsection{Magnetic Flux Evolution}\label{sec:magflux}

Existence of the transition zone significantly affect the magnetic flux evolution in PPDs.
From our results, the following important conclusion can be drawn;
{\it the magnetic fluxes cannot drift from the coherent zone to the active zone.}

The existence of the transition zone affects on the evolution of the active zone.
The magnetic flux initially possessed by the active zone falls to the central star or is 
transported to the transition zone. 
The magnetic flux transported to the transition zone does not return to the active zone.
This indicates that no magnetic flux may be supplied to the active zone from $R>\Rmri$.
If this is the case, the magnetic flux in $R\le \Rmri$ decreases with time, 
leading to a decrease in the gas accretion rates onto the central star.
As the process regulates the magnetic flux accretion onto the central star, it can be important to 
determine magnetization of young stellar objects.

The outer-edge of the region where $B_z\sim 0$ moves following the drift velocity of the magnetic flux in the coherent zone as shown in Figure \ref{fig:vB_trans_long}.
We here define the transition zone as the region where $B_z\sim 0$. 
If $v_\mathrm{B}$ is positive as shown in our simulations and the previous simulations \citep{BaiStone2017,Bai2017,Gressel2020}, 
the outer edge of the transition zone moves outwards. 
Since the gas accretion will be suppressed in the region where $B_z\sim 0$, 
the inner-edge of the coherent zone where the surface gas accretion occurs moves outward.
If $v_\mathrm{B}$ is negative, the outer edge of the transition zone will not move and the magnetic flux 
will be accumulated just outside the transition zone.

Long-term evolution of magnetic flux transport in the coherent zone remains elusive.
The drift velocities have about an order of magnitude differences between the previous studies \citep{BaiStone2017,Bai2017,Gressel2020}.
\citet{Lesur2021} shows that the radial drift velocities strongly depend on the spatial 
distributions of $\etaO$ and $\etaA$ using simple self-similar solutions.
Although their results show only outward drifts, we need to investigate how drift velocities 
depend on the disk parameters and the size distribution of dust grains.
If the drift velocity can be negative in a parameter range, 
the magnetic flux will be accumulated at the outer edge of the transition zone.
We will investigate how drift velocities depend on the disk parameters and 
the size of dust grains in the future.

\subsection{Implication for Dynamics of Dust Grains}\label{sec:accdust}

The inner dead-zone edge is thought to be a possible site where 
dust grains accumulate since a pressure bump is created owing 
to outward viscous expansion of the inner active zone \citep[e.g.,][]{Kretke2009}.
So far, accumulation of dust grains at the inner dead-zone edge was 
investigated mainly by using one-dimensional (1D) model taking into account 
viscous evolution of disks and coupling of the gas and dust grains 
\citep{Kretke2009,Pinilla2016,Ueda2019}.
The 1D models can compute the evolution for much longer periods than three-dimensional global disk simulations.

From our simulations considering realistic spatial distributions of $\etaO$ and $\etaA$, 
we found that the transition zone is created with a finite width at the inner dead-zone edge.
The gas accumulates from both inner and outer edges of the transition zone, 
producing the two ring structures at both the edges (Figure \ref{fig:soundwave}a).
The structure found in this study is quite different from that predicted 
by the 1D models that show a single pressure bump because 
the 1D models assume a step-function-like distribution of the viscous parameter $\alpha$ 
and neglect the gas surface accretion in the dead zone.

\citet{Ueda2019} derived a critical value of the $\alpha$ parameter $\alpha_\mathrm{dead}$ 
in the dead zone below which the dust pileup occurs,
\begin{equation}
    \alpha_\mathrm{dead} \sim 3\times 10^{-4} \left(\frac{v_\mathrm{frag}}{1~\mathrm{m~s^{-1}}}\right),
    \label{Ueda}
\end{equation}
where $v_\mathrm{frag}$ is the minimum collision speed that triggers fragmentation of dust grains.
For silicate aggregates, $v_\mathrm{frags}$ spans $1-10~\mathrm{m~s^{-1}}$ \citep{BlumWurm2000,Wada2013}.
The dead zone in \citet{Ueda2019} corresponds to the transition zone in our simulations.
We note that \citet{Ueda2019} considers $\alpha_\mathrm{dead}$ as the turbulent viscosity.
In our simulations, the velocity dispersion in the transition zone is generated by the sound waves originating from the active zone (Section \ref{sec:sound}).
The $\alpha$ parameter estimated from the Reynolds and Maxwell stresses, $\alpha_\mathrm{tr}$, cannot be 
substituted into $\alpha_\mathrm{dead}$ in Equation (\ref{Ueda}) 
because the velocity dispersion in the transition zone $\delta v_\mathrm{tr}$ does not satisfy $\delta v_\mathrm{tr}/c_\mathrm{s} \sim \sqrt{\alpha_\mathrm{tr}}$
, which is valid in the MRI turbulence.
Instead, $\alpha_\mathrm{dead}$ in Equation (\ref{Ueda}) should be considered as a 
measure of the random velocity dispersion, and 
is replaced with $(\delta v_\mathrm{tr}/\cs)^2$.
In terms of the velocity dispersion $\delta v_\mathrm{tr}$, 
Equation (\ref{Ueda}) is rewritten as 
\begin{equation}
    \frac{\delta v_\mathrm{tr}}{c_\mathrm{s}} \lesssim 0.02
    \left( \frac{v_\mathrm{frag}}{1~\mathrm{m~s^{-1}}}\right)^{1/2}.
    \label{vfrag}
\end{equation}
Equation (\ref{vfrag}) shows that in order for dust grains to accumulate without being radially 
diffused, 
the Mach number of the velocity dispersion in the transition zone is required to be lower than 
$0.02$ for $v_\mathrm{frag} = 1~\mathrm{m~s^{-1}}$. 

Our results show that two rings form at both the edges of the transition zone, or $R=\Rmri$ and $\Reta$ 
although the setting of \citet{Ueda2019} yields a single ring forms at the inner edge of the dead zone. 
Thus, accumulation of dust grains potentially occurs in the two rings.
Figure \ref{fig:soundwave}a
shows that the Mach numbers of the velocity dispersion $\avphi{\delta {\cal M}_\mathrm{tot}}$ at the mid-plane 
is as large as 0.1 at the inner edge of the transition zone, and decreases with radius to reach $\sim 0.01$ at the outer edge of the transition zone.
Comparing our results with Equation (\ref{vfrag})
shows that accumulation of dust grains may be suppressed in the ring 
near the inner edge of the transition zone. 
In the ring at the outer edge of the transition zone, 
dust grains may accumulate if $v_\mathrm{frag}$ is well above $1~\mathrm{m~s^{-1}}$.
The velocity dispersion in the transition zone is expected to be proportional to 
that in the active zone.
To accumulate dust grains in the rings, plasma beta larger than $10^4$ would 
be preferred.

\subsection{Implications for Protoplanetary Disks Around Solar-type Stars}\label{sec:solartype}

In this paper, PPDs around Herbig Ae stars were focused on.
It is worth discussing implications for PPDs around solar-type stars from our results.

Since the inner dead-zone edge $\Rmri$ is roughly determined by 
$T\sim 10^3~\mathrm{K}$ \citep{DeschTurner2015},
$\Rmri$ is predicted to be 
\begin{eqnarray}
    \Rmri &\sim & 0.6~\mathrm{\au} 
    \left(\frac{\kappa_\mathrm{R}}{5~\mathrm{cm^2~g^{-1}}}\right)^{2/9}
    \left(\frac{\dot{M}}{7\times 10^{-8}~M_\odot~\mathrm{yr}^{-1}}\right)^{4/9} \nonumber\\
    && 
    \left(\frac{{M}_*}{1~M_\odot}\right)^{1/3}
    \left(\frac{\alpha}{10^{-2}}\right)^{-2/9},
    \label{alphadisk}
\end{eqnarray}
where $\kappa_\mathrm{R}$ is the Rossland mean opacity, and 
$\kappa_\mathrm{R}=5~\mathrm{cm^2~g^{-1}}$ corresponds to 
the opacity of  dust grains with 0.1~$\mu$m and a dust-to-gas mass ratio of $10^{-2}$.
In derivation of Equation (\ref{alphadisk}), we adopt the standard $\alpha$ disk model 
\citep{ShakuraSunyaev1973,Gammie1996}
where the vertical optical depth is set to $\kappa_\mathrm{R}\Sigma /2$. 

{
\citet{Jankovic2021,Jankovic2022} investigated steady-state axisymmetric disk structures 
considering detailed physical and chemical processes.
They found that the temperature structure
is more complex than our adopted one (Equation (\ref{tem})).
This will affect the shape of the dead zone boundaries. 
They derived a similar $\Rmri$ as Equation (\ref{alphadisk}) even when 
they took into account the disk structures modified by convective instability.
}

\citet{MoriBaiOkuzumi2019} pointed out that $\Rmri$ derived from 
Equation (\ref{alphadisk}) assumes the existence of MRI turbulence.
If no MRI turbulence exists, the disk is laminar. 
Joule heating is much less effective than viscous heating 
\citep{MoriBaiOkuzumi2019}, and 
the gas temperature becomes comparable to that determined by
the stellar irradiation \citep{ChiangGoldreich1997}.
Thus, the MRI should be self-sustained in a sense that 
the MRI turbulent heating provides a sufficient amount of charged particles to drive the MRI.
It is however unclear whether this self-sustaining cycle works stably.
If the MRI is suppressed, the inner edge of the dead zone moves closer to the central star.

The existence of the transition zone 
should be critical in global disk evolution as discussed in Section \ref{sec:longterm}.
Although criteria to form the transition zone are unclear, 
if the active zone exists stably, 
Section \ref{sec:loopform} shows that the formation of the transition zone 
requires $\etaO$ to increase rapidly across $R\sim \Rmri$ and $\etaA$ to be large enough to amplify 
$B_R$ and $B_\phi$ around the inner edge of the dead zone.
Further investigations are required to reveal 
whether the transition zone exists and whether the active zone can exist stably by using 
non-ideal MHD numerical simulations with radiative transfer.

\section{Summary}\label{sec:summary}

In this paper, we performed global non-ideal MHD simulations 
of inner regions of a protoplanetary disk around an intermediate star
focusing on the behaviors of the inner dead zone boundaries by taking into account 
the Ohmic resistivity (OR) and ambipolar diffusion (AD).
The radial extent spans from $R=0.1~$au to $R\sim 10~$au.

The three characteristic radii, $\Rmri$, $R_\eta$, and $\Rch$, are defined 
from the spatial distribution of $\etaO$ and $\etaA$.
The first radius $\Rmri=0.5~$au 
corresponds to the radius outside of which the MRI is suppressed around the mid-plane.
The second radius $R_\eta=1.2$~au shows the radius around which 
$\eta_\mathrm{O}$ and $\etaA$ are maximized.
In the disk where $R<R_\eta$, the thermal 
ionization of metals determine the ionization degree.
Beyond the last radius $\Rch=3~\au$,
the diffusion coefficients are small enough for the accretion flow to be driven at the mid-plane.

We found that 
the dead zone identified by $\LambdaO < 1$ and $\LambdaA<1$ 
can be divided into two regions separated by $R_\eta$.
The transition zone in $\Rmri<R<\Reta$ is discovered in this paper and 
has different properties than those of the conventional dead zones while
the coherent zone in $R\ge \Reta$ shares the main characteristics with the conventional dead zone.

We summarize our work as follows:

\begin{enumerate}
    \item 
The overall physical properties of the active zone ($R<\Rmri$) are consistent with those found in the literature.
\begin{itemize}
\item Prominent ring structures develop in the long-term evolution (Figure \ref{fig:Ring}),
independent of resolution.
The vertical field strength is anti-correlated with the surface density.
Although the driving mechanisms are uncertain, the viscous instability \citep{Suzuki2023}
would be promising  (Sections \ref{sec:structureform_active} and \ref{sec:ring_discuss}).
The development of ring structures in MRI disks is an open question that requires further investigation.

\item The magnetic flux transport in the active zone depends both on time and radius (Section \ref{sec:mag_active}).
After the ring structures and flux concentrations have developed, 
the magnetic flux drifts inward.
A typical drift velocity in the active zone is about $\sim 10^{-4}~v_\mathrm{K}$.

\end{itemize}

\item The physical properties of the coherent zone ($R\ge \Reta$) are also consistent with those found in the literature.

\begin{itemize}
    \item The behaviors of the angular momentum transfer are characterized by $R=\Rch$. 
       For $R_\eta< R <\Rch$, surface gas accretion is driven just above the OR dead zone.
       The mass loss rate of the disk wind is comparable to the mass accretion rate. 
       For $R\ge \Rch$, $\etaO$ is low enough for the toroidal field to be amplified inside the disk, 
       leading to mid-plane gas accretion (Sections \ref{sec:overall} and \ref{sec:angmomtrans}).

    \item The drift velocity of the magnetic flux is determined by 
    the inward drag owing to the surface gas accretion and outward diffusion owing to AD (Section \ref{sec:mag_dead}).
    Our simulations show that the latter dominates over the former, and 
    the magnetic flux drift outwards at a speed of $\sim 10^{-3}v_\mathrm{K}$.
    
\end{itemize}

\item The transition zone appears in $\Rmri\le R\le R_\eta$. 
    Although it belongs to the conventional dead zone defined by Els\"asser nubmers,
    their physical properties are completely different from those in the conventional dead zone.
    Our findings regarding the transition zone are as follows:
    \begin{itemize}
        \item The vertical magnetic flux is 
              rapidly transported outward both by OR and AD,
              forming a magnetic gap where the vertical field is almost zero
              (Section \ref{sec:loopform}).
              Without AD, the rapid flux transport would not occur.
              
        \item Unlike the conventional dead zone \citep{BaiStone2013}, 
              the lack of the vertical field suppresses the surface gas accretion.
              Since neither MRI nor coherent magnetic fields extract the angular momentum, 
              gas accretion is largely suppressed in the transition zone.

        \item The mass of the transition zone monotonically 
        increases with time by the gas inflow both from the 
        inner- and outer-edges of the transition zone during the calculated period.
        As a result, rings form in both the edges of the transition zone.
        The outflow launched around the inner edge of the transition zone fails to 
        escape the disk but falls onto 
        the outer part of the transition zone, resulting in 
        zero net vertical mass flux.

    \end{itemize}

\end{enumerate}

Our simulations have demonstrated that the complicated structures formed by 
the non-ideal MHD effects affect the mass, angular momentum, and magnetic flux transport as well 
as dust accumulation.
We are planning to perform wider parameter search to investigate such structures systematically.

{
There are two main caveats in our study.
One is that the Hall effect is neglected for simplicity.
In our model setup, the diffusion coefficient of the Hall effect 
is as large as $\etaO$ and $\etaA$ near the dead-zone inner boundaries.
We will include the Hall effect because 
it can change the disk structure significantly \citep{BaiStone2017,Bethune2017, MartelLesur2022}.
The other is that we use the locally isothermal equation of state with 
a simple temperature profile (Equation (\ref{tem})).
The temperature profile is more complicated near the dead-zone 
inner boundaries when the radiative transfer is 
considered \citep[e.g.,][]{Flock2017,Gressel2020}.
This can change the spatial distributions of the diffusion coefficients and 
the thermal evolution of the gas. 
}





\begin{ack}
Numerical computations were carried out on 
supercomputer Fugaku provided by the RIKEN Center 
for Computational Science (Project ids: hp210164, hp220173), 
Oakforest-PACS at Joint Center for Advanced High Performance Computing (Project ids: hp190088, hp200046, hp210113),
and  Cray XC50 at the CfCA of the   National Astronomical Observatory of Japan.
\end{ack}

\section*{Funding}
This work was supported in part by the Ministry of Education, Culture, Sports, Science and Technology (MEXT), Grants-in-Aid for Scientific Research, JP19K03929 (K.I.), JP16H05998 (K.T. and K.I.), JP21H04487, JP22KK0043 (K.T., K.I., and S.T.), JP22K14074 (S.T.), JP20H00182, JP23H00143, JP23K25923, JP23H01227 (S.O.), JP17H01105, JP21H00033, and JP22H01263 (T.K.S.). This research was also supported by MEXT as ``Exploratory Challenge on Post-K computer'' (Elucidation of the Birth of Exoplanets [Second Earth]  and the Environmental Variations of Planets in the Solar System) and “Program for Promoting Researches on the Supercomputer Fugaku” (Toward a unified view of the universe: from large scale structures to planets, JPMXP1020200109).

\appendix 

\section{Resolution Dependence of the $\alpha$ Parameter}\label{sec:QF}

We investigate whether the resolution of our simulations is sufficient to resolve the MRI.
In our simulations, the scale height $H(R)$ is resolved with $N=17(R/0.3~\mathrm{au})^{1/4}$ for LowRes run
and $N=34(R/0.3~\mathrm{au})^{1/4}$ for HighRes run.

\citet{Noble2010} introduces the so-called quality factors, 
which are defined as the number of grid cells resolving the fastest MRI growing mode
measured by amplified magnetic fields,
\begin{equation}
    Q_r = \frac{2\pi v_{\mathrm{A},r}}{\Omega \Delta r},\;\;\;
    Q_\theta = \frac{2\pi v_{\mathrm{A},\theta}}{\Omega r\Delta \theta},\;\;\;
    Q_\phi = \frac{2\pi v_{\mathrm{A},\phi}}{\Omega r\sin\theta\Delta \phi},
\end{equation}
where $v_{\mathrm{A},r,\theta,\phi} = B_{r,\theta,\phi}/\sqrt{4\pi \rho}$.
\citet{Sorathia2012} shows that in order to obtain the converged result, 
$Q_\theta$ needs to be larger than $10-15$ if $Q_\phi\sim 10$, and 
$Q_\theta$ can be smaller if $Q_\phi > 25$ \citep[see also][]{Hawley2013}.

\begin{figure}
\begin{center}
    \includegraphics[width=8cm]{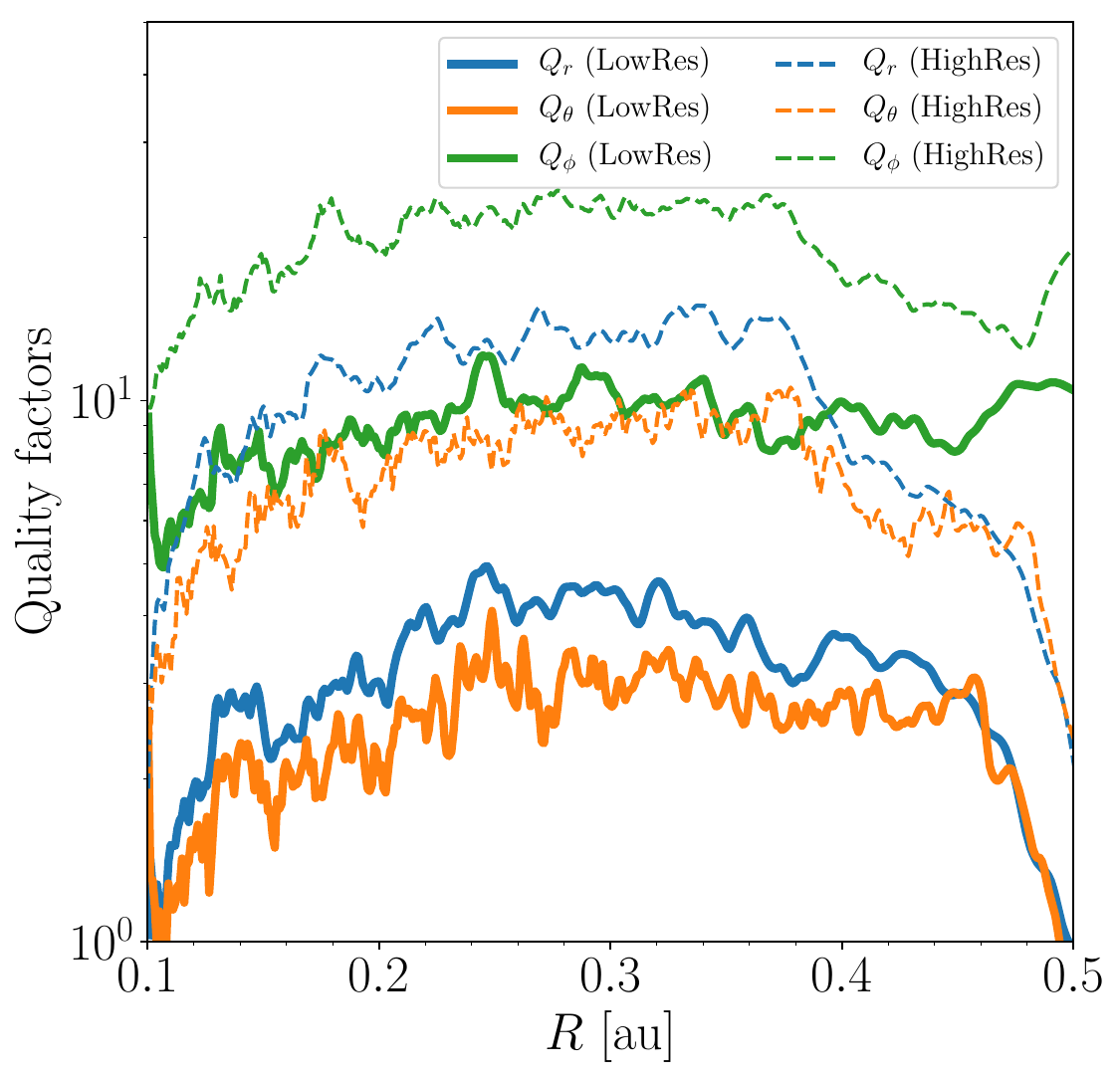}
\end{center}
    \caption{
        Radial distribution of the quality factors $Q_r$, $Q_\theta$, and $Q_\phi$ 
        averaged over $0\le\phi\le 2\pi$ and $|z|\le H$ at $t=300\tK$.
        The thick-solid and thin-dashed lines correspond to the results of LowRes and HighRes, respectively.
    }
    \label{fig:QC}
\end{figure}
Figure \ref{fig:QC} shows the radial profiles of $Q_r$, $Q_\theta$, and $Q_\phi$ averaged over 
$0\le \phi\le 2\pi$ and $|z|<H$ both for LowRes and HighRes runs.
HighRes run barely satisfies the condition derived by \citet{Sorathia2012}.
By contrast, for LowRes run, the fastest growing mode is not resolved in our simulation although 
the MRI at longer wavelengths can be captured because 
$Q_\phi$ is around 10 for $r>0.2~$au and  $Q_\theta$ is around 3.

\begin{figure}
    \begin{center}
    \includegraphics[width=8cm]{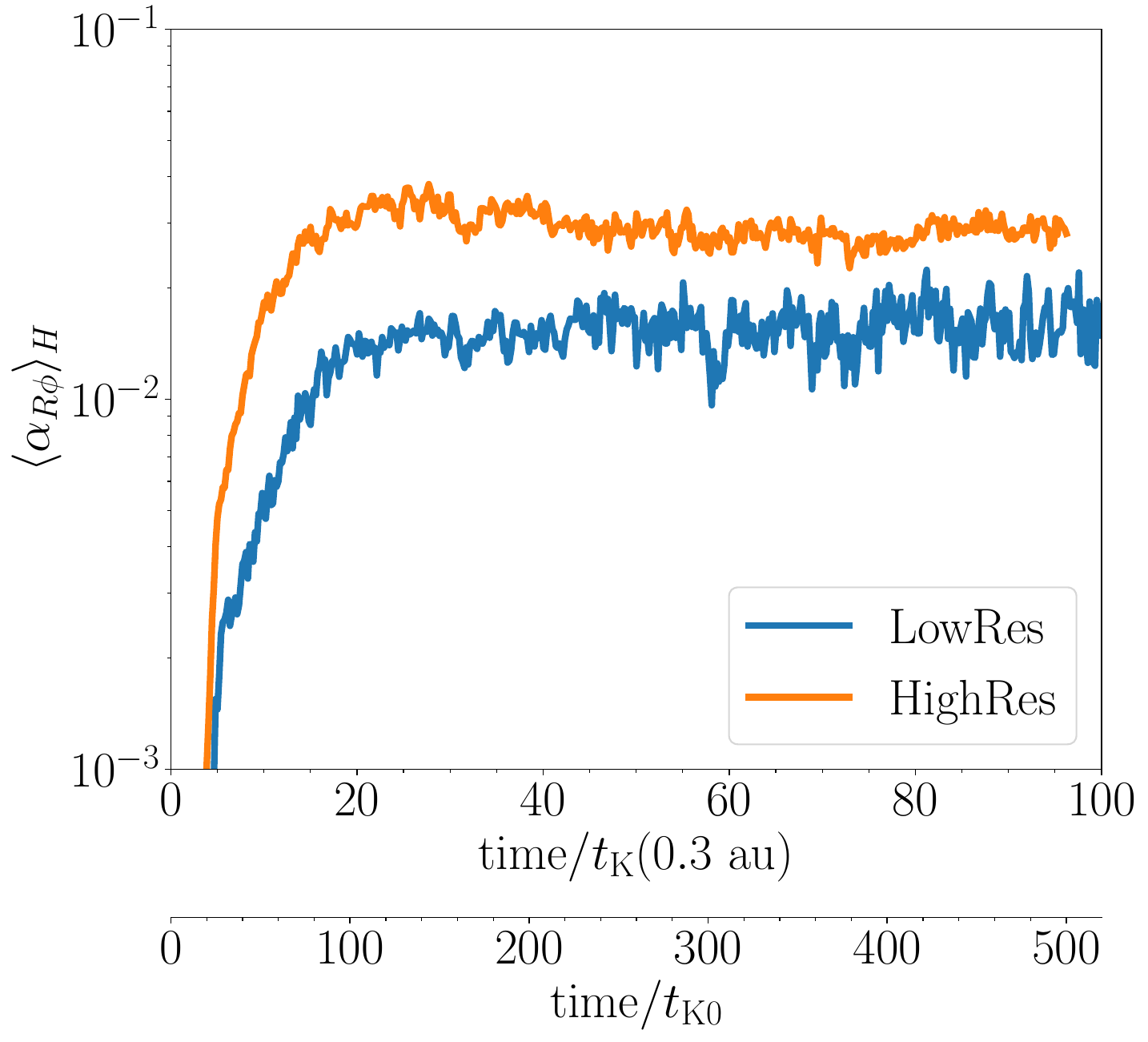}
    \end{center}
    \caption{
    Time evolution of $\avphi{\alpha_{R\phi}}_H$ at $R=0.3~\au$ for LowRes and HighRes.
    }
    \label{fig:resolution}
\end{figure}

Figure \ref{fig:resolution} compares the time evolution of $\avphi{\alpha_{R\phi}}_H$ 
(Equation (\ref{alpha}) between LowRes and HighRes runs.
Both models show that the MRI turbulence is saturated after $t\sim 20 t_\mathrm{K}(R=0.3~\au)$.
The saturation level of $\avphi{\alpha_{R\phi}}_H$ is 
about twice larger for HighRes than for LowRes runs,
consistent with the fact that LowRes run does not resolve the MRI turbulence sufficiently.
The saturation level 
$\avphi{\alpha_{R\phi}}_H \sim 
2.8\times 10^{-2}$ 
for HighRes run is comparable to 
those found in local shearing-box simulations \citep[e.g.,][]{Hawley1995,Sano2004}. 
We confirmed that HighRes run is expected to provide converged results by measuring 
the so-called quality factors \citep{Noble2010}.

\section{MRI Turbulence in the Active Zone}\label{app:MRIturbulence}

\begin{figure*}
    \begin{center}
    \includegraphics[width=17cm]{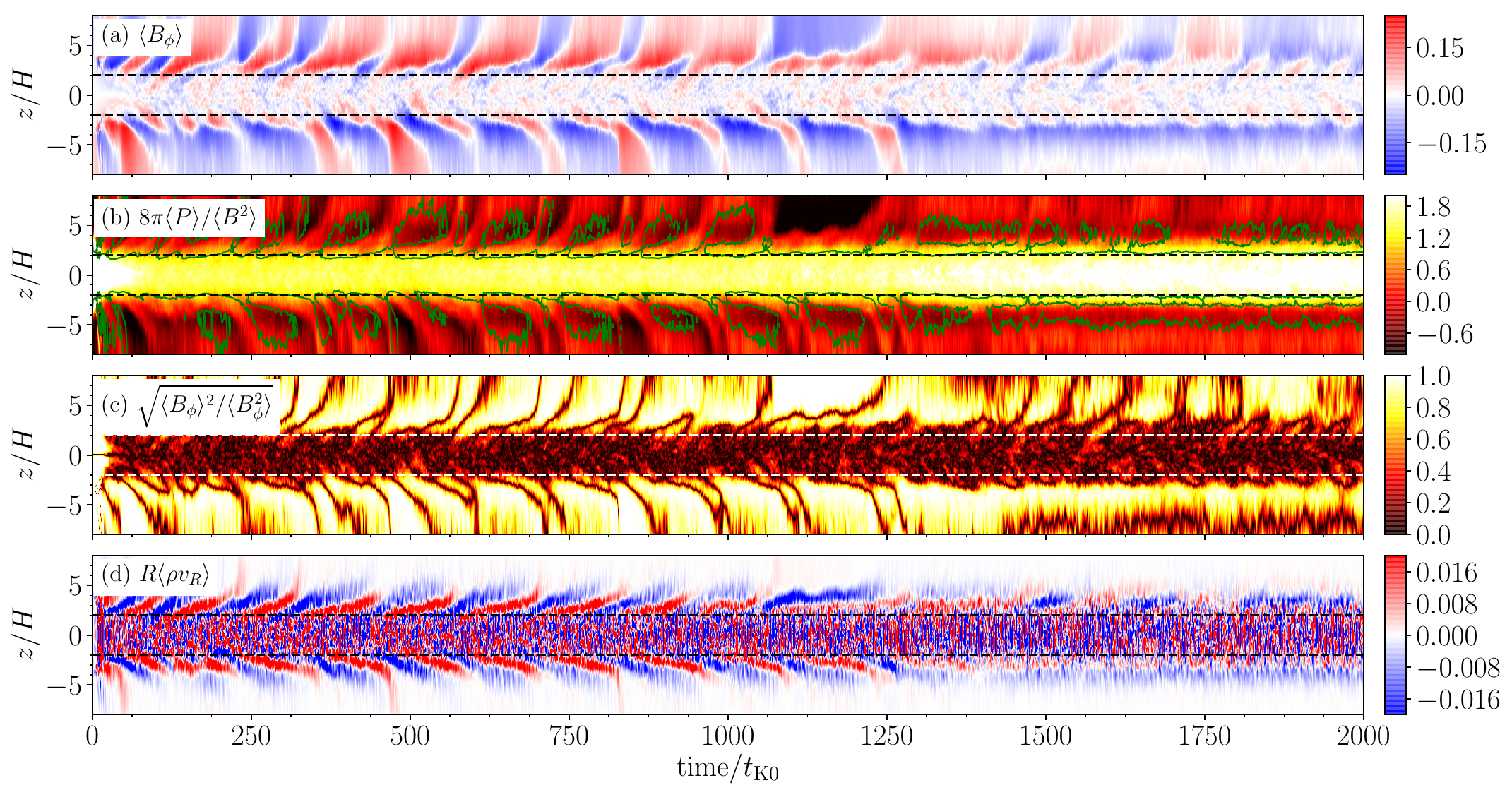}
    \end{center}
    \caption{
        Spacetime diagram for 
        (a) $\avphi{B_\phi}$, (b) $8\pi \avphi{P}/\avphi{B^2}$, (c) 
        $\sqrt{\avphi{B_\phi}^2/\avphi{B_\phi^2}}$, (d) $R\avphi{\rho v_R}$ at 
        $R=0.3$~au in the active zone.
        In Panel (b), the contours of $8\pi \avphi{P}/\avphi{B^2}$ at $R=0.3~\au$ are displayed.
        In each panel, the two horizontal dashed lines show the positions of $z=\pm 2H$.
    }
    \label{fig:active_ver}
\end{figure*}

Evolution of the MRI-driven turbulence is
characterized by magnetic-field 
amplification owing to the channel flows 
followed by dissipation \citep{SanoInutsuka2001}.
The time evolution of the vertical structures 
at $R=0.3~$au is shown in Figure \ref{fig:active_ver}.

Figure \ref{fig:active_ver}a shows that 
the vertical structure of $\avphi{B_\phi}$ changes 
around $|z|\sim 2H$ \citep{Suzuki2009}.
Around the mid-plane for $|z|< 2H$, the gas pressure 
dominates over the magnetic pressure, and the magnetic 
fields are amplified by  the MRI dynamo activity.
At $|z|\sim 2H$, the magnetic pressure becomes comparable to the 
gas pressure (Figure \ref{fig:active_ver}b), and 
the maximum growth scale of the MRI is comparable to $H$ since it is expressed as $2\pi v_\mathrm{A}/\Omega_0$.
As a result, the magnetic fields are amplified by the channel flow 
whose scale is comparable to $H$, and are dissipated by magnetic reconnection.
The $t$-$z$ diagram of $B_\phi$ shows a clear so-called butterfly structure where 
$B_\phi$ flips its sign quasi-periodically and drifts toward upper atmospheres.
The timescale of the quasi-periodic activity is about a few tens rotations,
which is consistent with local shearing-box simulations with stratified density structures.

Figure \ref{fig:active_ver}c shows $\sqrt{\avphi{B_\phi}^2/\avphi{B_\phi^2}_\phi}$ which is an 
indicator of the contribution of the coherent component of $B_\phi$ to the total toroidal field strength.
The coherent field component dominates in the upper atmosphere for $|z|>2H$ while the field structure 
is controlled by the MRI turbulence in the disk ($|z|<2H$).

The gas motion is affected by the magnetic field structures.
As shown in Figure \ref{fig:active_ver}d, $R\avphi{\rho v_R}$ stochastically changes owing to the MRI turbulence for $|z|<2H$, 
it has a coherent quasi-periodic structure in the upper atmosphere.
The time variability of $R\avphi{\rho v_R}$ is correlated with that of $\avphi{B_\phi}$.
This suggests that the magnetic torque of the coherent magnetic field drifting upwards for $|z|>2H$ drives the coherent gas motion
near the surface layers.

In the long term evolution, 
Figure \ref{fig:active_ver} shows that 
MRI dynamo at $R=0.3~\au$ is somewhat weakened and the plasma $\beta$ starts to increase
when $t\gtrsim 1300t_\mathrm{K0}$ corresponding to $250 t_\mathrm{K}(R=0.3~\au)$.
This is related to the formation of ring structures shown in Section \ref{sec:structureform_active}.

\bibliographystyle{aasjournal}
\bibliography{ms} 

\begin{thebibliography}{}
\expandafter\ifx\csname natexlab\endcsname\relax\def\natexlab#1{#1}\fi
\providecommand{\url}[1]{\href{#1}{#1}}
\providecommand{\dodoi}[1]{doi:~\href{http://doi.org/#1}{\nolinkurl{#1}}}
\providecommand{\doeprint}[1]{\href{http://ascl.net/#1}{\nolinkurl{http://ascl.net/#1}}}
\providecommand{\doarXiv}[1]{\href{https://arxiv.org/abs/#1}{\nolinkurl{https://arxiv.org/abs/#1}}}

\bibitem[{{Armitage}(1998)}]{Armitage1998}
{Armitage}, P.~J. 1998, \apjl, 501, L189

\bibitem[{{Bai}(2013)}]{Bai2013}
{Bai}, X.-N. 2013, \apj, 772, 96

\bibitem[{{Bai}(2017)}]{Bai2017}
---. 2017, \apj, 845, 75

\bibitem[{{Bai} \& {Goodman}(2009)}]{BaiGoodman2009}
{Bai}, X.-N., \& {Goodman}, J. 2009, \apj, 701, 737

\bibitem[{{Bai} \& {Stone}(2011)}]{BaiStone2011}
{Bai}, X.-N., \& {Stone}, J.~M. 2011, \apj, 736, 144

\bibitem[{{Bai} \& {Stone}(2013)}]{BaiStone2013}
---. 2013, \apj, 769, 76

\bibitem[{{Bai} \& {Stone}(2014)}]{BaiStone2014}
---. 2014, \apj, 796, 31

\bibitem[{{Bai} \& {Stone}(2017)}]{BaiStone2017}
---. 2017, \apj, 836, 46

\bibitem[{{Bai} {et~al.}(2016){Bai}, {Ye}, {Goodman}, \& {Yuan}}]{Baietal2016}
{Bai}, X.-N., {Ye}, J., {Goodman}, J., \& {Yuan}, F. 2016, \apj, 818, 152

\bibitem[{{Balbus} \& {Hawley}(1991)}]{BalbusHawley1991}
{Balbus}, S.~A., \& {Hawley}, J.~F. 1991, \apj, 376, 214

\bibitem[{{Barker} \& {Latter}(2015)}]{BarkerLatter2015}
{Barker}, A.~J., \& {Latter}, H.~N. 2015, \mnras, 450, 21

\bibitem[{{Beckwith} {et~al.}(2009){Beckwith}, {Hawley}, \&
  {Krolik}}]{Beckwith2009}
{Beckwith}, K., {Hawley}, J.~F., \& {Krolik}, J.~H. 2009, \apj, 707, 428

\bibitem[{{B{\'e}thune} {et~al.}(2017){B{\'e}thune}, {Lesur}, \&
  {Ferreira}}]{Bethune2017}
{B{\'e}thune}, W., {Lesur}, G., \& {Ferreira}, J. 2017, \aap, 600, A75

\bibitem[{{Blaes} \& {Balbus}(1994)}]{BlaesBalbus1994}
{Blaes}, O.~M., \& {Balbus}, S.~A. 1994, \apj, 421, 163

\bibitem[{{Blandford} \& {Payne}(1982)}]{BlandfordPayne1982}
{Blandford}, R.~D., \& {Payne}, D.~G. 1982, \mnras, 199, 883

\bibitem[{{Blum} \& {Wurm}(2000)}]{BlumWurm2000}
{Blum}, J., \& {Wurm}, G. 2000, icarus, 143, 138

\bibitem[{{Brandenburg} \& {Zweibel}(1994)}]{BrandenburgZweibel1994}
{Brandenburg}, A., \& {Zweibel}, E.~G. 1994, \apjl, 427, L91

\bibitem[{{Chandrasekhar}(1961)}]{Cha1961}
{Chandrasekhar}, S. 1961, {Hydrodynamic and hydromagnetic stability}

\bibitem[{{Chiang} \& {Goldreich}(1997)}]{ChiangGoldreich1997}
{Chiang}, E.~I., \& {Goldreich}, P. 1997, \apj, 490, 368

\bibitem[{{Cui} \& {Bai}(2021)}]{CuiBai2021}
{Cui}, C., \& {Bai}, X.-N. 2021, \mnras, 507, 1106

\bibitem[{{Cui} \& {Lin}(2021)}]{CuiLin2021}
{Cui}, C., \& {Lin}, M.-K. 2021, \mnras, 505, 2983

\bibitem[{{Desch} \& {Turner}(2015)}]{DeschTurner2015}
{Desch}, S.~J., \& {Turner}, N.~J. 2015, \apj, 811, 156

\bibitem[{{Dzyurkevich} {et~al.}(2010){Dzyurkevich}, {Flock}, {Turner},
  {Klahr}, \& {Henning}}]{Dzyurkevich2010}
{Dzyurkevich}, N., {Flock}, M., {Turner}, N.~J., {Klahr}, H., \& {Henning}, T.
  2010, \aap, 515, A70

\bibitem[{{Evans} \& {Hawley}(1988)}]{EH1988}
{Evans}, C.~R., \& {Hawley}, J.~F. 1988, \apj, 332, 659

\bibitem[{{Ferreira}(1997)}]{Ferreira1997}
{Ferreira}, J. 1997, \aap, 319, 340

\bibitem[{{Finocchi} \& {Gail}(1997)}]{FinocchiGail1997}
{Finocchi}, F., \& {Gail}, H.~P. 1997, \aap, 327, 825

\bibitem[{{Fleming} \& {Stone}(2003)}]{FlemingStone2003}
{Fleming}, T., \& {Stone}, J.~M. 2003, \apj, 585, 908

\bibitem[{{Flock} {et~al.}(2017){Flock}, {Fromang}, {Turner}, \&
  {Benisty}}]{Flock2017}
{Flock}, M., {Fromang}, S., {Turner}, N.~J., \& {Benisty}, M. 2017, \apj, 835,
  230

\bibitem[{{Fukuhara} {et~al.}(2023){Fukuhara}, {Okuzumi}, \&
  {Ono}}]{Fukuhara2023}
{Fukuhara}, Y., {Okuzumi}, S., \& {Ono}, T. 2023, \pasj, 75, 233

\bibitem[{{Gammie}(1996)}]{Gammie1996}
{Gammie}, C.~F. 1996, \apj, 457, 355

\bibitem[{{Gardiner} \& {Stone}(2008)}]{Gardiner2008}
{Gardiner}, T.~A., \& {Stone}, J.~M. 2008, Journal of Computational Physics,
  227, 4123

\bibitem[{{Gressel} {et~al.}(2020){Gressel}, {Ramsey}, {Brinch}, {Nelson},
  {Turner}, \& {Bruderer}}]{Gressel2020}
{Gressel}, O., {Ramsey}, J.~P., {Brinch}, C., {et~al.} 2020, \apj, 896, 126

\bibitem[{{Gressel} {et~al.}(2015){Gressel}, {Turner}, {Nelson}, \&
  {McNally}}]{Gressel2015}
{Gressel}, O., {Turner}, N.~J., {Nelson}, R.~P., \& {McNally}, C.~P. 2015,
  \apj, 801, 84

\bibitem[{{Guilet} \& {Ogilvie}(2012)}]{GuiletOgilvie2012}
{Guilet}, J., \& {Ogilvie}, G.~I. 2012, \mnras, 424, 2097

\bibitem[{{Hawley}(2001)}]{Hawley2001}
{Hawley}, J.~F. 2001, \apj, 554, 534

\bibitem[{{Hawley} {et~al.}(1995){Hawley}, {Gammie}, \& {Balbus}}]{Hawley1995}
{Hawley}, J.~F., {Gammie}, C.~F., \& {Balbus}, S.~A. 1995, \apj, 440, 742

\bibitem[{{Hawley} {et~al.}(2011){Hawley}, {Guan}, \& {Krolik}}]{Hawley2011}
{Hawley}, J.~F., {Guan}, X., \& {Krolik}, J.~H. 2011, \apj, 738, 84

\bibitem[{{Hawley} {et~al.}(2013){Hawley}, {Richers}, {Guan}, \&
  {Krolik}}]{Hawley2013}
{Hawley}, J.~F., {Richers}, S.~A., {Guan}, X., \& {Krolik}, J.~H. 2013, \apj,
  772, 102

\bibitem[{{Hirose} \& {Turner}(2011)}]{HiroseTurner2011}
{Hirose}, S., \& {Turner}, N.~J. 2011, \apjl, 732, L30

\bibitem[{{Igea} \& {Glassgold}(1999)}]{IgeaGlassgold1999}
{Igea}, J., \& {Glassgold}, A.~E. 1999, \apj, 518, 848

\bibitem[{{Jacquemin-Ide} {et~al.}(2021){Jacquemin-Ide}, {Lesur}, \&
  {Ferreira}}]{Jacquemin-Ide2021}
{Jacquemin-Ide}, J., {Lesur}, G., \& {Ferreira}, J. 2021, \aap, 647, A192

\bibitem[{{Jankovic} {et~al.}(2022){Jankovic}, {Mohanty}, {Owen}, \&
  {Tan}}]{Jankovic2022}
{Jankovic}, M.~R., {Mohanty}, S., {Owen}, J.~E., \& {Tan}, J.~C. 2022, \mnras,
  509, 5974

\bibitem[{{Jankovic} {et~al.}(2021){Jankovic}, {Owen}, {Mohanty}, \&
  {Tan}}]{Jankovic2021}
{Jankovic}, M.~R., {Owen}, J.~E., {Mohanty}, S., \& {Tan}, J.~C. 2021, \mnras,
  504, 280

\bibitem[{{Jin}(1996)}]{Jin1996}
{Jin}, L. 1996, \apj, 457, 798

\bibitem[{{Kretke} {et~al.}(2009){Kretke}, {Lin}, {Garaud}, \&
  {Turner}}]{Kretke2009}
{Kretke}, K.~A., {Lin}, D.~N.~C., {Garaud}, P., \& {Turner}, N.~J. 2009, \apj,
  690, 407

\bibitem[{{Kunz} \& {Balbus}(2004)}]{KunzBalbus2004}
{Kunz}, M.~W., \& {Balbus}, S.~A. 2004, \mnras, 348, 355

\bibitem[{{Latter} \& {Kunz}(2022)}]{LatterKunz2022}
{Latter}, H.~N., \& {Kunz}, M.~W. 2022, \mnras, 511, 1182

\bibitem[{{Latter} \& {Papaloizou}(2018)}]{LatterPapaloizou2018}
{Latter}, H.~N., \& {Papaloizou}, J. 2018, \mnras, 474, 3110

\bibitem[{{Lesur}(2021)}]{Lesur2021}
{Lesur}, G. R.~J. 2021, \aap, 650, A35

\bibitem[{{Lightman} \& {Eardley}(1974)}]{LightmanEardley1974}
{Lightman}, A.~P., \& {Eardley}, D.~M. 1974, \apjl, 187, L1

\bibitem[{{Lovelace} {et~al.}(1999){Lovelace}, {Li}, {Colgate}, \&
  {Nelson}}]{Lovelace1999}
{Lovelace}, R.~V.~E., {Li}, H., {Colgate}, S.~A., \& {Nelson}, A.~F. 1999,
  \apj, 513, 805

\bibitem[{{Lynden-Bell} \& {Pringle}(1974)}]{Lynden-BellPringle1974}
{Lynden-Bell}, D., \& {Pringle}, J.~E. 1974, \mnras, 168, 603

\bibitem[{{Lyra} \& {Mac Low}(2012)}]{Lyra2012}
{Lyra}, W., \& {Mac Low}, M.-M. 2012, \apj, 756, 62

\bibitem[{{Martel} \& {Lesur}(2022)}]{MartelLesur2022}
{Martel}, {\'E}., \& {Lesur}, G. 2022, \aap, 667, A17

\bibitem[{{Meyer} {et~al.}(2014){Meyer}, {Balsara}, \& {Aslam}}]{Meyers2014}
{Meyer}, C.~D., {Balsara}, D.~S., \& {Aslam}, T.~D. 2014, Journal of
  Computational Physics, 257, 594

\bibitem[{{Miyoshi} \& {Kusano}(2005)}]{MK2005}
{Miyoshi}, T., \& {Kusano}, K. 2005, Journal of Computational Physics, 208, 315

\bibitem[{{Montesinos} {et~al.}(2009){Montesinos}, {Eiroa}, {Mora}, \&
  {Mer{\'\i}n}}]{Montesinos2009}
{Montesinos}, B., {Eiroa}, C., {Mora}, A., \& {Mer{\'\i}n}, B. 2009, \aap, 495,
  901

\bibitem[{{Mori} {et~al.}(2019){Mori}, {Bai}, \&
  {Okuzumi}}]{MoriBaiOkuzumi2019}
{Mori}, S., {Bai}, X.-N., \& {Okuzumi}, S. 2019, \apj, 872, 98

\bibitem[{{Nakano} {et~al.}(2002){Nakano}, {Nishi}, \&
  {Umebayashi}}]{Nakano2002}
{Nakano}, T., {Nishi}, R., \& {Umebayashi}, T. 2002, \apj, 573, 199

\bibitem[{{Nelson} {et~al.}(2013){Nelson}, {Gressel}, \&
  {Umurhan}}]{Nelson2013}
{Nelson}, R.~P., {Gressel}, O., \& {Umurhan}, O.~M. 2013, \mnras, 435, 2610

\bibitem[{{Noble} {et~al.}(2010){Noble}, {Krolik}, \& {Hawley}}]{Noble2010}
{Noble}, S.~C., {Krolik}, J.~H., \& {Hawley}, J.~F. 2010, \apj, 711, 959

\bibitem[{{Okuzumi}(2009)}]{Okuzumi2009}
{Okuzumi}, S. 2009, \apj, 698, 1122

\bibitem[{{Okuzumi} \& {Hirose}(2011)}]{OkuzumiHirose2011}
{Okuzumi}, S., \& {Hirose}, S. 2011, \apj, 742, 65

\bibitem[{{Ono} {et~al.}(2016){Ono}, {Muto}, {Takeuchi}, \& {Nomura}}]{Ono2016}
{Ono}, T., {Muto}, T., {Takeuchi}, T., \& {Nomura}, H. 2016, \apj, 823, 84

\bibitem[{{Ono} {et~al.}(2018){Ono}, {Muto}, {Tomida}, \& {Zhu}}]{Ono2018}
{Ono}, T., {Muto}, T., {Tomida}, K., \& {Zhu}, Z. 2018, \apj, 864, 70

\bibitem[{{Pelletier} \& {Pudritz}(1992)}]{PelletierPudritz1992}
{Pelletier}, G., \& {Pudritz}, R.~E. 1992, \apj, 394, 117

\bibitem[{{Perez-Becker} \& {Chiang}(2011)}]{Perez-BeckerChian2011}
{Perez-Becker}, D., \& {Chiang}, E. 2011, \apj, 735, 8

\bibitem[{{Pinilla} {et~al.}(2016){Pinilla}, {Flock}, {Ovelar}, \&
  {Birnstiel}}]{Pinilla2016}
{Pinilla}, P., {Flock}, M., {Ovelar}, M. d.~J., \& {Birnstiel}, T. 2016, \aap,
  596, A81

\bibitem[{{Pucci} {et~al.}(2021){Pucci}, {Tomida}, {Stone}, {Takasao}, {Ji}, \&
  {Okamura}}]{Pucci2021}
{Pucci}, F., {Tomida}, K., {Stone}, J., {et~al.} 2021, \apj, 907, 13

\bibitem[{{Riols} \& {Lesur}(2018)}]{RiolsLesur2018}
{Riols}, A., \& {Lesur}, G. 2018, \aap, 617, A117

\bibitem[{{Riols} \& {Lesur}(2019)}]{RoilsLesur2019}
---. 2019, \aap, 625, A108

\bibitem[{{Salmeron} \& {Wardle}(2003)}]{SalmeronWardle2003}
{Salmeron}, R., \& {Wardle}, M. 2003, \mnras, 345, 992

\bibitem[{{Salvesen} {et~al.}(2016){Salvesen}, {Simon}, {Armitage}, \&
  {Begelman}}]{Salvesen2016}
{Salvesen}, G., {Simon}, J.~B., {Armitage}, P.~J., \& {Begelman}, M.~C. 2016,
  \mnras, 457, 857

\bibitem[{{Sano} \& {Inutsuka}(2001)}]{SanoInutsuka2001}
{Sano}, T., \& {Inutsuka}, S. 2001, \apjl, 561, L179

\bibitem[{{Sano} {et~al.}(2004){Sano}, {Inutsuka}, {Turner}, \&
  {Stone}}]{Sano2004}
{Sano}, T., {Inutsuka}, S., {Turner}, N.~J., \& {Stone}, J.~M. 2004, \apj, 605,
  321

\bibitem[{{Sano} \& {Miyama}(1999)}]{Sano1999}
{Sano}, T., \& {Miyama}, S.~M. 1999, \apj, 515, 776

\bibitem[{{Scepi} {et~al.}(2018){Scepi}, {Lesur}, {Dubus}, \&
  {Flock}}]{Scepi2018A&A...620A..49S}
{Scepi}, N., {Lesur}, G., {Dubus}, G., \& {Flock}, M. 2018, \aap, 620, A49

\bibitem[{{Shakura} \& {Sunyaev}(1973)}]{ShakuraSunyaev1973}
{Shakura}, N.~I., \& {Sunyaev}, R.~A. 1973, \aap, 24, 337

\bibitem[{{Sorathia} {et~al.}(2012){Sorathia}, {Reynolds}, {Stone}, \&
  {Beckwith}}]{Sorathia2012}
{Sorathia}, K.~A., {Reynolds}, C.~S., {Stone}, J.~M., \& {Beckwith}, K. 2012,
  \apj, 749, 189

\bibitem[{{Steinacker} \& {Papaloizou}(2002)}]{Steinacker2002}
{Steinacker}, A., \& {Papaloizou}, J. C.~B. 2002, \apj, 571, 413

\bibitem[{{Stone} {et~al.}(2008){Stone}, {Gardiner}, {Teuben}, {Hawley}, \&
  {Simon}}]{Stone2008}
{Stone}, J.~M., {Gardiner}, T.~A., {Teuben}, P., {Hawley}, J.~F., \& {Simon},
  J.~B. 2008, \apjs, 178, 137

\bibitem[{{Stone} {et~al.}(2020){Stone}, {Tomida}, {White}, \&
  {Felker}}]{Stone2020}
{Stone}, J.~M., {Tomida}, K., {White}, C.~J., \& {Felker}, K.~G. 2020, \apjs,
  249, 4

\bibitem[{{Suriano} {et~al.}(2018){Suriano}, {Li}, {Krasnopolsky}, \&
  {Shang}}]{Suriano2018}
{Suriano}, S.~S., {Li}, Z.-Y., {Krasnopolsky}, R., \& {Shang}, H. 2018, \mnras,
  477, 1239

\bibitem[{{Suriano} {et~al.}(2019){Suriano}, {Li}, {Krasnopolsky}, {Suzuki}, \&
  {Shang}}]{Suriano2019}
{Suriano}, S.~S., {Li}, Z.-Y., {Krasnopolsky}, R., {Suzuki}, T.~K., \& {Shang},
  H. 2019, \mnras, 484, 107

\bibitem[{{Suzuki}(2023)}]{Suzuki2023}
{Suzuki}, T.~K. 2023, \apj, 957, 99

\bibitem[{{Suzuki} \& {Inutsuka}(2009)}]{Suzuki2009}
{Suzuki}, T.~K., \& {Inutsuka}, S. 2009, \apjl, 691, L49

\bibitem[{{Suzuki} \& {Inutsuka}(2014)}]{Suzuki2014}
---. 2014, \apj, 784, 121

\bibitem[{{Suzuki} {et~al.}(2010){Suzuki}, {Muto}, \& {Inutsuka}}]{Suzuki2010}
{Suzuki}, T.~K., {Muto}, T., \& {Inutsuka}, S.-i. 2010, \apj, 718, 1289

\bibitem[{{Takasao} {et~al.}(2018){Takasao}, {Tomida}, {Iwasaki}, \&
  {Suzuki}}]{Takasao2018}
{Takasao}, S., {Tomida}, K., {Iwasaki}, K., \& {Suzuki}, T.~K. 2018, \apj, 857,
  4

\bibitem[{{Takasao} {et~al.}(2019){Takasao}, {Tomida}, {Iwasaki}, \&
  {Suzuki}}]{Takasao2019}
---. 2019, \apjl, 878, L10

\bibitem[{{Turner} \& {Sano}(2008)}]{TurnerSano2008}
{Turner}, N.~J., \& {Sano}, T. 2008, \apjl, 679, L131

\bibitem[{{Ueda} {et~al.}(2019){Ueda}, {Flock}, \& {Okuzumi}}]{Ueda2019}
{Ueda}, T., {Flock}, M., \& {Okuzumi}, S. 2019, \apj, 871, 10

\bibitem[{{Umebayashi} \& {Nakano}(1981)}]{Umebayashi1981}
{Umebayashi}, T., \& {Nakano}, T. 1981, \pasj, 33, 617

\bibitem[{{Umurhan} {et~al.}(2016){Umurhan}, {Nelson}, \&
  {Gressel}}]{UmurhanNelsonGressel2016}
{Umurhan}, O.~M., {Nelson}, R.~P., \& {Gressel}, O. 2016, \aap, 586, A33

\bibitem[{{Urpin} \& {Brandenburg}(1998)}]{UrpinBrandenburg1998}
{Urpin}, V., \& {Brandenburg}, A. 1998, \mnras, 294, 399

\bibitem[{{van der Marel}(2023)}]{vanderMarel2023}
{van der Marel}, N. 2023, European Physical Journal Plus, 138, 225

\bibitem[{{Velikhov}(1959)}]{Velikhov1959}
{Velikhov}, E.~P. 1959, Sov. Phys. J. Exp. Theor. Phys., 36, 1959

\bibitem[{{Wada} {et~al.}(2013){Wada}, {Tanaka}, {Okuzumi}, {Kobayashi},
  {Suyama}, {Kimura}, \& {Yamamoto}}]{Wada2013}
{Wada}, K., {Tanaka}, H., {Okuzumi}, S., {et~al.} 2013, \aap, 559, A62

\bibitem[{{Wardle}(1999)}]{Wardle1999}
{Wardle}, M. 1999, \mnras, 307, 849

\bibitem[{{Xu} \& {Bai}(2016)}]{XuBai2016}
{Xu}, R., \& {Bai}, X.-N. 2016, \apj, 819, 68

\bibitem[{{Zanni} {et~al.}(2007){Zanni}, {Ferrari}, {Rosner}, {Bodo}, \&
  {Massaglia}}]{Zanni2007}
{Zanni}, C., {Ferrari}, A., {Rosner}, R., {Bodo}, G., \& {Massaglia}, S. 2007,
  \aap, 469, 811

\bibitem[{{Zhu} \& {Stone}(2018)}]{ZhuStone2018}
{Zhu}, Z., \& {Stone}, J.~M. 2018, \apj, 857, 34

\end{thebibliography}

\end{document}